\documentclass{statsoc}
\usepackage{amsmath}
\usepackage{amssymb}
\usepackage{graphicx}
\usepackage{mathbbol}
\usepackage{animate}
\usepackage{url}
\usepackage{multirow}	
\usepackage{booktabs}	
\usepackage{algpseudocode}
\usepackage{setspace}
\usepackage{xr}
\usepackage{color} 

\usepackage{dsfont}		
\usepackage{booktabs}	

\makeatletter
\renewcommand{\@biblabel}[1]{\quad#1.}
\makeatother

\DeclareMathOperator*{\argmax}{argmax}
\DeclareMathOperator*{\Ind}{\mathds{1}}
\DeclareMathOperator*{\atanh}{atanh}
\DeclareMathOperator*{\Link}{\mathcal{L}}
\DeclareMathOperator*{\determinant}{det}
\DeclareMathOperator*{\MAP}{MAP}
\DeclareMathOperator*{\MLE}{MLE}

\DeclareMathOperator*{\KL}{KL}
\DeclareMathOperator*{\abc}{a\!b\!c}

\newcommand{\R}{\mathbb{R}}

\newcommand{\Ro}{R_0}
\newcommand{\birth}{\mu}
\providecommand{\trans}[1][]{{\beta_{#1}}}
\providecommand{\transtwo}[2]{{\beta^{#2}_{#1}}}

\providecommand{\fsu}[1][n]{s_k^{1:{{#1}_k}}}

\providecommand{\nuo}[1][]{\nu_{x#1}}
\providecommand{\mcnuo}[1][]{\hat{\nu}_{x#1}}
\providecommand{\mle}[1][]{\hat{\theta}^{\MLE}_{#1}}

\providecommand{\map}[1][]{\hat{\theta}^{\MAP}_{#1}}

\newcommand{\STP}{STP$_3$}
\newcommand{\dincub}{\phi}
\newcommand{\wane}{\gamma}
\newcommand{\di}{\nu}

\newcommand{\repR}{\omega}

\newcommand{\seas}{\varphi}
\newcommand{\trav}{m}

\newcommand{\snk}{{\scriptscriptstyle\downarrow}}
\newcommand{\src}{{\scriptscriptstyle\circlearrowright}}
\newcommand{\lkl}{\ell}
\newcommand{\ithresh}{\tau}
\newcommand{\ABCthresh}{c}

\newcommand{\rejABC}{ABC-r}
\newcommand{\nABC}{ABC$^\star$}
\newcommand{\rejnABC}{\nABC-r}

\newcommand{\figdir}{figs}

\newcommand{\sXATT}{aINC}
\newcommand{\sMATT}{aILI}
\newcommand{\sVATT}{fdILI}

\newcounter{assumptioncounter}
\newcommand{\assumption}{		\addtocounter{assumptioncounter}{1}A\arabic{assumptioncounter}		}

\newif\ifallstuff

\providecommand{\apostr}[1]{``#1"}

\providecommand{\abs}[1]{\lvert#1\rvert}	

\newenvironment{singlespaceddescription}{
\begin{description}
  \setlength{\itemsep}{1pt}
  \setlength{\parskip}{0pt}
  \setlength{\parsep}{0pt}
}{\end{description}}

\newtheorem{theorem}{Theorem}

\title[ABC based on modelling summary values]{Statistical modelling of summary values\\ leads to accurate Approximate Bayesian Computations}
\author[]{Oliver Ratmann}
\address{Department of Infectious Disease Epidemiology, Imperial College London, Norfolk Place, London W2 1PG, UK. oliver.ratmann@imperial.ac.uk}
\author[]{Anton Camacho}
\address{London School of Hygiene and Tropical Medicine, Keppel Street, London WC1E 7HT, UK}
\author[]{Adam Meijer}
\address{RIVM, National Institute for Public Health and the Environment, Centre for Infectious Disease Control, Bilthoven, The Netherlands}
\author[Ratmann O, Camacho A, Meijer A, Donker G]{G\'e Donker}
\address{NIVEL, Netherlands Institute for Health Services Research, P.O.Box 1568, 3500 BN Utrecht, The Netherlands}

\begin{document}
\begin{abstract} 
Approximate Bayesian Computation (ABC) methods rely on asymptotic arguments, implying that parameter inference can be systematically biased even when sufficient statistics are available.
We propose to construct the ABC accept/reject step from decision theoretic arguments on a suitable auxiliary space. This framework, referred to as \nABC, fully specifies which test statistics to use, how to combine them, how to set the tolerances and how long to simulate in order to obtain accuracy properties on the auxiliary space. Akin to maximum-likelihood indirect inference, regularity conditions establish when the \nABC\ approximation to the posterior density is accurate on the original parameter space in terms of the Kullback-Leibler divergence and the maximum a posteriori point estimate. Fundamentally, escaping asymptotic arguments requires knowledge of the distribution of test statistics, which we obtain through modelling the distribution of summary values, data points on a summary level. Synthetic examples and an application to time series data of influenza A (H3N2) infections in the Netherlands illustrate \nABC\ in action. 
\end{abstract}
%




\section{Introduction}
Approximate Bayesian Computation (ABC) methods have become a popular technique throughout the applied sciences because they
can handle posterior densities
\begin{equation}\label{e:exactposterior}
\pi(\theta|x) \propto \lkl(x|\theta) \pi(\theta)
\end{equation}
that arise under high-dimensional stochastic processes with intractable likelihoods. ABC methods circumvent computations of the likelihood $\lkl(x|\theta)$ by comparing the observed data $x$ to simulated data $y$ in terms of many, lower-dimensional summary statistics $S_k$. Observed and simulated summary statistics are compared with distances $d_k(S_k(y), S_{k}(x))= S_k(y) - S_{k}(x)$, $k=1,\dotsc,K_S$ and then combined and weighted either under a Mahalanobis approach \citep{Beaumont2002}
\begin{subequations}\label{e:MahaOrIntersect}
	\begin{align}	
	&\Ind\Big\{\:\:\abs{(d_1,\dotsc,d_{K_S})\Sigma^{-1}(d_1,\dotsc,d_{K_S})^t}\leq \ABCthresh\:\:\Big\}\label{e:Mahalanobis}\\
	\intertext{or under an intersection approach \citep{Pritchard1999}}
	&\prod_{k=1}^{K_S} \Ind\Big\{\:\:\ABCthresh^-_k\leq d_k\leq \ABCthresh^+_k\:\:\Big\}\label{e:intersection}
	\end{align}	
\end{subequations}
with a user-defined tolerance $c\geq 0$ (or $c_k^-\leq 0$,$c_k^+\geq 0$ for all $k$). This $0$/$1$ weight replaces the likelihood term in Monte Carlo algorithms as in the following simple rejection sampler based on \eqref{e:intersection}:\\[-2mm]

\hspace{-0.52cm}{\bf standard ABC rejection sampler (\rejABC)}\\[-5mm]
\begin{algorithmic}[1]
\For{$i = 1 \dotsc N$}
	\State Propose $\theta^\prime\sim\pi$ and simulate $y^\prime\sim \lkl(\,\cdot\,|\,\theta^\prime)$.
	\State Compute summary statistics $S_k(y^\prime)$, $k=1,\dotsc,K_S$.
	\State Compute summary errors $z^\prime_k=d_k\big(S_k(y^\prime),S_k(x)\big)$ for all $k$.
	\State Accept $(\theta^\prime,z^\prime)$ if for all $k\quad \ABCthresh^-_k\leq {z_k}^\prime\leq\ABCthresh^+_k$. Go to line 2.		
\EndFor\vspace{2mm}
\end{algorithmic}

The ABC likelihood approximation based on the above intersection approach is 
\begin{equation}\label{e:abclikelihood}
\lkl(x|\theta)\approx\:\lkl_{\abc}(x|\theta)=\:\int \prod_{k=1}^{K_S} \Ind\Big\{\:\:\ABCthresh^-_k\leq d_k\big(S_k(y),S_k(x)\big) \leq \ABCthresh^+_k\:\:\Big\} \: \lkl(dy|\theta).
\end{equation}
Even if the summary statistics are sufficient for $\theta$, no ABC theory exists to establish the accuracy of $\pi_{\abc}(\theta|x)\propto \lkl_{\abc}(x|\theta)\pi(\theta)$ in terms of point estimates (such as the maximum a posteriori (MAP) estimate) or distributional properties (such as the Kullback-Leibler (KL) divergence) in the classical, computationally feasible setting that $c>0$ or $\ABCthresh^-_k<\ABCthresh^+_k$. All available ABC methods invoke asymptotic arguments that involve either the $\ABCthresh^-_k<\ABCthresh^+_k$ or the number of data points $n$. This has lead to the notion that ABC is noisy \citep{Fearnhead2012}, despite notable recent advances \citep{Grelaud2009, Dean2011, Drovandi2011, Barnes2012, Filippi2013, Marin2013}.

\begin{table}
	\caption{\label{t:uabc} Accuracy of parameter estimates with standard ABC and \nABC}	
	\hspace{-0.9cm}
	\begin{tabular}{ll@{\hskip 0.05in}l@{\hskip 0.05in}l@{\hskip 0.05in}ll@{\hskip 0.05in}l@{\hskip 0.05in}l@{\hskip 0.05in}l@{\hskip 0.05in}l}										
				\multicolumn{10}{l}{}\\[0mm]
				\multicolumn{10}{l}{Inference with standard ABC on simulated pseudo data}\\[1mm]
				\toprule				
										&\multicolumn{4}{l}{Normal variance}					&\multicolumn{4}{l}{Moving average}				&Influenza time series\\[1mm]
				$d_k$					&\multicolumn{4}{l}{$\text{var}(y)-\text{var}(x)$}				&\multicolumn{4}{l}{$\text{var}(y)-\text{var}(x),$}		&log ratio of sample means$^\P$\\[0mm]
										&\multicolumn{4}{l}{}									&\multicolumn{4}{l}{$\text{cor}(y)-\text{cor}(x)$}			&\\[0mm]
				$n=m$					&\multicolumn{4}{l}{$60$}								&\multicolumn{4}{l}{$(150, 149)$}					&$(15, 14, 15)$\\[0mm]
				$\ABCthresh^+$, $\ABCthresh^-=-\ABCthresh^+$	&$0.8$ &$0.4$	&$0.2$ &$0.05$&$0.89$	&$0.71$ &$0.59$ &$0.39$ 						&$(0.35, 0.35, 0.03)$\\[0mm]
										&&&&									&$0.18$	&$0.13$ &$0.09$ &$0.03$ 						&\\[0mm]
				\midrule
				$\theta_0$				&\multicolumn{4}{l}{$1$}								&\multicolumn{4}{l}{(1, 0.1)}						&$(3.5, 10, 0.08)$\\[1mm]
				MAP m.~s.~e.				&0.202 		&0.057 	&0.012 	&0.002$^\dagger$		& 0.27 & 0.014 & 0.019 & 0.017$^\dagger$ 			&1.88$^{\ddagger,\Box}$\\[0mm]					
				KL divergence				& 4.84 		& 0.41 & 0.11 	& 0.007$^\dagger$			& 2.76 & 0.31 & 0.15 & 0.07$^\dagger$ 				& -	\\[0mm]					
				mean acc \%				& 43 			& 22 		& 11 		& 3$^\dagger$			& 20 & 10 & 5 & 0.5$^\dagger$ 					& 5$^\ddagger$	\\[1mm]				
				\bottomrule				
				\multicolumn{10}{l}{}\\[2mm]
				\multicolumn{10}{l}{Inference with \nABC\ on simulated pseudo data}\\[1mm]
				\toprule				
										&\multicolumn{4}{l}{Normal variance}			&\multicolumn{4}{l}{Moving average}						&Influenza time series	\\[1mm]
				$T_k$					&\multicolumn{4}{l}{$\chi^2$}					&\multicolumn{4}{l}{$\chi^2$, Z	$^\clubsuit$}					&TOST test statistics$^\spadesuit$		\\[0mm]
				$n$						&\multicolumn{4}{l}{$60$}						&\multicolumn{4}{l}{$(150,149)$}							&$(8,7,7,6,8,7)$			\\[0mm]
				$m$						&\multicolumn{4}{l}{$108$}					&\multicolumn{4}{l}{$(297,278)$}							&variable, e.g. Figure~\ref{f:SEIIRS_simCali}A\\[0mm]	 		
				$\ithresh^-$				&\multicolumn{4}{l}{$0.589$}					&\multicolumn{4}{l}{$(0.705, -0.239)$}						&$-\ithresh^+$			\\[0mm]
				$\ithresh^+$				&\multicolumn{4}{l}{$1.752$}					&\multicolumn{4}{l}{$(1.432, 0.239)$}						&variable, e.g. Figure~\ref{f:SEIIRS_simCali}B\\[2mm]
				\midrule
				$\theta_0$				&\multicolumn{4}{l}{1}						&\multicolumn{4}{l}{(1, 0.1)}								&(3.5, 10, 0.08)			\\[1mm]
				MAP m.~s.~e.				&\multicolumn{4}{l}{0.002$^\dagger$}			&\multicolumn{4}{l}{0.006$^\dagger$}						&0.071$^{\ddagger,\Box}$		\\[0mm]		
				KL divergence				&\multicolumn{4}{l}{0.007$^\dagger$}			&\multicolumn{4}{l}{0.07$^\dagger$}						&-			\\[0mm]		
				mean acc \%				&\multicolumn{4}{l}{13$^\dagger$}				&\multicolumn{4}{l}{5$^\dagger$}							&14$^\ddagger$		\\[1mm]		
				\bottomrule
				\multicolumn{10}{l}{$^\dagger$Computed over 4000 replicate pseudo data sets $x$. $^\ddagger$Computed over 50 replicate pseudo data sets.}\\[-1mm]				
				\multicolumn{10}{l}{$^\P$The summary statistics are the samples means of the observed and simulated \sMATT, \sVATT\ and \sXATT.}\\[-1mm]									
				\multicolumn{10}{l}{$^\spadesuit$The summary values are the odd and even subsets of the \sMATT, \sVATT\ and \sXATT\ time series, and one }\\[-1mm]	
				\multicolumn{10}{l}{test was used for each set of observed and simulated summary values. $^\clubsuit$For the configuration that}\\[-1mm]	
				\multicolumn{10}{l}{ignores correlations in $\fsu$. $^\Box$Exact posterior density not available. Here, the mean squared error was}\\[-1mm]
				\multicolumn{10}{l}{computed against $\theta_0$ instead of $\map$.}								 						
	\end{tabular}		
\end{table}

\paragraph{Example.} Suppose that $x^{1:n}=(x_1,\dotsc,x_n)$ are identically and independently $\mathcal{N}(0,\sigma^2)$ distributed, with $n=60$ and unknown $\sigma^2$. The true value to be re-estimated is $\sigma^2_0=1$ and we have the prior density $\pi(\sigma^2)\propto\Ind\{0.2<\sigma^2<4\}$ so that a simple rejection sampler will suffice. The MAP estimate of the exact posterior density is $\map=\frac{1}{n} S^2(x)$, where $S^2(x)$ denotes the sum of squares. To employ \rejABC, we decide to simulate data $y^{1:m}\sim\mathcal{N}(0,\sigma^2)$,  with $m=60$ set arbitrarily, compute $S^2(y^{1:m})$, consider (arbitrarily) the difference $z=\frac{1}{m-1}S^2(y)-\frac{1}{n-1}S^2(x)$ as well as (again arbitrarily) the tolerances $c^+=0.4$ and $c^-=-c^+$.
To evaluate the accuracy of the standard ABC approximation, we also considered $c^+=0.8,0.2,0.05$ and ran \rejABC\ on 4000 pseudo data sets $x$. 
The KL divergence of $\pi_{\abc}(\theta|x)$ to $\pi(\theta|x)$ and the mean square error between the ABC MAP estimate $\map[\abc]$ and $\map$ were initially very large and approached zero as predicted \citep{Beaumont2002}, with decreasing tolerances at decreasing acceptance probability (top left in Table~\ref{t:uabc} and Figure~\ref{f:scaledFpower}D below). 

We present a statistical framework that fully specifies the free ABC parameters so that the resulting ABC approximation is accurate in that $\map=\map[\abc]$ and that the KL divergence of $\pi_{\abc}(\theta|x)$ to $\pi(\theta|x)$ is very small. We define the ABC accept/reject step as a statistical decision test with well-understood frequency properties on an empirically validated auxiliary probability space, and then calibrate the free ABC parameters before ABC is run such that the above two approximation criteria are met. This basic procedure is called \nABC. Our paper is structured as follows. In Section~\ref{s:suvalues}, we suggest to construct a suitable auxiliary probability space from independent and approximately sufficient \apostr{summary values} whose distribution is relatively simple. These summary values must be identified either by the user or through computational methods and form the interface between the rigorous and automated part of \nABC\ with a broad range of inference problems. Section~\ref{s:equivstat} develops our notation and the \nABC\ approximation on the auxiliary space. Sections~\ref{s:unitest}-\ref{s:multitest} present an equivalence hypothesis testing framework under which the ABC approximation emerges as the power function of particular hypothesis test statistics $T$, first for the univariate and then for the multivariate case. We then demonstrate in Section~\ref{s:calibrations} that the power function can be matched to the likelihood of the summary values on the auxiliary space by calibrating all free ABC parameters. The appropriate test statistics and calibrations can be analytically derived for relatively simple auxiliary probability models on the distribution of the summary values, which explains part of the required regularity conditions. The remaining conditions ensure that indirect inference on the auxiliary space is appropriate for estimating the original parameters of interest, as we outline in Section~\ref{s:indirectABC}. In particular, we identify that the link function between the auxiliary and original parameter space must be bijective and sufficiently regular for \nABC\ inference to be accurate, and demonstrate how this can be verified from Monte Carlo output. Finally, \nABC\ inference is illustrated at synthetic examples, and its broader applicability is demonstrated on an influenza A (H3N2) virus transmission model.

\section{Accurate \nABC\ parameter inference}

\subsection{Modelling summary values to specify the ABC auxiliary space}\label{s:suvalues}
We construct a theoretical framework for accurate ABC that is based on modelling distributions of \apostr{summary values}, data points on a summary level. We consider multiple sets of summary values, $k= 1,\dotsc,K$, and denote each observed set by $\fsu= (s^1_k,\dotsc,s^{n_k}_k)$ and each simulated set by $\fsu[m]$. The $k^\prime=1,\dotsc,K_S$ summary statistics in standard ABC are a function thereof, for example $S_{k^\prime}(\fsu)=\frac{1}{n_k}\sum_{i=1}^{n_k}s^i_k$ and $S_{k^\prime+1}(\fsu)=\frac{1}{n-1}\sum_{i=1}^n(s^i_k-\bar{s}_k)^2$ (Figure~\ref{f:notation}). The number of summary values can differ for each $k$ and between observation and simulation ($n_k\neq m_k$) but is strictly $>1$. We require that summary values can be found such that 
\begin{singlespaceddescription}
\item[\assumption] they are sufficient for $\theta$ and
\item[\assumption] their distribution can be modelled so that test statistics $T_k$ exist and can be calibrated (A2.1-A2.4 below).
\end{singlespaceddescription}
Such sets of summary values can be identified in practice as we illustrate in Section~\ref{s:applications}, although not necessarily in all previous applications of standard ABC. A2 holds, for example, when both the $\fsu$ and $\fsu[m]$ follow a normal, lognormal, $\chi^2$ or similar simple distribution. A2 also validates the use of a particular auxiliary probability model based on the available data. The simplest algorithm we consider is:\\[-2mm] 

\hspace{-0.52cm}{\bf calibrated \nABC\ rejection sampler (\rejnABC)}\\[-5mm]
\begin{algorithmic}[1]
\For{$i = 1 \dotsc N$}
	\State Propose $\theta^\prime\sim\pi$ and simulate data $y^\prime\sim \lkl(\,\cdot\,|\,\theta^\prime)$..
	\State Extract summary values $\fsu[m](y^\prime)$ from $y^\prime$ for all $k=1,\dotsc,{K}$.
	\State Compute $z^\prime_k=T_k\big(\fsu[m](y^\prime),\fsu(x)\big)$ for all $k$.
	\State Accept $(\theta^\prime,z^\prime)$ if for all $k\quad \ABCthresh^-_k\leq {z_k}^\prime\leq\ABCthresh^+_k$. Go to line 2.
\EndFor\vspace{2mm}
\end{algorithmic}

$\hspace{-5mm}$where the $T_k$'s are {\it specific} hypothesis test statistics, and the $\ABCthresh^-_k$, $\ABCthresh^+_k$ are the boundary points of the critical region of these hypothesis tests. The distribution of the summary values determines which $T_k$ is to be used. The $T_k$'s implicitly specify the summary statistics in standard ABC, and generally $K_S\geq K$. More complicated Monte Carlo samplers for \nABC\ are detailed in the supplementary online material (SOM), section~\ref{a:mcmc}. 

\subsection{ABC approximation on the auxiliary space}\label{s:equivstat}
\begin{figure}[t]
	\centering	
	\begin{minipage}[t]{\textwidth}		
		{\bf A}\hspace{9cm}{\bf B}
		\begin{center}
			\includegraphics[type=pdf,ext=.pdf,read=.pdf,width=\textwidth]{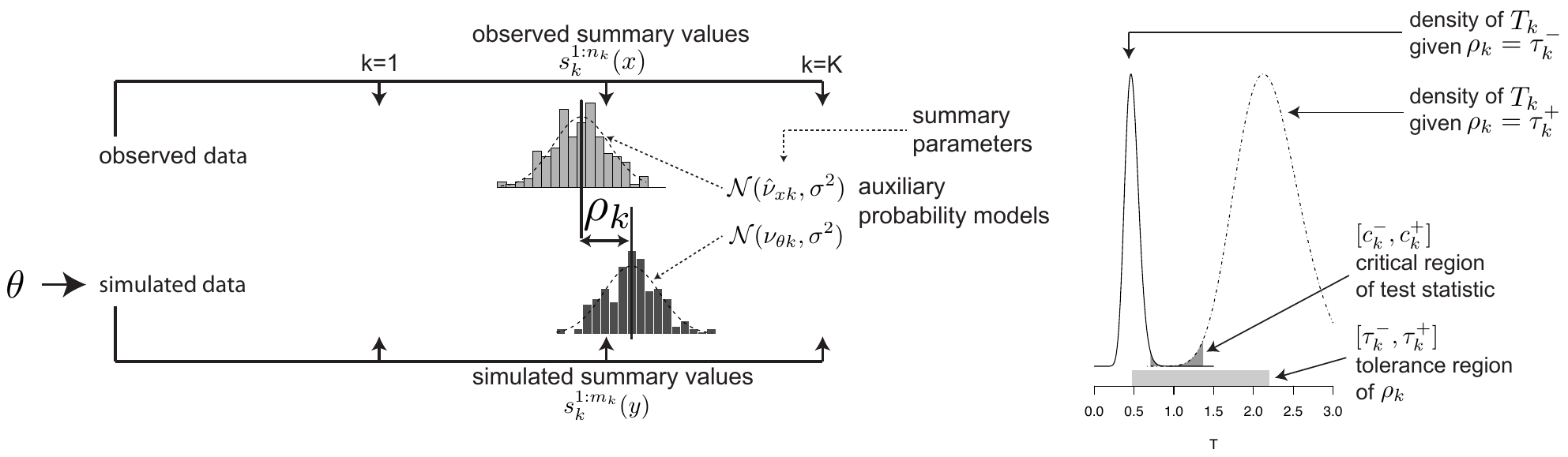}			
		\end{center}
	\end{minipage}	
	\caption{
{\bf Notation for \nABC\ parameter inference.}
(A) Notation associated with the auxiliary probability models. 
In the $\mathcal{N}(\nu_{\theta k},\sigma^2_{\theta k})$ basic case, $\rho_k$ corresponds to the difference in the population means $\nu_{\theta k}-\nuo[k]$ and $\rho^\star_k$ equals zero. (B) Notation to construct equivalence tests, illustrated with the probability densities of the equivalence test statistic $T$, $[\ithresh^-,\ithresh^+]$ and $[\ABCthresh^-,\ABCthresh^+]$ in the $\mathcal{N}(0,\sigma^2)$ example. In this example, $\rho_k$ corresponds to the ratio $\nu_{\theta k}/\mcnuo[k]$ and $\rho^\star_k$ equals one.
}\label{f:notation}
\end{figure}

Figure~\ref{f:notation}A illustrates our notation to write the \nABC\ accept/reject step as a decision problem on the summary space. As the most basic case of A2, we consider $\fsu[m](y)\sim\mathcal{N}(\nu_{\theta k},\sigma^2_{\theta k})$, $\fsu[n](x)\sim\mathcal{N}(\nuo[k],\sigma^2_{x k})$. We are interested in testing if the unknown $\nu_{\theta k}$ and $\nuo[k]$ are equal or similar. We refer to the scalar parameters that are subject to testing as the \apostr{summary parameters} $\nu_k$, potentially in the presence of \apostr{hidden parameters} such as the $\sigma^2$'s above. The summary parameters of the simulated data are denoted by $\nu_{\theta k}$ to reflect their dependence on $\theta$. The summary parameters of the observed data are denoted by $\nuo[k]$. The $\nuo[k]$ are unknown. Since we consider the data fixed, these are replaced by maximum likelihood estimates $\mcnuo[k]$. The simulated and observed summary parameters are then compared with a distance function $\rho_k=\delta_k\big(\nu_{\theta k},\mcnuo[k]\big)$ that takes on values in a subset $\Delta_k$ of the real line. The point of equality is denoted by $\rho^\star_k=\delta_k\big(\nu,\nu\big)$. This setup defines the link function
\begin{equation}\label{e:linkfunction}
\Link:\Theta\subseteq\R^D\to\Delta\subseteq\R^{K},\quad\theta\to\rho,\quad \rho=(\rho_1,\dotsc,\rho_{K}).
\end{equation} 
Let us first suppose there are no hidden parameters. If A1-A2 hold and if 
\begin{singlespaceddescription}
\item[\assumption] the link function $\Link$ is bijective and continuously differentiable,
\end{singlespaceddescription}
then the exact posterior density can be written as
\begin{equation}\label{e:trafo}
\pi(\theta|x)\propto \lkl(\{\fsu(x), k=1,\dotsc,K\}|\rho) \:\:  \pi_\rho(\rho) \:\: \abs{\partial\!\Link(\theta)},
\end{equation}
where $\abs{\partial\!\Link(\theta)}$ is the absolute value of the determinant of the Jacobian of \eqref{e:linkfunction}, $\lkl(\{\fsu(x), k=1,\dotsc,K\}|\rho)$ is the likelihood of the summary values under the validated auxiliary probability models, and the prior density $\pi_\rho$ is induced from $\pi(\theta)$ through the change of variables. Now, the summary likelihood is known analytically but the $\abs{\partial\!\Link(\theta)}$ is a priori not known. If $\Link$ was known, then inference could proceed directly via \eqref{e:trafo} without recursion to the \nABC\ algorithm. 

Under the particular auxiliary probability models in A2, the \nABC\ approximation to \eqref{e:trafo} can be written as 
\begin{equation}\label{e:abctrafo}
\pi_{\abc}(\theta|x)\propto P_{x}\big(R\:|\:\rho\big) \: \pi_\rho\big(\rho\big) \: \abs{\partial\!\Link(\theta)},
\end{equation}
where either
\begin{subequations}\label{e:abclklrho}
	\begin{align}
	&P_{x}(R|\rho)=\int \Ind\Big\{|T\big(s^{1:m}_{1:K}, s^{1:n}_{1:K}(x)\big)|\leq c\Big\}\:F_{\rho}(ds^{1:m}_{1:K})\label{e:abclklrhoMAHA}\\
	\intertext{in analogy to the Mahalanobis approach \eqref{e:Mahalanobis} for a multivariate test statistic $T$ or}
	&P_{x}(R|\rho)=\int \prod_{k=1}^{K} \Ind\Big\{\ABCthresh_k^-\leq T_k(\fsu[m],\fsu(x))\leq\ABCthresh_k^+\Big\}\:F_{\rho}(ds^{1:m}_{1:K})\label{e:abclklrhoIS}
	\end{align}
\end{subequations}
in analogy to the intersection approach \eqref{e:intersection}. The subscript $x$ emphasises that the observed summary values are fixed. The integral evaluates with respect to the distribution of simulated summary values for given $\rho$ and approximates the summary likelihood. Our key observation is that integrals of the form \eqref{e:abclklrho} arise in hypothesis testing theory and are well understood if the distribution of the summary values is sufficiently regular \citep{Wellek2003}. 

Additional hidden parameters of an auxiliary probability model complicate equations (\ref{e:trafo}-\ref{e:abctrafo}) slightly. Following composite hypothesis testing theory \citep{Lehmann2005}, we typically condition $\lkl(\{\fsu(x), k=1,\dotsc,K\}|\rho)$ and $P_{x}\big(R\:|\:\rho\big)$ on additional statistics $C$.

\subsection{Univariate equivalence tests on the auxiliary space}\label{s:unitest}
From a Neyman-Pearson {\it equivalence testing} perspective, $\rho\to P_{x}(R|\rho)$ is the power in {\it rejecting} the multivariate test statistic $T=(T_1,\dotsc,T_{K})$ at an error $\rho$ (hence \apostr{R} for reject). We begin with the univariate case. Tests for the point null hypothesis
\begin{equation*}
\tilde{H}_{0k}\colon\nu_{\theta k}=\mcnuo[k]\quad\quad\text{versus}\quad\quad\tilde{H}_{1k}\colon \nu_{\theta k}\neq\mcnuo[k]
\end{equation*}
are designed to declare $\nu_{\theta k}$ and $\mcnuo[k]$ unequal. Because we aim to match the simulated and observed summary values, the \nABC\ objective is to declare $\nu_{\theta k}$ and $\mcnuo[k]$ equal. With this objective, \cite{Schuirmann1981} noted that the appropriate hypothesis testing framework is
\begin{subequations}\label{e:ihyp}
\begin{align}
&H_{0k}\colon\quad\rho_k=\delta_k\big(\nu_{\theta k},\mcnuo[k]\big)\notin[\ithresh^-_k,\ithresh^+_k]\quad\text{versus}\\
&H_{1k}\colon\quad\rho_k=\delta_k\big(\nu_{\theta k},\mcnuo[k]\big)\in[\ithresh^-_k,\ithresh^+_k],\label{e:ihyp1}
\end{align}
\end{subequations}
for an equivalence region $[\ithresh^-_k,\ithresh^+_k]$ that contains the point of equality $\rho^\star_k$. Hereafter, we call the $\ithresh^-_k$ and $\ithresh^+_k$ the \apostr{tolerances}. Note that the null and alternative are flipped compared to $\tilde{H}_{0k}$ and $\tilde{H}_{1k}$. This means that {\it rejecting} an equivalence test for a particular set of simulated summary values $\fsu[m](y^\prime)$ implies that the proposed $\theta^\prime$ in an ABC iteration is {\it accepted}. The distance function $\delta_k$ and test statistic $T_k$ are defined in such a way that the univariate $P_x(R_k|\rho_k)$ can be evaluated. This is a key difference to standard ABC where $d_k$ is chosen in an ad-hoc fashion. \cite{Wellek2003} describes many univariate equivalence tests for a variety of distributional assumptions on the summary values. These tests are adapted to the one-sample case because the data $x$ are considered fixed in \nABC. The tolerances in standard ABC are the critical region $[\ABCthresh^-_k, \ABCthresh^+_k]$  of equivalence test statistics, and for given $\alpha$ obtained as the simultaneous solution to the constraints
\begin{equation}\label{e:abcthresh}
P_x(\ABCthresh^-_k\leq T_k\leq \ABCthresh^+_k | \ithresh^-_k)= \alpha= P_x(\ABCthresh^-_k\leq T_k\leq \ABCthresh^+_k | \ithresh^+_k)
\end{equation}
for continuous $T_k$ (see also Figure~\ref{f:notation}B). 
The size of the univariate test, $\sup_{\rho_k\in H_{0k}}P_x(\ABCthresh^-_k\leq T_k\leq \ABCthresh^+_k\:|\:\rho_k)$, is then equal to $\alpha$ and limits the probability to falsely accept $\rho_k$ at any \nABC\ iteration. The power of the test gives the probability to correctly accept $\rho_k$ at any \nABC\ iteration.

\paragraph{Example.} Consider the basic $\mathcal{N}(0,\sigma^2)$ example where the single parameter $\sigma^2$ is to be estimated. We will therefore only need one test statistic ($K=1$, and we drop $k$). The $n=60$ data points act as the observed summary values, the link function is the identity, and we let $m\geq n$. Since the summary values are normally distributed, we choose the scaled $\chi^2$ test and set $\rho=\delta(\sigma^2,\hat{\sigma}_x^2)=\sigma^2/\hat{\sigma}_x^2\in (0,\infty)$ where $\hat{\sigma}^2_x$ is the maximum likelihood estimate $\frac{1}{n}S^2(x)$ \citep{Wellek2003}. The point of equality is $\rho^\star=1$. We consider $\ithresh^-\leq \rho^\star\leq\ithresh^+$ and test $H_0\colon\rho\notin[\ithresh^-,\ithresh^+]$ versus $H_1\colon\rho\in[\ithresh^-,\ithresh^+]$ with the test statistic
\begin{equation}\label{ex:scaledChi2}
T(y)=\frac{S^2(y)}{S^2(x)}=
\frac{\sigma^2}{\hat{\sigma}^2_x}\:
\frac{1}{n}\sum_{i=1}^m\Big(\frac{y_i-\bar{y}}{\sigma}\Big)^2.
\end{equation}
$\frac{n}{\rho}T$ follows a $\chi^2$-distribution with $m-1$ degrees of freedom, $F_{\chi^2_{m-1}}$. We set $\alpha=0.01$. The size-$\alpha$ \nABC\ acceptance region $[\ABCthresh^-,\ABCthresh^+]$ is the solution of the equations
\begin{equation}\label{ex:scaledF_rej}
\begin{split}
\alpha&=F_{\chi^2_{m-1}}(n\ABCthresh^+/\ithresh^+)-F_{\chi^2_{m-1}}(n\ABCthresh^-/\ithresh^+)\\
\alpha&=F_{\chi^2_{m-1}}(n\ABCthresh^+/\ithresh^-)-F_{\chi^2_{m-1}}(n\ABCthresh^-/\ithresh^-),
\end{split}
\end{equation}
as illustrated in Figure~\ref{f:notation}B. Having solved for $\ABCthresh^-$ and $\ABCthresh^+$, the power is
\begin{equation}\label{ex:scaledF_pw}
P_x(\ABCthresh^-\leq T\leq \ABCthresh^+|\rho)=F_{\chi^2_{m-1}}(n\ABCthresh^+/\rho)-F_{\chi^2_{m-1}}(n\ABCthresh^-/\rho).
\end{equation}
This is the probability to accept a single \nABC\ iteration, conditional on the underlying population level variance of the simulated summary values being a multiple $\rho$ of $\hat{\sigma}_x^2$. It is thus possible to quantify the behaviour of the \nABC\ accept/reject step exactly for the particular distance function and test statistic above.

\begin{figure}[p]
	\centering
	\begin{minipage}[t]{0.9\textwidth}		
	\begin{minipage}[t]{0.32\textwidth}		
			\begin{minipage}[b]{0.001\textwidth}
				{\bf A}\newline\vspace{-0.2cm}

			\end{minipage}	
			\begin{minipage}[b]{0.99\textwidth}
				\includegraphics[type=pdf,ext=.pdf,read=.pdf,width=\textwidth]{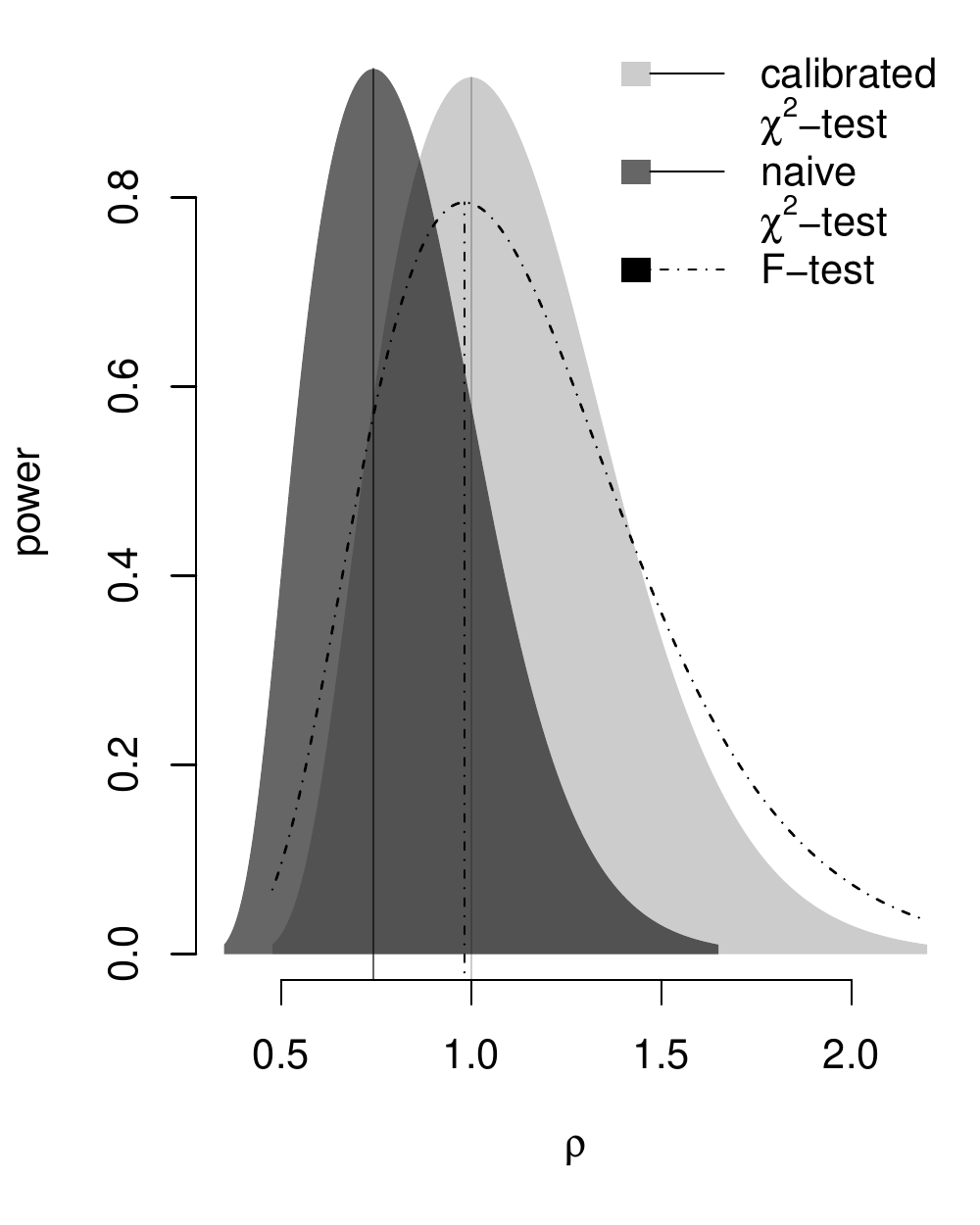}
			\end{minipage}	
	\end{minipage}			
	\begin{minipage}[t]{0.32\textwidth}		
			\begin{minipage}[b]{0.001\textwidth}
				{\bf B}\newline\vspace{-0.2cm}

			\end{minipage}	
			\begin{minipage}[b]{0.99\textwidth}
				\includegraphics[type=pdf,ext=.pdf,read=.pdf,width=\textwidth]{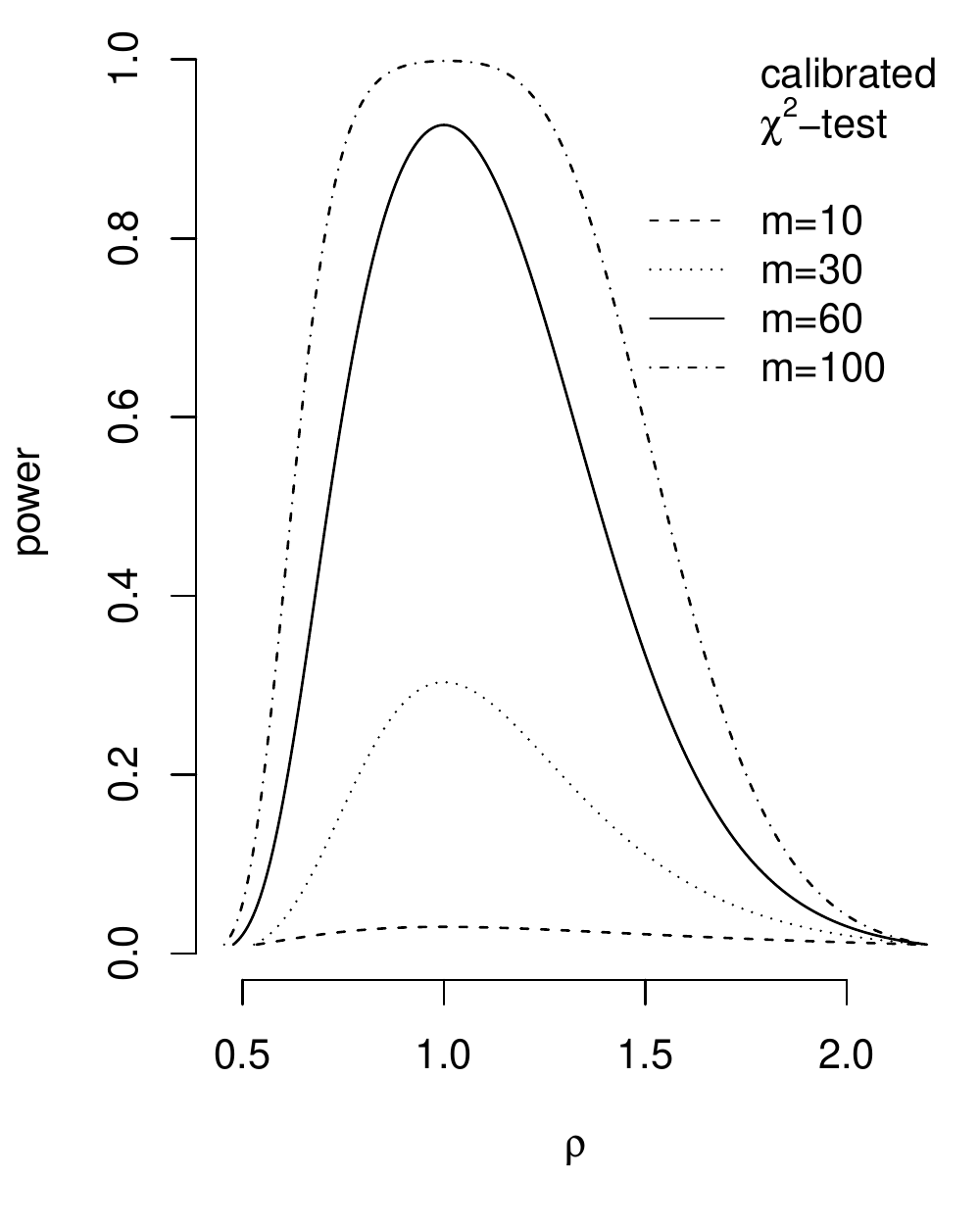}
		\end{minipage}	
	\end{minipage}	
	\begin{minipage}[t]{0.32\textwidth}		
			\begin{minipage}[b]{0.001\textwidth}
				{\bf C}\newline\vspace{-0.2cm}

			\end{minipage}	
			\begin{minipage}[b]{0.99\textwidth}
				\includegraphics[type=pdf,ext=.pdf,read=.pdf,width=\textwidth]{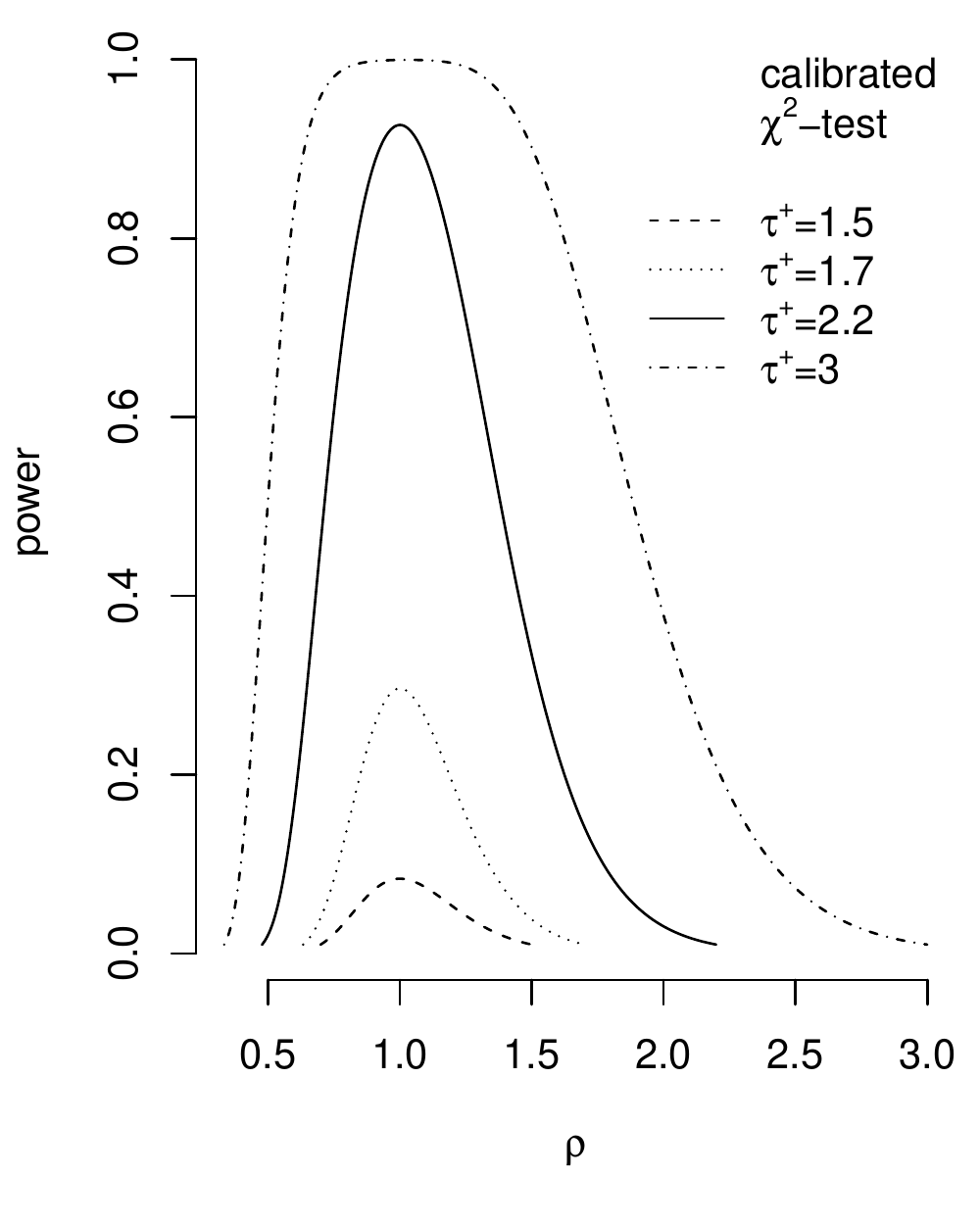}
		\end{minipage}	
	\end{minipage}	
	\begin{minipage}[t]{0.32\textwidth}
			\begin{minipage}[b]{0.001\textwidth}
				{\bf D}\newline\vspace{-0.4cm}

			\end{minipage}	
			\begin{minipage}[b]{0.99\textwidth}
				\includegraphics[type=pdf,ext=.pdf,read=.pdf,width=\textwidth]{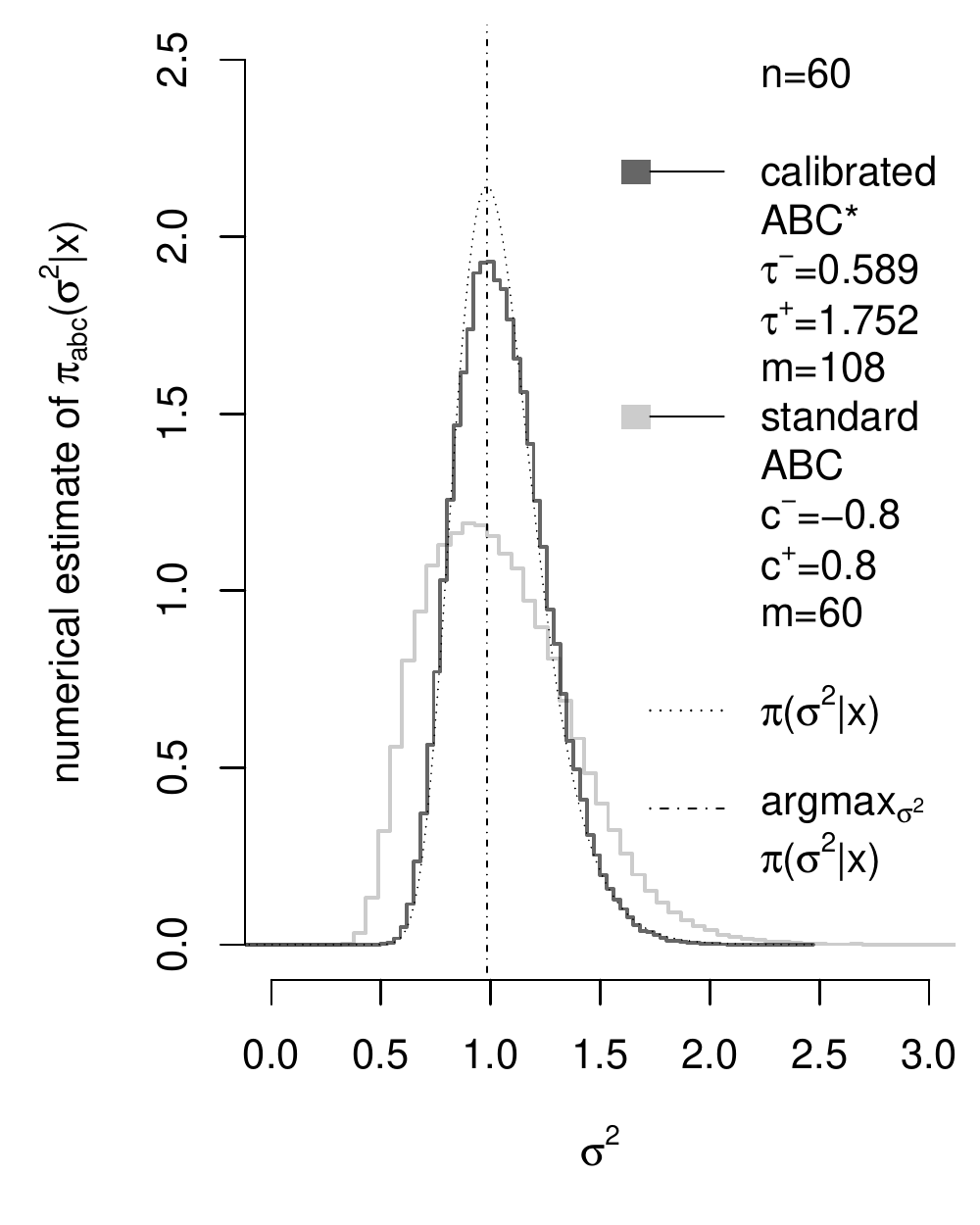}
			\end{minipage}						
	\end{minipage}	
	\begin{minipage}[t]{0.32\textwidth}		
			\begin{minipage}[b]{0.001\textwidth}
				{\bf E}\newline\vspace{-0.4cm}

			\end{minipage}	
			\begin{minipage}[b]{0.99\textwidth}
				\includegraphics[type=pdf,ext=.pdf,read=.pdf,width=\textwidth]{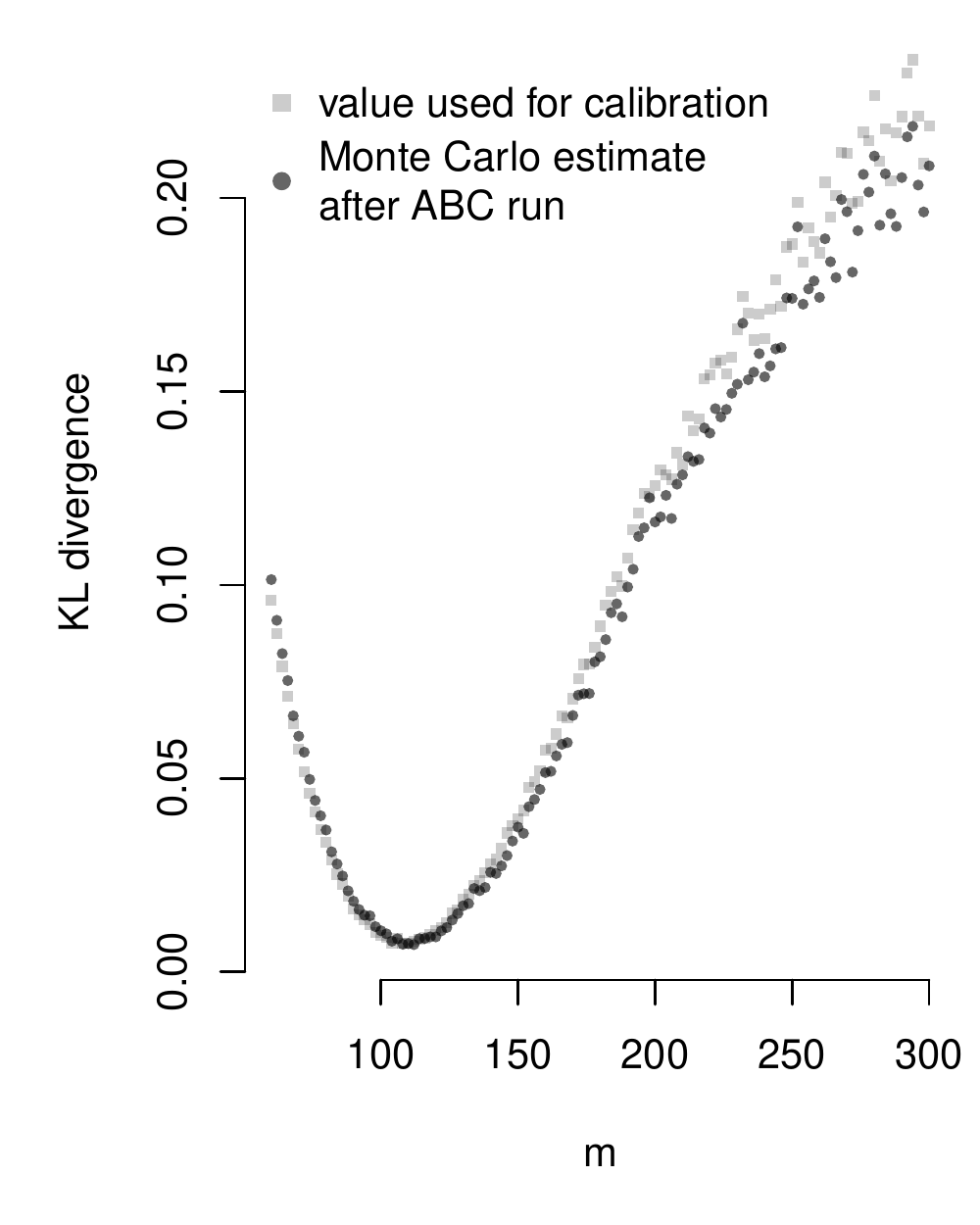}
			\end{minipage}							
	\end{minipage}	
	\begin{minipage}[t]{0.32\textwidth}				
			\begin{minipage}[b]{0.001\textwidth}
				{\bf F}\newline\vspace{-0.4cm}

			\end{minipage}	
			\begin{minipage}[b]{0.99\textwidth}
				\includegraphics[type=pdf,ext=.pdf,read=.pdf,width=\textwidth]{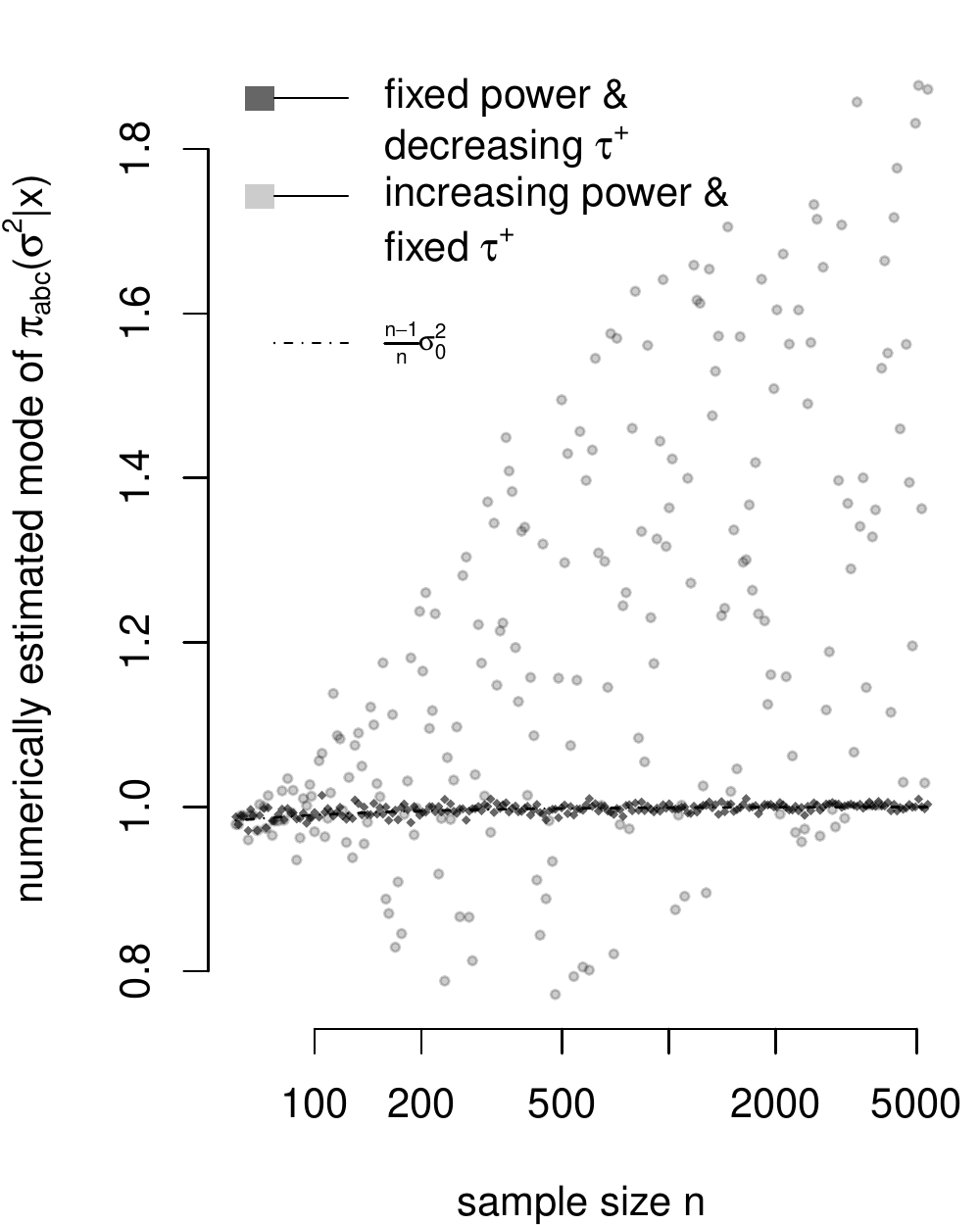}
			\end{minipage}			
	\end{minipage}		
	\end{minipage}		
\caption{{\bf Power function of the $\chi^2$ equivalence test statistic and \nABC\ inference of $\sigma^2$.} (A) Power function of the univariate $\chi^2$ test statistic for two different tolerance regions. For $n=m=60$ and $\alpha=0.01$, we considered first $[\ithresh^-,\ithresh^+]=[1-0.65,1+0.65]$ which lead to $[\ABCthresh^-,\ABCthresh^+]\approx [0.509,1.009]$ as the solution to \eqref{ex:scaledF_rej}. The power function (dark gray) has a unimodal shape and maximises at $\rho^{\text{max}}<\rho^\star= 1$. Next, we fixed $\ithresh^+=2.2$ and calibrated $\ithresh^-=0.477$ such that the power function maximises at $\rho^\star=1$ (light gray). In this case, $[\ABCthresh^-,\ABCthresh^+]= [0.704,1.368]$. Note that, if the observed summary values were considered random even though they are fixed in practice, then $T/\rho$ would follow an $\mathcal{F}$ distribution with $n-1$, $m-1$ degrees of freedom \citep{Wang1997} and the power function would differ from \eqref{ex:scaledF_pw} (dash-dot for same $[\ABCthresh^-,\ABCthresh^+]$). (B) Power of the $\chi^2$ test as $m=n$ increases for fixed $\ithresh^+=2.2$ and calibrated $\ithresh^-$ such that $\rho^{\text{max}}=\rho^\star$. (C) Power of the $\chi^2$ test as $\ithresh^+$ increases for fixed $m=n=60$ and calibrated $\ithresh^-$ such that $\rho^{\text{max}}=\rho^\star$.
(D) We then estimated $\sigma^2$ in $\mathcal{N}(0.\sigma^2)$ under the prior density specified in the main text with \nABC. (dotted) Exact posterior density $\pi(\sigma^2|x)$ and (dash-dot) the MAP estimate of $\sigma^2$ under $\pi(\sigma^2|x)$. (dark gray) Histogram of the ABC approximation $\pi_{\abc}(\sigma^2|x)$ for calibrated $\ithresh^-$, $\ithresh^+$, $m$. (light gray) Histogram of $\pi_{\abc}(\sigma^2|x)$ for the standard ABC approximation used in the introduction, with $z= S^2(y)-S^2(x)$, $[\ABCthresh^-,\ABCthresh^+]=[-0.8,0.8]$ and $m=60$. The \nABC\ approximation obtained with \rejnABC\ has a KL divergence of $0.007$ and an acceptance probability of 12\%, while \rejABC\ reaches similar accuracy for $[\ABCthresh^-,\ABCthresh^+]=[-0.05, 0.05]$ at 3\% acceptance probability (top and bottom left in Table~\ref{t:uabc}). (E) KL divergence for increasing $m$ as used for calibrations before ABC was run (light gray) and as estimated from Monte Carlo output after ABC was run (dark gray). (F) To assess numerical robustness, we compared \nABC\ for increasing $n=m$ when $\ithresh^+=2.2$ was fixed throughout to the case where $\ithresh^+$ was calibrated. In both cases, we numerically estimated the mode of  $\pi_{\abc}(\theta|x)$ as a function of $n$. Calibration of $\ithresh^+$ leads to numerically stable estimates of $\map[\abc]$.}\label{f:scaledFpower}	
\end{figure}

We focus on $\ithresh^-<\ithresh^+$ because for $\ithresh^-=1=\ithresh^+$ power is too small in practice, $P_x(\ABCthresh^-\leq T\leq \ABCthresh^+|\rho=1)=\alpha=0.01$. Then, the power function is larger or equal than $\alpha$ and maximised somewhere in $[\ithresh^-,\ithresh^+]$, see Figure~\ref{f:scaledFpower}A. Calibrating the values $\ithresh^-$, $\ithresh^+$, e.g. to $0.477$ and $2.2$, we can ensure that the value of $\rho$ with largest power coincides with $\rho^\star$. Figure~\ref{f:scaledFpower}B illustrates that the power increases with $m$. Figure~\ref{f:scaledFpower}C illustrates that the power also increases for increasing $\ithresh^+$. Here, $\ithresh^-$ was calibrated so that the largest power coincides with $\rho^\star$. 

A few points are worth emphasising. In the interesting case $\ithresh^-_k<\rho^\star_k<\ithresh^+_k$, the power function of two-sided equivalence hypothesis test statistics $\rho_k\to P_x(R_k|\rho_k)$ has a unimodal shape within $[\ithresh^-_k, \ithresh^+_k]$. We will make extensive use of this property, which does not hold for one-sided equivalence hypothesis tests or point null hypothesis tests. This limits all tests of interest to those of the form \eqref{e:ihyp}. Moreover, two-sample equivalence tests have different power functions, which limits interest further to one-sample tests of the form \eqref{e:ihyp} for fixed data $x$ (Figure~\ref{f:scaledFpower}A). Finally, equation~\eqref{ex:scaledF_rej} implies that the calibrated $[\ABCthresh^-,\ABCthresh^+]$ for which power is maximised at $\rho^\star=1$ do not necessarily include the point $T=1$. This is especially so when $m\gg n$. The correct critical regions used in \nABC\ are not necessarily what might be naively expected at first sight.

\subsection{Multivariate equivalence tests on the auxiliary space}\label{s:multitest}
\paragraph{Composite approach.}
As we aim for a bijective link function, we typically need to handle several tests simultaneously and need to characterise the multivariate $\rho\to P_x(R|\rho)$. In the equivalence testing framework, sufficiently regular multivariate tests are straightforward to construct from disparate univariate tests. For all $k=1,\dotsc,K$, suppose the distribution of summary values is such that 
\begin{singlespaceddescription}
\item[\,\,A2.1] there are level-$\alpha$ test statistics $T_k$ with rejection regions $R_k=[\ABCthresh_k^-,\ABCthresh_k^+]$ for univariate equivalence hypotheses \eqref{e:ihyp} with a unimodal power function $\rho_k\to P_x(R_k|\rho_k)$,
\item[\assumption\negthinspace] the family of distributions induced by $T_k$ is independent of $\rho_j$ for $j\neq k$.
\end{singlespaceddescription}
Then, as shown by \cite{Berger1996}, the multivariate statistic $T=(T_1,\dotsc,T_K)$ with the hypercube rejection region $R=\bigcap_{k=1}^K[\ABCthresh_k^-,\ABCthresh_k^+]$ is level-$\alpha$
for
\begin{equation}\label{e:jihyp}
H_0= \bigcup_{k=1}^K H_{0k}\text{ versus }H_{1}=\bigcap_{k=1}^KH_{1k}.
\end{equation}
The power of the multivariate $T$ has a unique mode $\rho^{\max}$ and decreases monotonically in each direction from $\rho^{\max}$, which follows from A2.1 because A4 implies $P_x( R\:|\:\rho)=\prod_{k=1}^KP_x( R_{k}\:|\:\rho)=\prod_{k=1}^KP_x( R_{k}\:|\:\rho_k)$. The resulting ABC accept/reject step is exactly \eqref{e:intersection}, so we can interpret \eqref{e:intersection} as the outcome of a composite testing approach in which all tests are assumed to be independent. No downwards adjustment on the level of the individual tests is necessary to obtain a multivariate level-$\alpha$ equivalence test statistic. Based on the argument in Figure~\ref{f:tpr}, we fix $\alpha=0.01$. The main appeal of the composite approach is that we can combine different univariate tests in a flexible manner. A2.1 is met when the distribution of summary values is continuous in $\rho$ and strictly totally positive of order 3  (\STP) \cite[p369]{Wellek2003}, but also in other cases. The main limitation is A4, which is clear by considering the equivalent statement to the summary likelihood,
\begin{singlespaceddescription}
\item[\assumption] $\lkl\big(\{ \fsu(x), k=1,\dotsc,K\}\:|\:\rho \big)= \prod_{k=1}^K \lkl( \fsu(x)\:|\:\rho_k)$.
\end{singlespaceddescription}

\paragraph{Mahalanobis approach.}
The alternative Mahalanobis approach \eqref{e:abclklrhoMAHA} is highly used in ABC. It corresponds to a multivariate equivalence test of $K$ population means of $K$ sets of multivariate Gaussian summary values \citep[p221ff]{Wellek2003}. So it is only applicable in this basic case. There is only one scalar tolerance parameter $\ithresh>0$, $\ithresh^-=-\ithresh$, $n_k=n$, $m_k=m$ and the resulting critical region is described by $\ABCthresh>0$ as in \eqref{e:abclklrhoMAHA}. A2.1 is not met exactly because the test is not exactly level-$\alpha$ but alternatives exist at least in the univariate case \citep{Berger1996}. Importantly, the summary values can be correlated in this case (A4-A5 not required). 

\subsection{Calibration}\label{s:calibrations}
\paragraph{Composite approach.}
Our next observation is that the free parameters of the equivalence tests can be calibrated such that $P_x(R|\rho)$ is close to $\lkl(\{s_1^{1:n_1}(x),\dotsc,s_K^{1:n_K}(x)\}\:|\:\rho)$ on the auxiliary space. 
For each $k$, we further assume
\begin{singlespaceddescription}
\item[\,\,A2.2] $P(R_k|\rho_k)$ is continuous in $\ithresh^-_k$ and $\ithresh^+_k$
\item[\,\,A2.3] $\argmax_{\rho_k} P(R_k|\rho_k)$ shifts to the right (left) when $\ithresh_k^-$ decreases (increases)
\item[\,\,A2.4] the equivalence test induced by $T_k$ is consistent,
\end{singlespaceddescription}
and proceed in three hierarchical steps. At the lowest level, we calibrate $\ithresh_k^-$ for given $\ithresh_k^+$ and $m_k$ such that the modes of the univariate $\rho_k\to P_x(R_k|\rho_k)$ and $\rho_k\to \lkl(\fsu(x)|\rho_k)$ coincide. We discussed this possibility already on the $\mathcal{N}(0,\sigma^2)$ example (Figure~\ref{f:scaledFpower}A). 
Next, we calibrate  $\ithresh_k^+$ for given $m_k$ for numerical stability. This will include repeat calls to the first calibration step. Estimating the mode of $\rho_k\to P_x(R_k|\rho_k)$ from Monte Carlo output is numerically more stable when this function is not too flat around $\rho^\star_k$. Power functions are flat if they are close to zero {\it or} one for all $\rho_k$. We saw in Figure~\ref{f:scaledFpower}C that the flatness around $\rho^\star_k$ can be adjusted by changing the width of the tolerance region, and this is why we calibrate $\ithresh_k^+$.
At the top level, we calibrate $m_k$ such that the KL divergence of $P_x(R_k|\rho_k)$ to $\lkl(\fsu(x)|\rho_k)$ is minimised. This will involve calls to the first two calibration steps. As power concentrates with increasing $m_k$ under A2.4, we are able to offset the broadening effect of the ABC tolerances by choosing $m_k>n_k$. Using A4-A5, the multi-dimensional end-product is a numerically stable ABC approximation that minimises its KL divergence to the exact summary likelihood, subject to the constraint that their modes coincide. Full details are found in SOM~\ref{a:calibration} and calibration routines are available from \url{https://github.com/olli0601/abc.star}. 
\paragraph{Mahalanobis approach.}
If the assumptions on the $\fsu$ are met so that the test is applicable, calibrations proceed as for a single $k$ under the composite approach. A4-A5 are not required and A2.2-A2.4 are met.

\subsection{\nABC\ parameter inference}\label{s:indirectABC}
The remaining question is when these calibrations on the auxiliary parameter space improve the accuracy of the \nABC\ approximation $\pi_{\abc}(\theta|x)$ to the exact posterior density $\pi(\theta|x)$. We need to limit considerations to the case where a model can explain the data in terms of the summary values. To this end, we require that the support of 
the prior density $\pi_\rho$ must extend beyond the tolerance region. In particular, it must include the point of equality so that there is $\hat{\theta}$ such that $\Link(\hat{\theta})=\rho^\star$. The basis for indirect inference is that maximum likelihood estimators are invariant under parameter transformation \citep{Gourieroux1993}. In \nABC, the rationale for calibrating with respect to the KL divergence is as well that the KL divergence between two densities is invariant under parameter transformation.

\begin{theorem}{{\bf (Minimal KL divergence)}}\label{th:accurate}
Suppose that the prior density on the auxiliary space is flat. (i) For the composite $T=(T_1,\dotsc,T_K)$, suppose that summary values can be constructed such that A1-A5 hold.
Calibrate $\ithresh^-_k$, $\ithresh^+_k$, $m_k$. Then
\begin{equation}\label{e:accurateabc}
\KL\big( \pi(\theta|x) |\!| \pi_{\abc}(\theta|x) \big)=\sum_{k=1}^K \varepsilon_k, 
\end{equation}
where $\varepsilon_k=\KL\big( \lkl(\fsu | \rho_k(x)) |\!| P_x(R_k | \rho_k)\big)$. (ii) For the Mahalanobis approach, suppose that the summary values are multivariate normal and that A1, A3 hold. Calibrate $\ithresh$, $m$. Then $\KL\big( \pi(\theta|x) |\!| \pi_{\abc}(\theta|x) \big)=\KL\big( \lkl(\{s_1^{1:n_k}(x)\}_{1:K}\:|\:\rho) |\!| P_x(R | \rho) \big)= \varepsilon$.
\end{theorem}

The proof is given in Appendix~\ref{a:proofs}. We can also control the accuracy of the MAP point estimate of $\pi_{\abc}(\theta|x)$. Considering the maximum likelihood estimate $\mle=\argmax_\theta \lkl(x|\theta)$, we have $\Link(\mle)=\argmax_\rho \lkl(x|\rho)$. With the calibrations in Section~\ref{s:calibrations}, we also have $\Link(\mle)=\rho^\star=\argmax_\rho P_x(R|\rho)$. With A3, we obtain $\mle= \argmax_\theta P_x(R|\theta)= \mle[\abc]$. To extend this fact to the MAP estimate of $\pi_{\abc}(\theta|x)\propto P_x(R|\rho) \:  \pi_\rho(\rho) \: \abs{\partial\!\Link(\theta)}$, we note that the influence of the typically non-constant $\abs{\partial\!\Link(\theta)}$ can be mitigated if $P_{x}(R|\rho)$ peaks on a small area around $\rho^\star$. We require that
\begin{singlespaceddescription}
\item[\assumption] $\pi_{\abc}(\rho|x)$ decays faster around $\rho^\star$ than $\abs{\partial\!\Link(\theta)}$ increases around $\theta_x=\Link^{-1}(\rho^\star)$ and in this case we say that the {\it data control the change of variables}. 
\end{singlespaceddescription} 
\begin{theorem}{{\bf (Exact MAP estimate)}}\label{th:map}
Suppose that the prior density on the auxiliary space does not change the location of $\argmax_\rho P_x(R|\rho)$. (i) For the composite $T=(T_1,\dotsc,T_K)$, suppose that summary values can be constructed such that A1-A5 hold and calibrate $\ithresh^-_k$, $\ithresh^+_k$, $m_k$. (ii) For the Mahalanobis approach, suppose that the summary values are multivariate normal and that A1, A3 hold, and calibrate $\ithresh$, $m$. In both cases,
\begin{equation}\label{e:exactMAP}
\map[\abc]= \argmax_\theta \pi_{\abc}(\theta|x)= \argmax_\theta \pi(\theta|x).
\end{equation}
\end{theorem}

\paragraph{Example.} We consider one more time the basic $\mathcal{N}(0,\sigma^2)$ example, for which we already constructed the $\chi^2$ test statistic and the critical region $[\ABCthresh^-,\ABCthresh^+]$ appropriate for testing dispersion equivalence of normally distributed summary values (A2.1 met). These $T$ and $[\ABCthresh^-,\ABCthresh^+]$ are used to construct the \nABC\ algorithm for given $n$ and fixed $\alpha=0.01$. Inspecting \eqref{ex:scaledF_pw}, we see that A2.2-A2.4 are met. A1, A3, A6 are not needed because the data can be directly taken as the summary values. A4-A5 are not needed in this univariate example. We can therefore apply the calibrations described in Section~\ref{s:calibrations}. For these calibrations, we need to know the likelihood density on $\rho$-space, which follows from $S^2(x)/\sigma^2\sim \chi^2_{n-1}$.

We calibrated the free ABC parameters to $m=108$,  $[\ithresh^-,\ithresh^+]=[0.589,1.752]$ and $[\ABCthresh^-,\ABCthresh^+]=[1.41,2.22]$ before ABC was run. The point $T=1$ is not included in the calibrated $[\ABCthresh^-,\ABCthresh^+]$. For the purpose of illustration, we considered a pseudo data set $x$ such that $\hat{\sigma}^2_x$ was exactly $\theta_0=1$ (except for Figure~\ref{f:scaledFpower}E-F). We then ran \nABC\ for the flat prior distribution $\pi(\sigma^2)=\mathcal{U}(0.2,4)$. As predicted by Theorem~\ref{th:map}, $\map[\abc]$ is very close to the mode of the posterior density (Figure~\ref{f:scaledFpower}D), with a mean absolute error of $0.002$ over 4000 pseudo data sets $x$ (bottom left in Table~\ref{t:uabc}). After the ABC run, we estimated the KL divergence of $\pi_{\abc}(\sigma^2|x)$ to the posterior density from Monte Carlo output at $0.007$. Figure~\ref{f:scaledFpower}E shows that the KL divergence is indeed minimised for values of $m$ that are larger than $n$. Increasing $m$ beyond $108$, $\pi_{\abc}(\sigma^2|x)$ concentrates more tightly at $\frac{1}{n}S^2(x)$ than $\pi(\sigma^2|x)$.

Finally, to illustrate the importance of calibrating $\ithresh^+$, we considered observed and simulated data sets that increase in size from $n=m=60$ to $5000$. First, we fixed $\ithresh^+=2.2$ throughout (and hence also $\ithresh^-$), so that power increases with size and $P_x(\ABCthresh^-\leq T\leq\ABCthresh^+|\rho)$ becomes more flat around $\rho^\star$ (Figure~\ref{f:scaledFpower}C). Second, we calibrated $\ithresh^+$ (and hence also $\ithresh^-$) for every $m$ as described in Section~\ref{s:calibrations} (Figure~\ref{f:nABCscaledFb}). We ran \nABC\ as above for increasing $n$ and both sets of tolerance values. In each case, we estimated the mode of $\pi_{\abc}(\theta|x)$ from Monte Carlo output. For fixed $\ithresh^+=2.2$, the Monte Carlo estimate of $\map[\abc]$ can be any value such that the corresponding $\rho$ lies on the flat part of $P_x(\ABCthresh^-\leq T\leq\ABCthresh^+|\rho)$. By contrast, for calibrated $\ithresh^+$, the Monte Carlo estimate of $\map[\abc]$ remains close to $\map$ (Figure~\ref{f:scaledFpower}F). 

\section{Application to time series dynamics}\label{s:applications}
\subsection{Synthetic moving average time series}
\begin{figure}[!t]				
	\begin{minipage}[c]{0.32\textwidth}		
		\begin{minipage}[b]{0.001\textwidth}
			{\bf A}\newline\vspace{-0.5cm}

		\end{minipage}	
		\begin{minipage}[b]{0.99\textwidth}
			\includegraphics[type=pdf,ext=.pdf,read=.pdf,width=\textwidth]{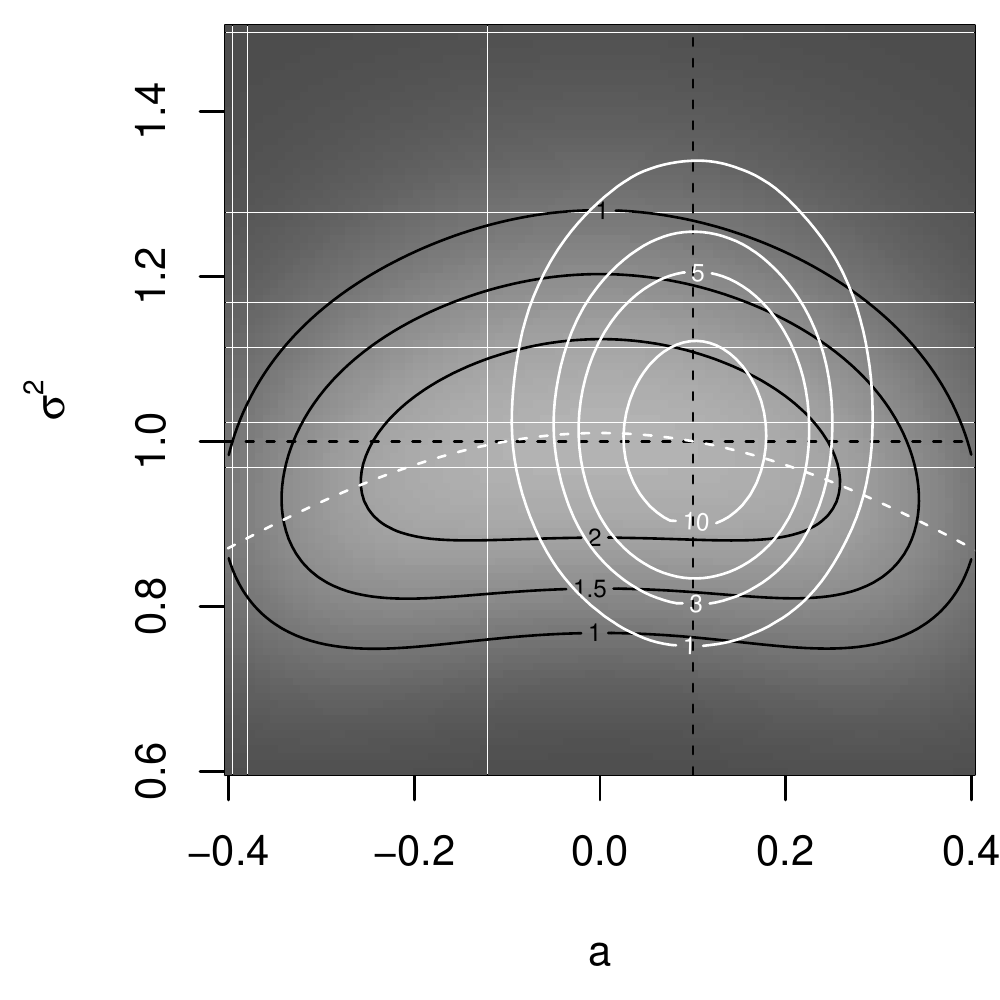}
		\end{minipage}	
	\end{minipage}	
	\begin{minipage}[c]{0.32\textwidth}		
		\begin{minipage}[b]{0.001\textwidth}
			{\bf B}\newline\vspace{-0.5cm}

		\end{minipage}	
		\begin{minipage}[b]{0.99\textwidth}
			\includegraphics[type=pdf,ext=.pdf,read=.pdf,width=\textwidth]{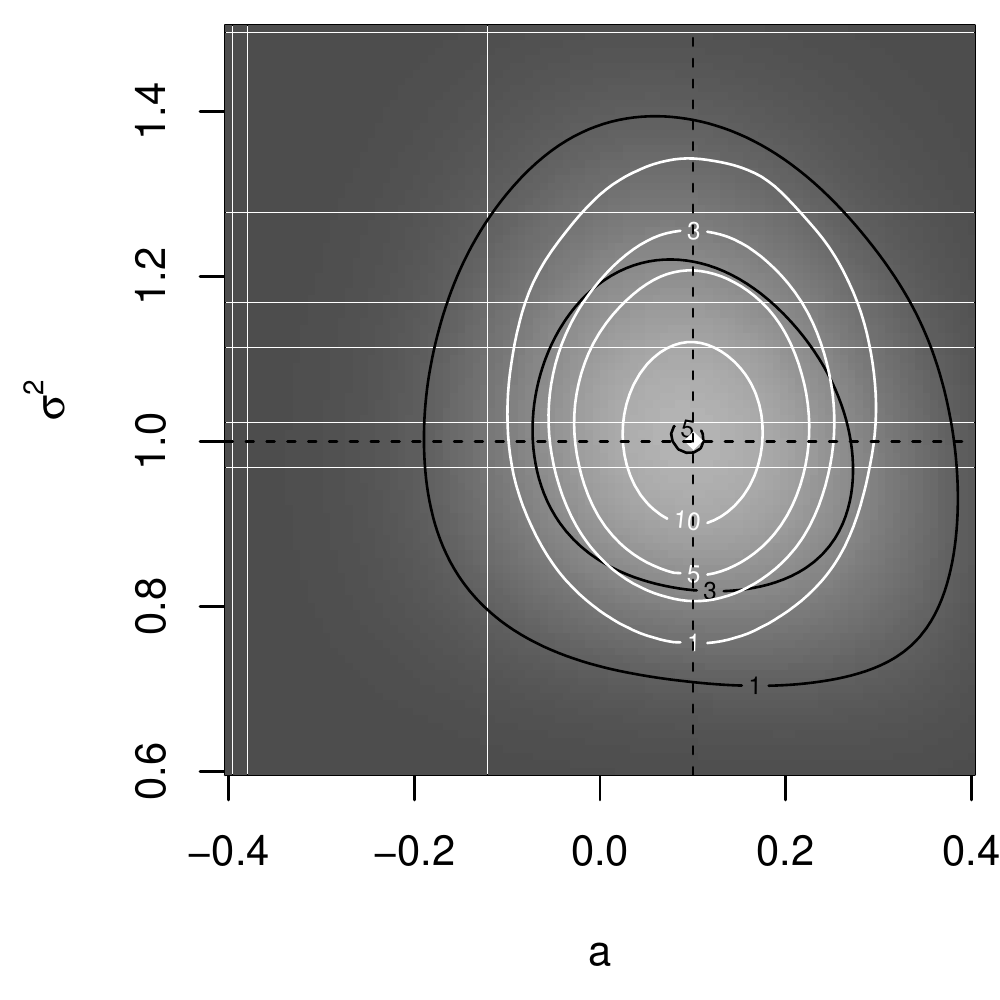}
		\end{minipage}			
	\end{minipage}	
	\begin{minipage}[c]{0.32\textwidth}		
			\begin{minipage}[b]{0.001\textwidth}
				{\bf C}\newline\vspace{-0.4cm}

			\end{minipage}	
			\begin{minipage}[b]{0.99\textwidth}
				\includegraphics[type=pdf,ext=.pdf,read=.pdf,width=\textwidth]{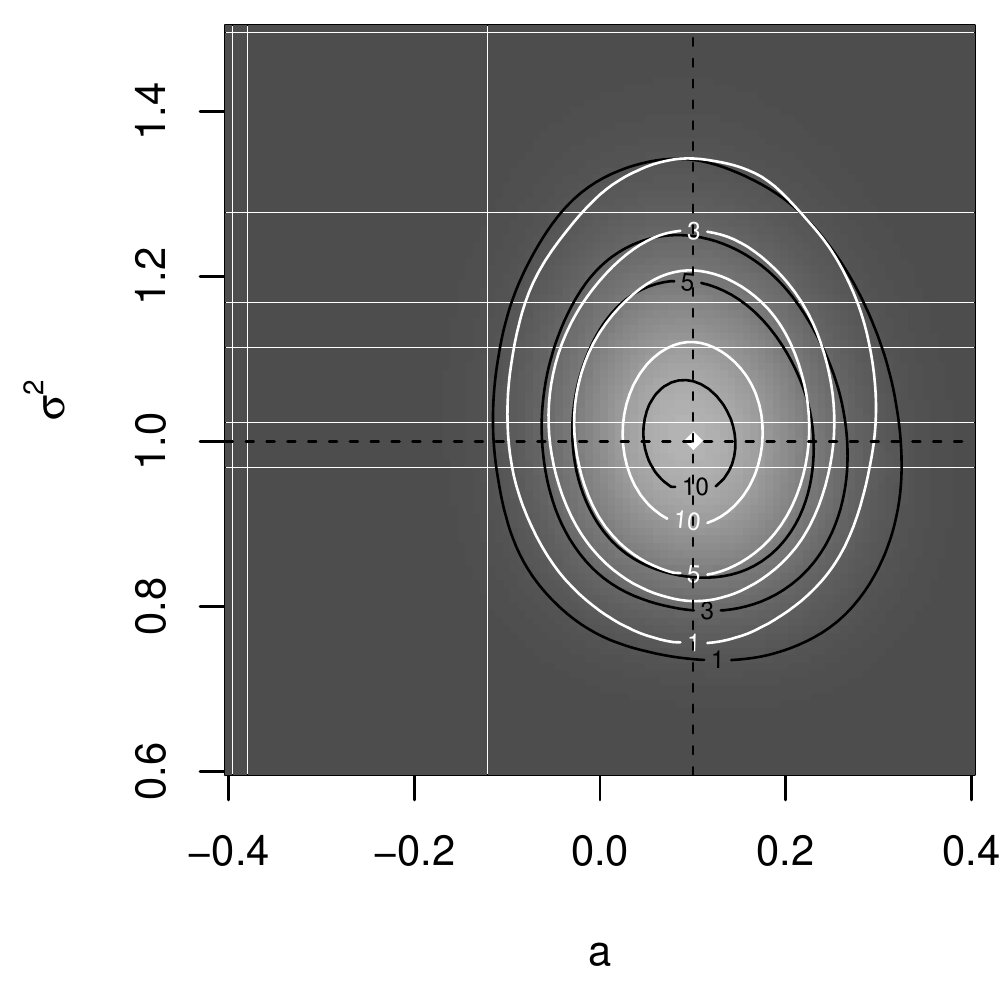}
		\end{minipage}	
	\end{minipage}	
	\begin{minipage}[c]{0.32\textwidth}		
			\begin{minipage}[b]{0.001\textwidth}
				{\bf D}\newline\vspace{-0.4cm}

			\end{minipage}	
			\begin{minipage}[b]{0.99\textwidth}
				\includegraphics[type=pdf,ext=.pdf,read=.pdf,width=\textwidth]{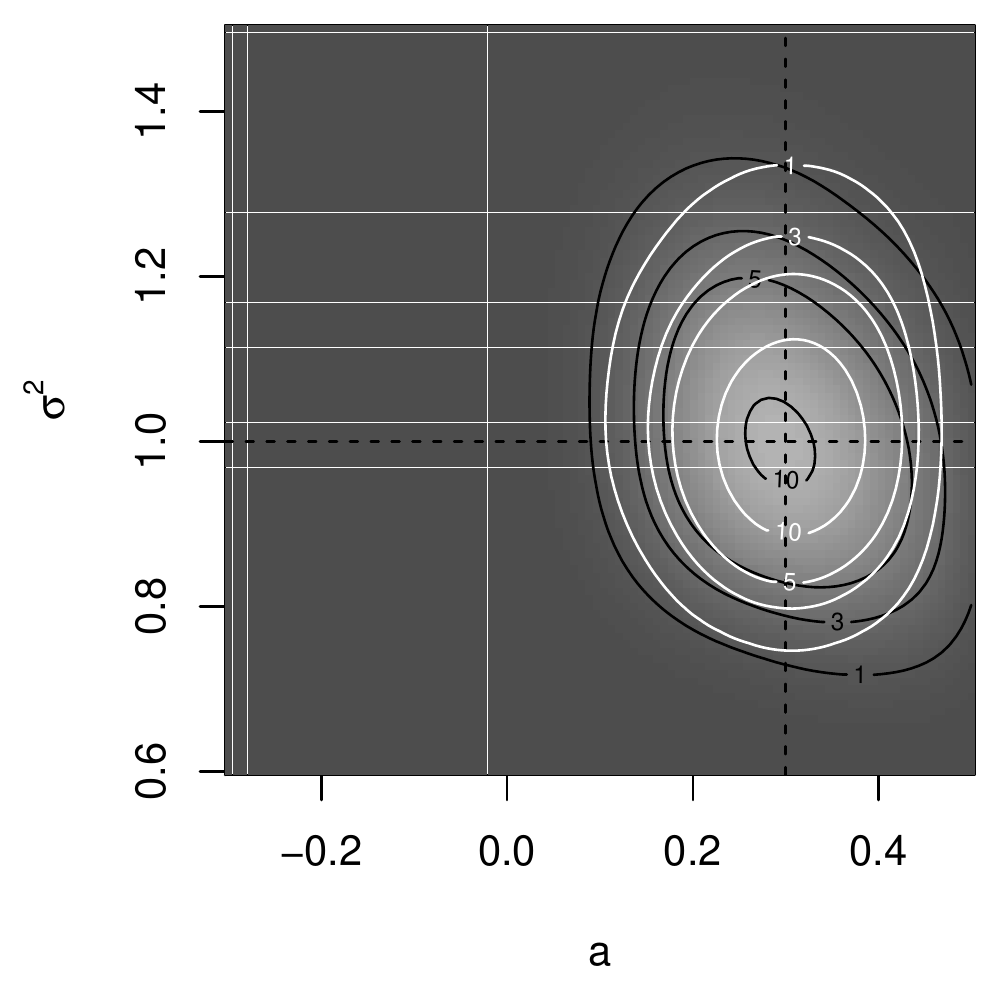}
		\end{minipage}	
	\end{minipage}		
	\begin{minipage}[c]{0.32\textwidth}		
			\begin{minipage}[b]{0.001\textwidth}
				{\bf E}\newline\vspace{-0.4cm}

			\end{minipage}	
			\begin{minipage}[b]{0.99\textwidth}
				\includegraphics[type=pdf,ext=.pdf,read=.pdf,width=\textwidth]{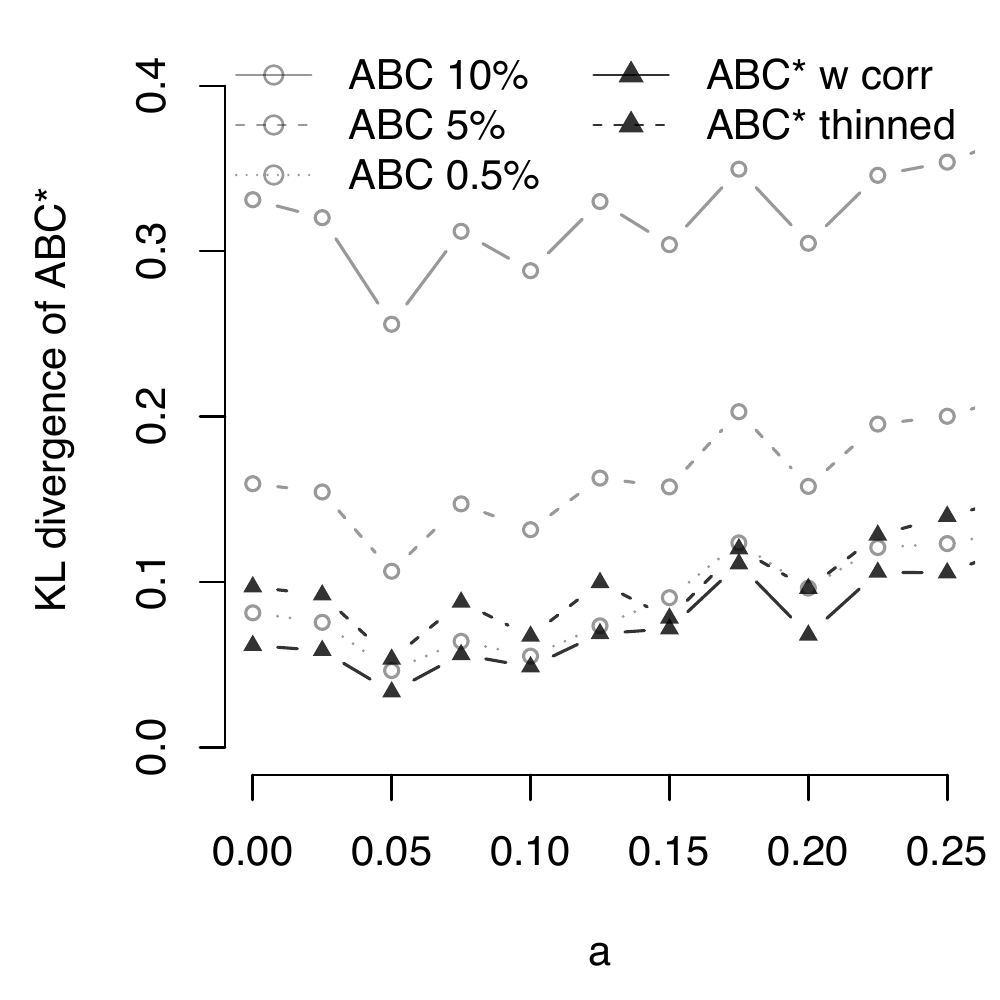}
		\end{minipage}	
	\end{minipage}	
	\begin{minipage}[c]{0.32\textwidth}		
		\begin{minipage}[b]{0.001\textwidth}
			{\bf F}\newline\vspace{-0.4cm}

		\end{minipage}	
		\begin{minipage}[b]{0.99\textwidth}
			\includegraphics[type=pdf,ext=.pdf,read=.pdf,width=\textwidth]{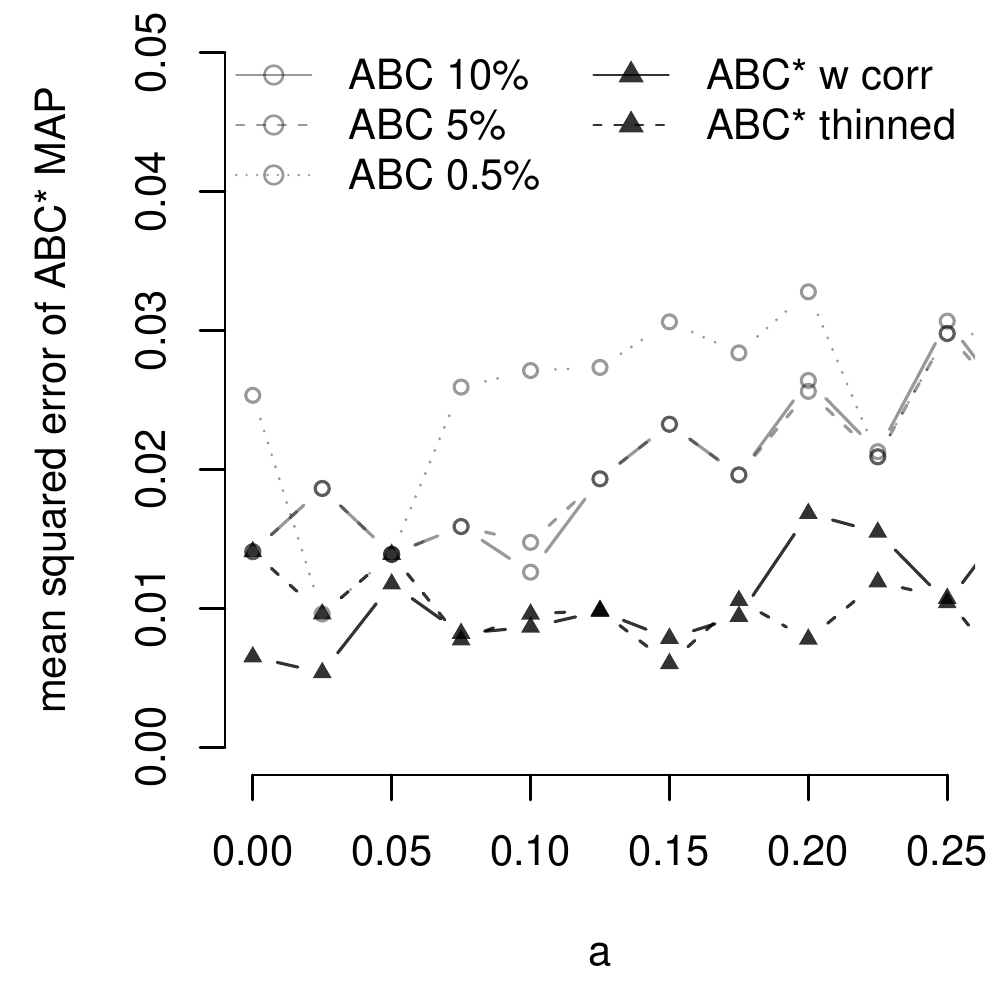}
		\end{minipage}	
	\end{minipage}		
\caption{{\bf \nABC\ inference for the MA(1) model.} We considered the prior density $\pi(\theta)$ induced by a uniform prior on $\rho=(\rho_1,\rho_2)$ and $a_0=0.1$, $\sigma^2_0=1$. The exact posterior density (white contours) was estimated with MCMC. (A) 2D histogram of the estimated \nABC\ posterior density, using only the dispersion test and ignoring autocorrelations. The \nABC\ approximation aligns around the level set. (B) 2D histogram of the estimated \nABC\ posterior density, using the dispersion and correlation test on the subsets $s_{1a}$ and $s_{2a}$ defined in the main text. The \nABC\ approximation is broader than $\pi(\theta | x_{1:n})$. (C) 2D histogram of the estimated \nABC\ posterior density, using two dispersion and three correlation tests so that A1, A3 are met.  The \nABC\ approximation is very close to $\pi(\theta|x)$. (D) Same as (C) but for $a_0=0.3$. (E) Comparison of the KL divergence of $\pi_{\abc}(\theta|x_{1:n})$ to $\pi(\theta|x_{1:n})$ for \rejABC\ with decreasing acceptance probabilities (gray) and \rejnABC\ for the configuration in subfigure C and the case where autocorrelations in the $s_{1}$ and $s_{2}$ are ignored. \rejABC\ must be run at very small acceptance probabilities of $0.5\%$ in order to match the accuracy in KL divergence of \rejnABC, which had a 5\% acceptance probability. (F) Same as Figure~\ref{f:MA1_inf}C for the mean squared error between $\map[\abc]$ and $\map$.
}\label{f:MA1_inf}	
\end{figure}

The challenge in applying \nABC\ to complex inferential problems lies in identifying and modelling summary values such that A1-A3 are met. We illustrate with data generated by a moving average process of order one (MA(1)) for which consecutive observations in time are correlated and sufficient statistics other than the full data cannot be found. Let $x_t= u_t + a_0u_{t-1}$ where $u_t\sim\mathcal{N}(0,\sigma_0^2)$ for $t=1,\dotsc,n$ and $a_0=0.1$, $\sigma_0^2=1$, $n=150$. The time series has variance $\nu_{x1}=(1+a_0^2)\sigma^2_0$ and autocorrelation $\nu_{x2}=a_0 / (1+a_0^2)$. We exploit this relationship to re-estimate $\theta=(a,\sigma^2)$. Tests for the equivalence of the auxiliary variance and autocorrelation are available (SOM~\ref{a:univariateequivalencestatistics}) but require independent normally distributed data points. To proceed, we defined summary values as thinned subsets of the time series, e.~g. $s^i_{1a}=x^{2i-1}$ and $s^i_{1b}=x^{2i}$, $i=1,\dotsc,n/2$; or simply ignored autocorrelations in the summary values, e.~g.~ $s^i_1=x^i$, $i=1,\dotsc,n$ for the variance test (see also Figure~\ref{f:MA1a}). Given these tests, we have $\Link\colon \theta=(a,\sigma^2)\to\rho=(\rho_1,\rho_2)$ with
\begin{equation}\label{e:MAlink}
\begin{split}
&\rho_1=(1+a^2)\sigma^2 / \hat{\nu}_{x1}\\
&\rho_2=\atanh(a/(1+a^2))-\atanh(\hat{\nu}_{x2}),
\end{split}
\end{equation}
which is bijective. Its Jacobian $\partial\!\Link$ is
\begin{equation*}
\left( \begin{array}{cc}
\frac{\partial\!\Link_1}{\partial a} & \frac{\partial\!\Link_1}{\partial \sigma^2}\\[1mm]
\frac{\partial\!\Link_2}{\partial a} & \frac{\partial\!\Link_2}{\partial \sigma^2}\\
\end{array} \right) =
\left( \begin{array}{cc}
2a\sigma^2/\hat{\nu}_{x1}		& (1+a^2)/\hat{\nu}_{x1} \\
(1-a^2)/(1+a^2+a^4)			& 0	\\
\end{array} \right),
\end{equation*}
and the rate of change is $\abs{\determinant\partial\!\Link}= (1-a^4)/((1+a^2+a^4)\hat{\nu}_{x1})$ (Figure~\ref{f:MA1_infs}A). The points of equality are $\rho^\star_1=1$ and $\rho^\star_2=0$. In more complicated applications, $\Link$ is unknown. The MA(1) example with $\Link$ known serves to illustrate the influence of $\Link$ on $\pi_{\abc}(\theta|x_{1:n})$. For this purpose, we also constructed from \eqref{e:MAlink} a prior density $\pi(\theta)$ that is uniform on the auxiliary space (SOM~\ref{a:ma}). This prior was used throughout. To assess the \nABC\ approximation, we estimated the exact $\pi(\theta|x_{1:n})$ with an MCMC sampler using $2\times 10^6$ iterations (white contours in Figure~\ref{f:MA1_inf}, SOM~\ref{a:ma}). For the purpose of illustration, we chose $x$ such that $\map$ of $\pi(\theta|x_{1:n})$ was exactly $\theta_0=(a_0, \sigma^2_0)$.

First, we used {\it the dispersion test only} on the $s_1$ defined above and {\it ignored autocorrelations} in the time series ($n_1=150$; A1-A3 not met; A4-A5 not required). We calibrated in analogy to the $\mathcal{N}(0,\sigma^2)$ example and then ran \rejnABC\ for $2\times 10^6$ iterations.
The estimated $\pi_{\abc}(\theta|x_{1:n})$ aligned around the level set ${\Link}_1^{-1}(\rho_1^\star)=\big\{a,\sigma^2\:|\:\sigma^2(1+a^2)=\hat{\nu}_{x1}\big\}$ because A3 was not met, and was not accurate (Figure~\ref{f:MA1_inf}A; KL divergence was very large, $1.08$).
Second, we used the dispersion and the autocorrelation tests under the composite testing approach on {\it thinned subsets} of $x_{1:n}$, the $s_{1a}$ defined above and $s_{2a}= (x_1,x_2),(x_4,x_5),\dotsc$ ($n_1=75$, $n_2=49$; A1 not met; A2-A5 met). We calibrated as described in Section~\ref{s:calibrations}.
The resulting link function is bijective. The \nABC\ posterior density resolved the identifiability issue but was still broader than $\pi(\theta|x_{1:n})$ (Figure~\ref{f:MA1_inf}B; KL divergence was large, $0.41$).
Third, we combined two dispersion tests using the $s^i_{1a}$ and $s^i_{1b}$ and three correlation tests (see SOM~\ref{a:ma}) so that {\it the full time series data was used} (A1-A3 met; A4-A5 not met). The resulting \nABC\ accuracy error in the MAP estimate and the KL divergence were small ($0.005$ and $0.07$ respectively) at an acceptance probability of 5\% for different values of $a_0$ (Figure~\ref{f:MA1_inf}C-E). 
Interestingly, we obtained very similar accuracy when the two sets of summary values $s_{1}$ and $s_{2}=(x_1,x_2),(x_2,x_3),\dotsc$ were used and autocorrelations were simply ignored (A1, A3-A5 met; A2 not met; bottom centre in Table~\ref{t:uabc} and Figure~\ref{f:MA1_inf}E; KL divergence was small, $0.07$).

We also compared the calibrated \nABC\ posterior density to a standard ABC approximation. Many choices for the free ABC parameters are conceivable. We considered the difference in the variances and autocorrelations, set $\ABCthresh^-_k=-\ABCthresh^+_k$ and chose $\ABCthresh^+_k$ such that the final acceptance probability was $10\%$, $5\%$ and as small as possible before Monte Carlo error became noticeable, $0.5\%$ after  $2\times 10^6$ iterations (Figure~\ref{f:MA1_infs}E).  The standard ABC MAP estimate was not accurate and the KL divergence was comparable to that of \nABC\ at an acceptance probability of $0.05\%$, which corresponds to a 10-fold loss in efficiency relative to \nABC\ (top centre in Table~\ref{t:uabc} and Figure~\ref{f:MA1_inf}E).


To summarise, we applied \nABC\ to a time series example that is complex enough so that no sufficient summary statistics exist, but still simple enough so that the link function is known analytically and $\pi(\theta|x_{1:n})$ can be computed. The model parameters $\theta=(a,\sigma^2)$ are only identifiable with \nABC\ when the link function is bijective. It was most important to meet A3. Thinning the time series lead to irreversible loss of information, and it was important to meet A1. Simply ignoring the autocorrelations in $s_1$ and $s_2$ lead to comparable results as with the most complex subsetting procedure, and both configurations gave overall better accuracy results than standard ABC at a 10-fold higher acceptance probability. It was least important to meet the independence assumption in A2. Nonetheless, A2 cannot be neglected entirely because the predicted KL divergence based on the calibrations was $\approx 10^{-3}$ and substantially smaller than the actual KL divergence of the \nABC\ approximation.

\subsection{Influenza A (H3N2) time series data}

\begin{figure}[p]
	\begin{minipage}[t]{\textwidth}		
	\centering	
	\begin{minipage}[c]{0.45\textwidth}				
		\begin{minipage}[t]{\textwidth}				
			\begin{minipage}[b]{0.001\textwidth}
				{\bf A}\newline\vspace{2.9cm}

			\end{minipage}	
			\begin{minipage}[b]{0.99\textwidth}
				\includegraphics[type=pdf,ext=.pdf,read=.pdf,width=\textwidth]{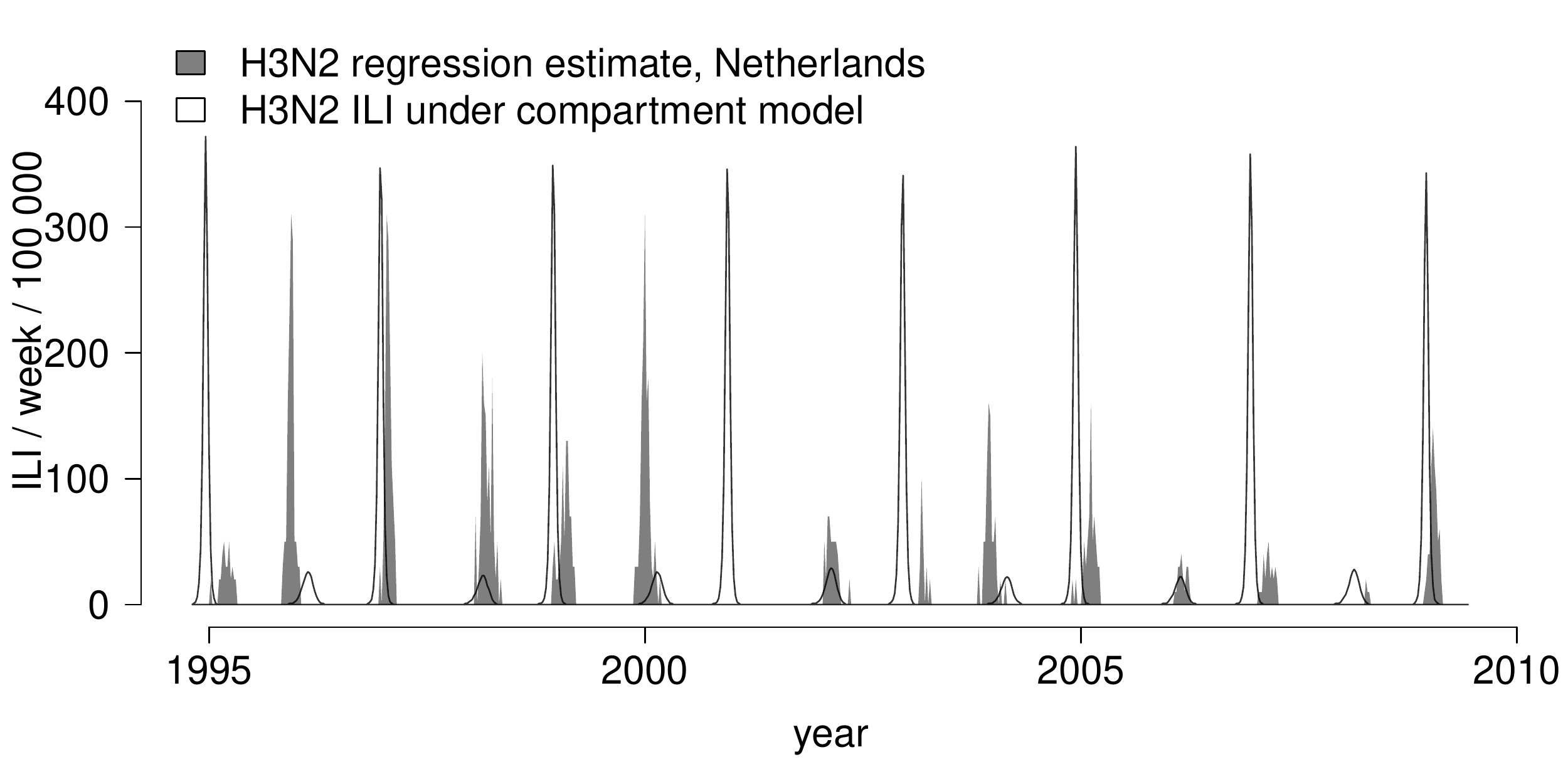}
			\end{minipage}	
		\end{minipage}	
		\begin{minipage}[t]{\textwidth}		
			\begin{minipage}[b]{0.001\textwidth}
				{\bf B}\newline\vspace{2.2cm}

			\end{minipage}	
			\begin{minipage}[b]{0.99\textwidth}
				\includegraphics[type=pdf,ext=.pdf,read=.pdf,width=\textwidth]{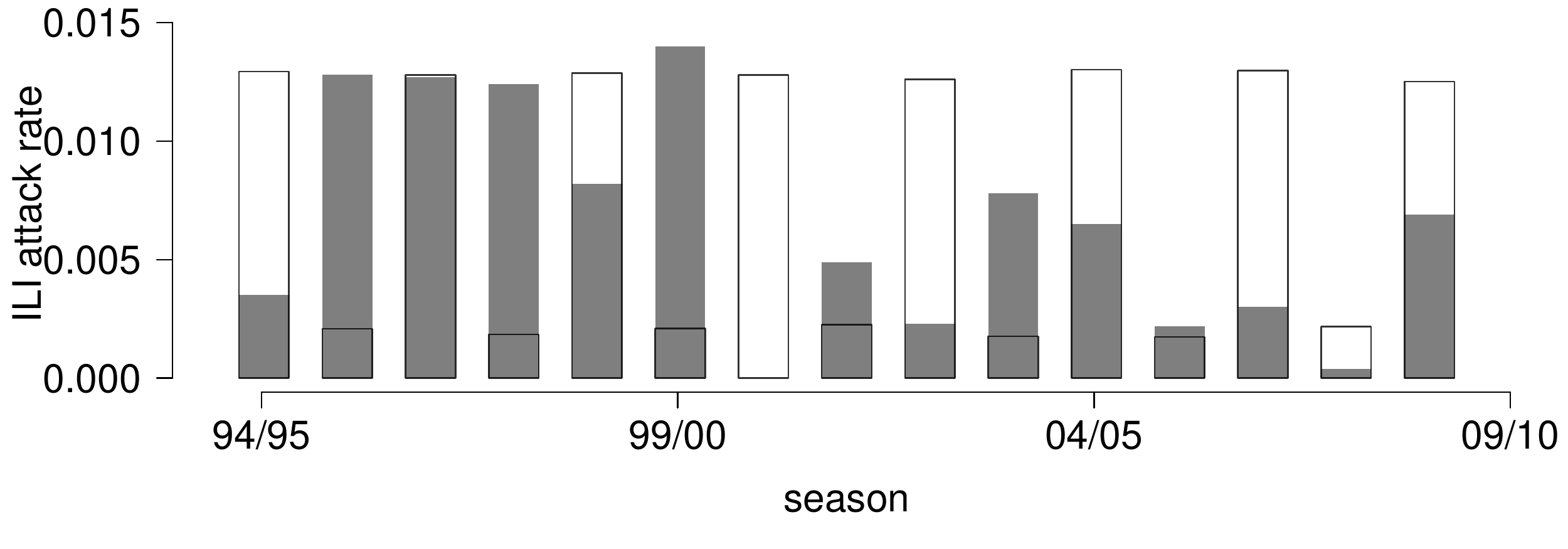}
			\end{minipage}	
		\end{minipage}	
		\begin{minipage}[t]{\textwidth}				
			\begin{minipage}[b]{0.001\textwidth}
				{\bf C}\newline\vspace{2.2cm}

			\end{minipage}	
			\begin{minipage}[b]{0.99\textwidth}
				\includegraphics[type=pdf,ext=.pdf,read=.pdf,width=\textwidth]{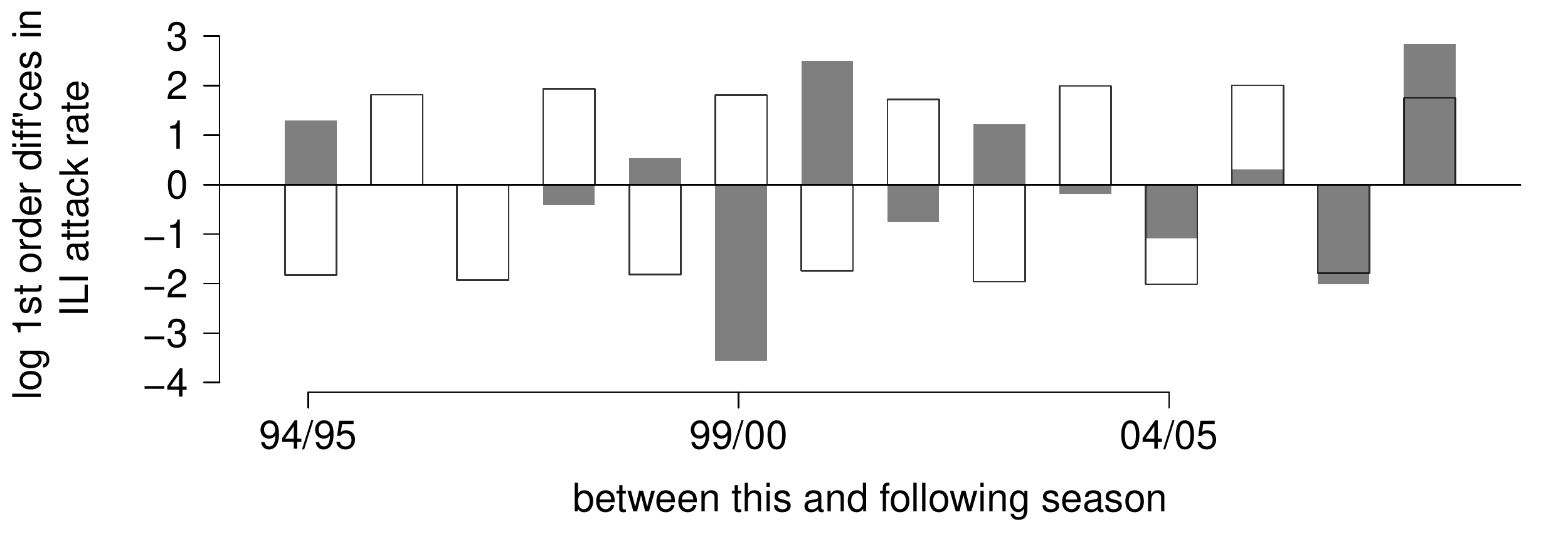}
			\end{minipage}	
		\end{minipage}		
		\begin{minipage}[t]{\textwidth}				
			\begin{minipage}[c]{0.99\textwidth}				
			\begin{minipage}[b]{0.001\textwidth}
				{\bf D}\newline\vspace{1.8cm}

			\end{minipage}	
			\begin{minipage}[b]{0.99\textwidth}	
				\begin{minipage}[b]{0.32\textwidth}
					\includegraphics[type=pdf,ext=.pdf,read=.pdf,width=\textwidth]{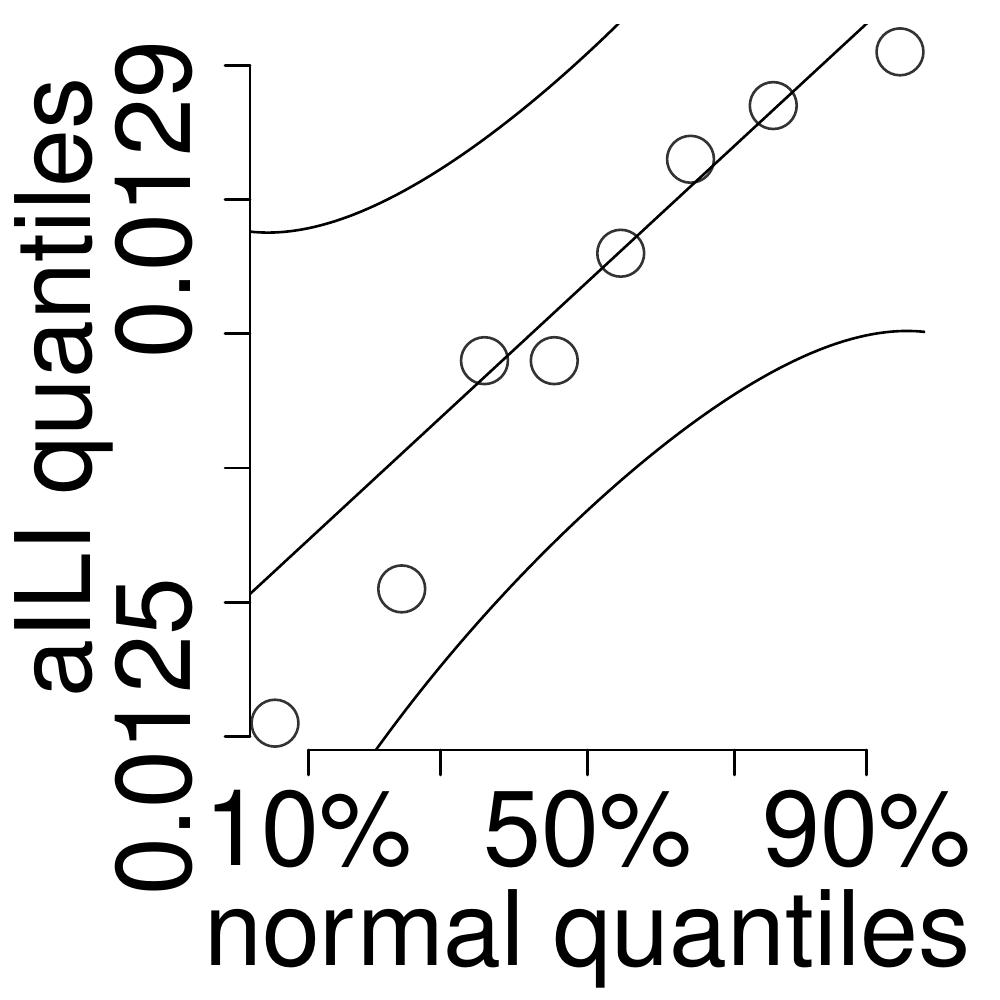}
				\end{minipage}	
				\begin{minipage}[b]{0.32\textwidth}
					\includegraphics[type=pdf,ext=.pdf,read=.pdf,width=\textwidth]{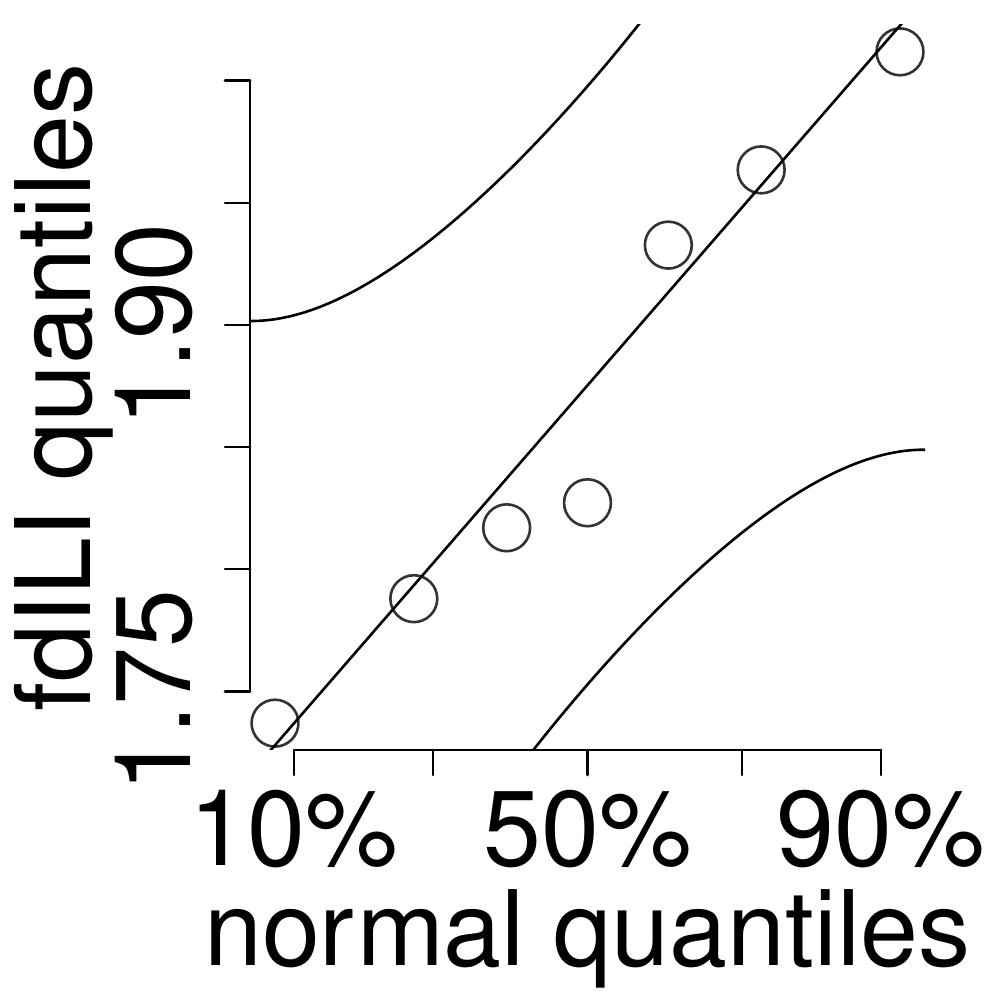}
				\end{minipage}	
				\begin{minipage}[b]{0.32\textwidth}
					\includegraphics[type=pdf,ext=.pdf,read=.pdf,width=\textwidth]{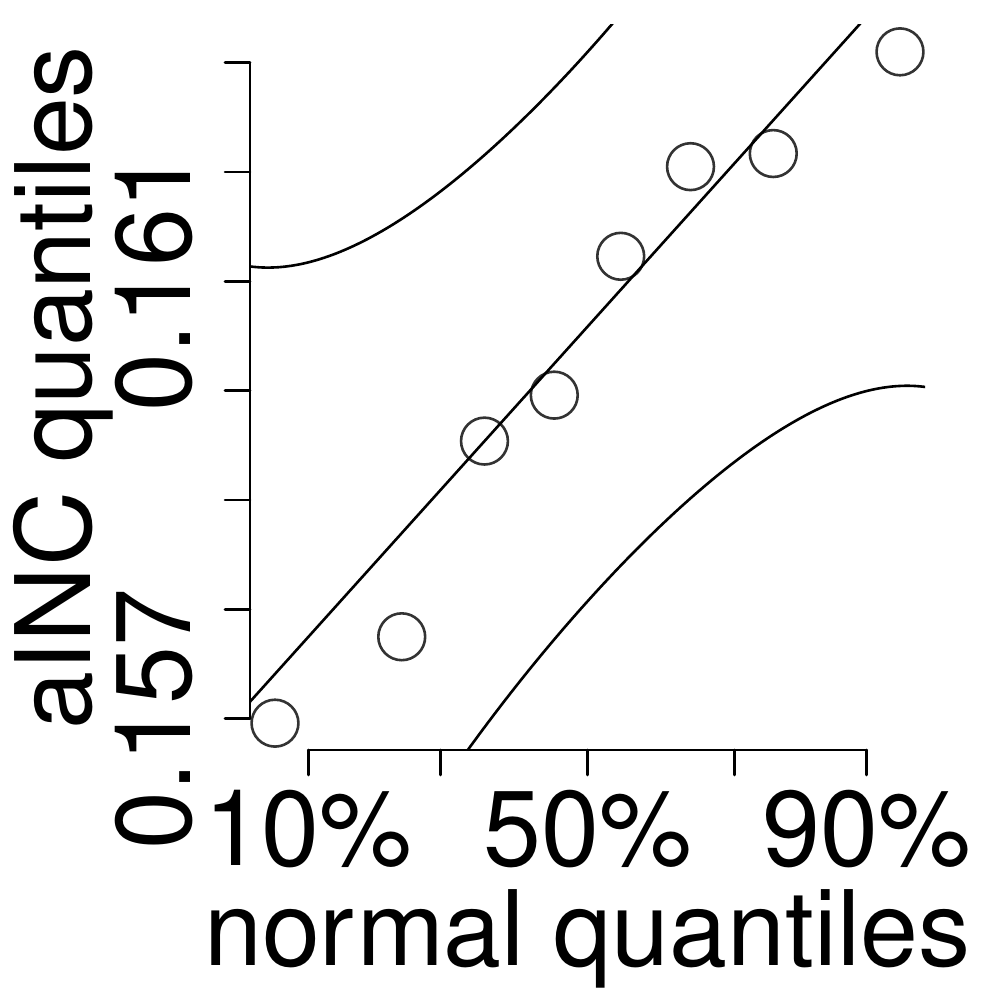}
				\end{minipage}	
			\end{minipage}					
		\end{minipage}	
		\end{minipage}			
	\end{minipage}	
	\begin{minipage}[c]{0.54\textwidth}	
		\begin{minipage}[c]{0.99\textwidth}				
			\begin{minipage}[c]{0.99\textwidth}	
				\begin{minipage}[b]{0.04\textwidth}
					{\bf E}\newline\vspace{2.7cm}

				\end{minipage}	
				\begin{minipage}[b]{0.99\textwidth}
					\animategraphics[controls,autoplay,palindrome,buttonsize=1em,width=0.49\textwidth]{12}{video/simu_AMED.FD.ATTR-}{0}{41}
					\animategraphics[controls,autoplay,palindrome,buttonsize=1em,width=0.49\textwidth]{12}{video/simu_X.ATTR-}{0}{41}
				\end{minipage}			
			\end{minipage}						
			\vspace{0.5cm}
		\end{minipage}	
		\begin{minipage}[c]{0.99\textwidth}				
				\begin{minipage}[b]{0.04\textwidth}
					{\bf F}\newline\vspace{3.05cm}

				\end{minipage}	
				\begin{minipage}[b]{0.99\textwidth}
					\hspace{-0.3cm}\includegraphics[type=pdf,ext=.pdf,read=.pdf,width=0.33\textwidth]{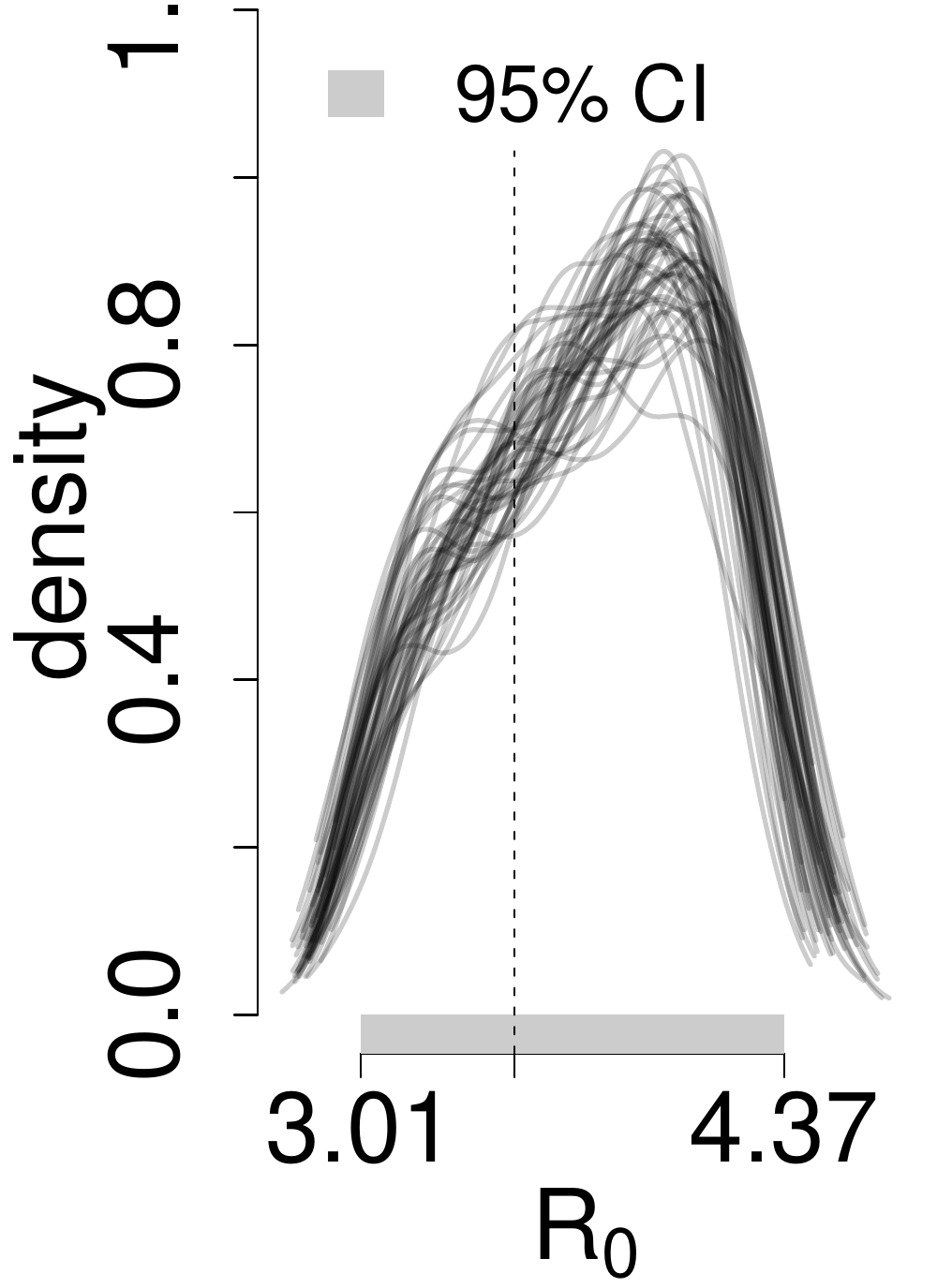}
					\includegraphics[type=pdf,ext=.pdf,read=.pdf,width=0.33\textwidth]{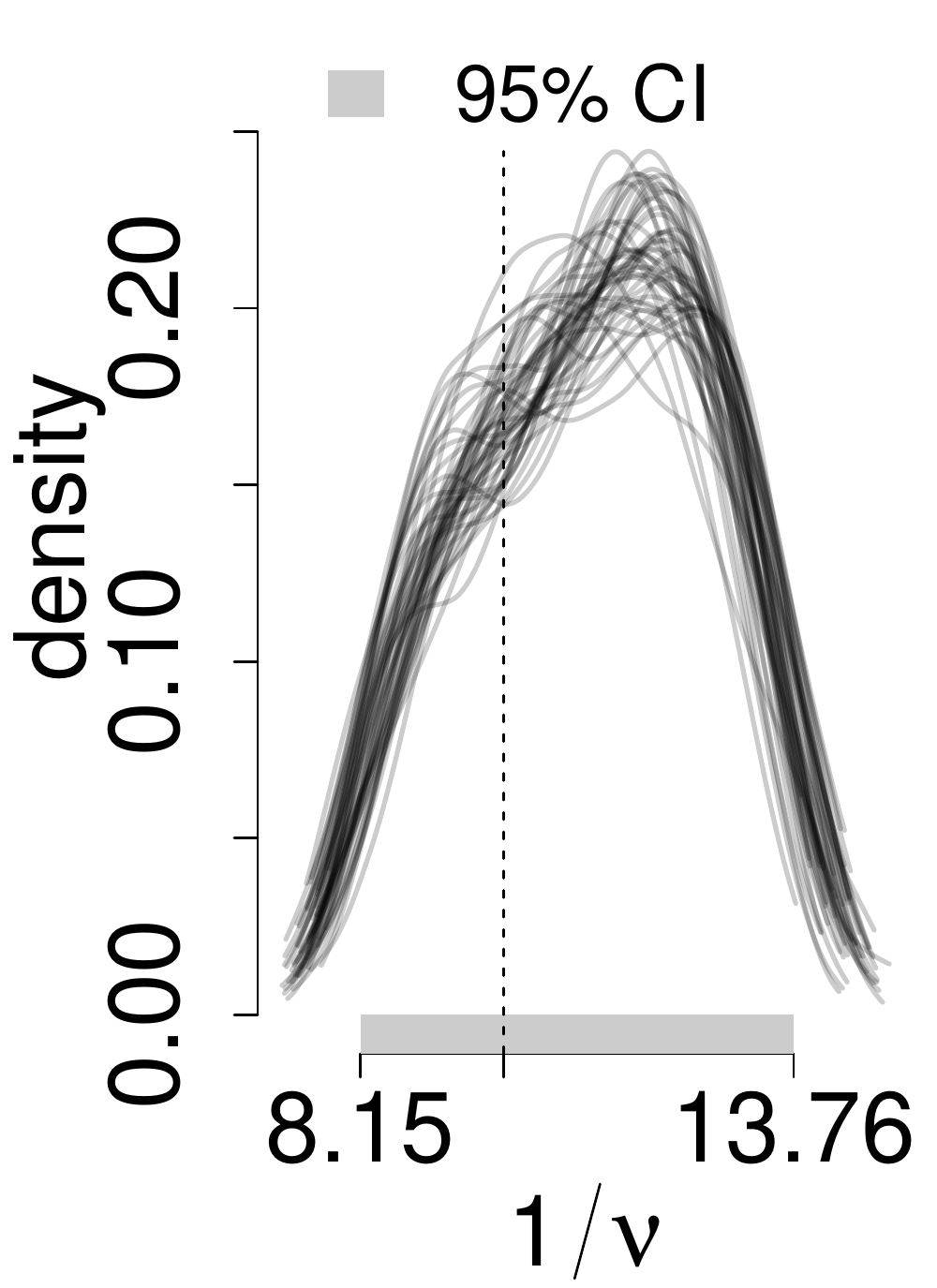}
					\includegraphics[type=pdf,ext=.pdf,read=.pdf,width=0.33\textwidth]{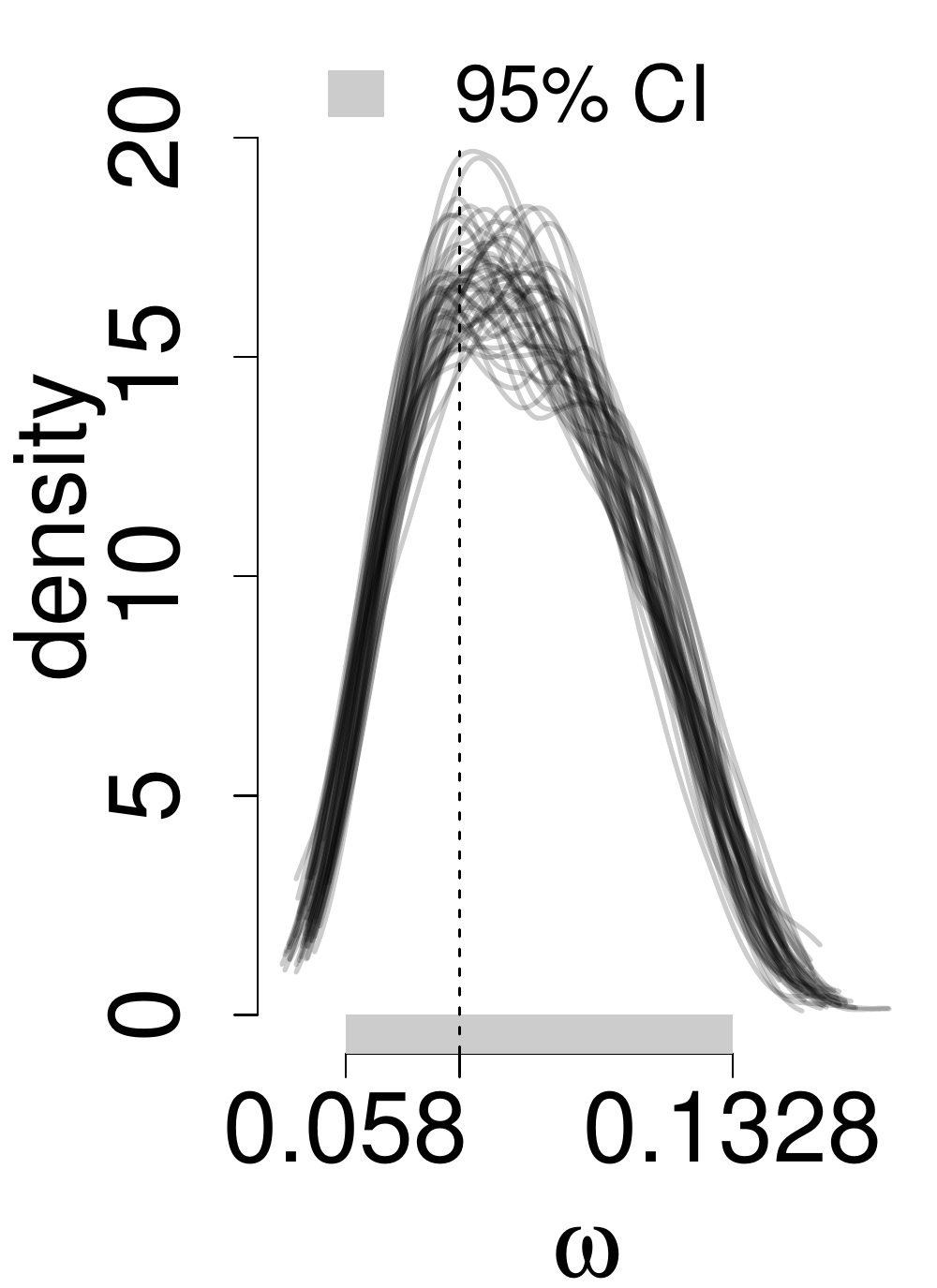}
				\end{minipage}								
		\end{minipage}		
		\begin{minipage}[c]{0.99\textwidth}				
				\begin{minipage}[b]{0.04\textwidth}
					{\bf G}\newline\vspace{3.05cm}

				\end{minipage}	
				\begin{minipage}[b]{0.99\textwidth}
					\hspace{-0.3cm}\includegraphics[type=pdf,ext=.pdf,read=.pdf,width=0.33\textwidth]{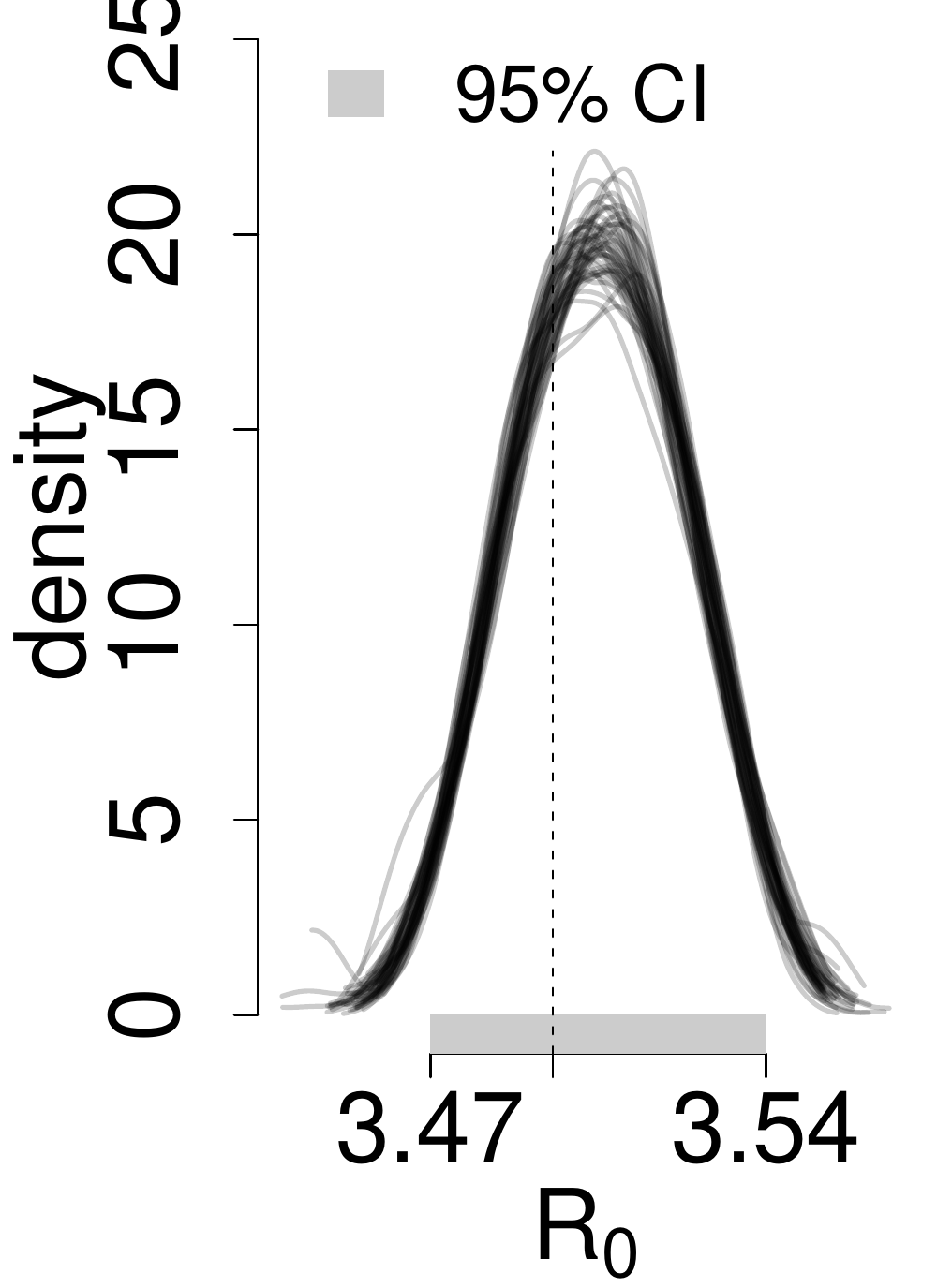}
					\includegraphics[type=pdf,ext=.pdf,read=.pdf,width=0.33\textwidth]{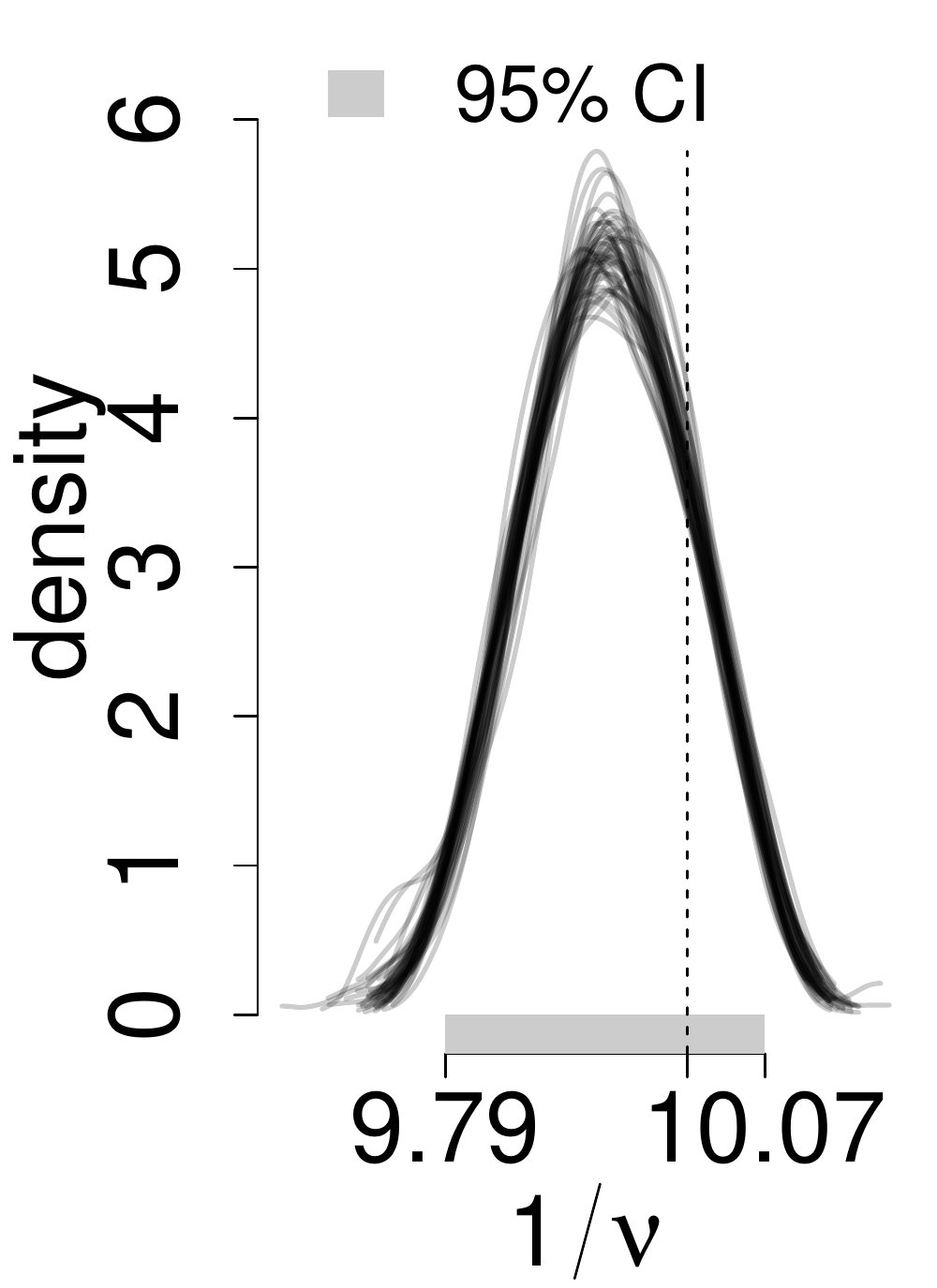}
					\includegraphics[type=pdf,ext=.pdf,read=.pdf,width=0.33\textwidth]{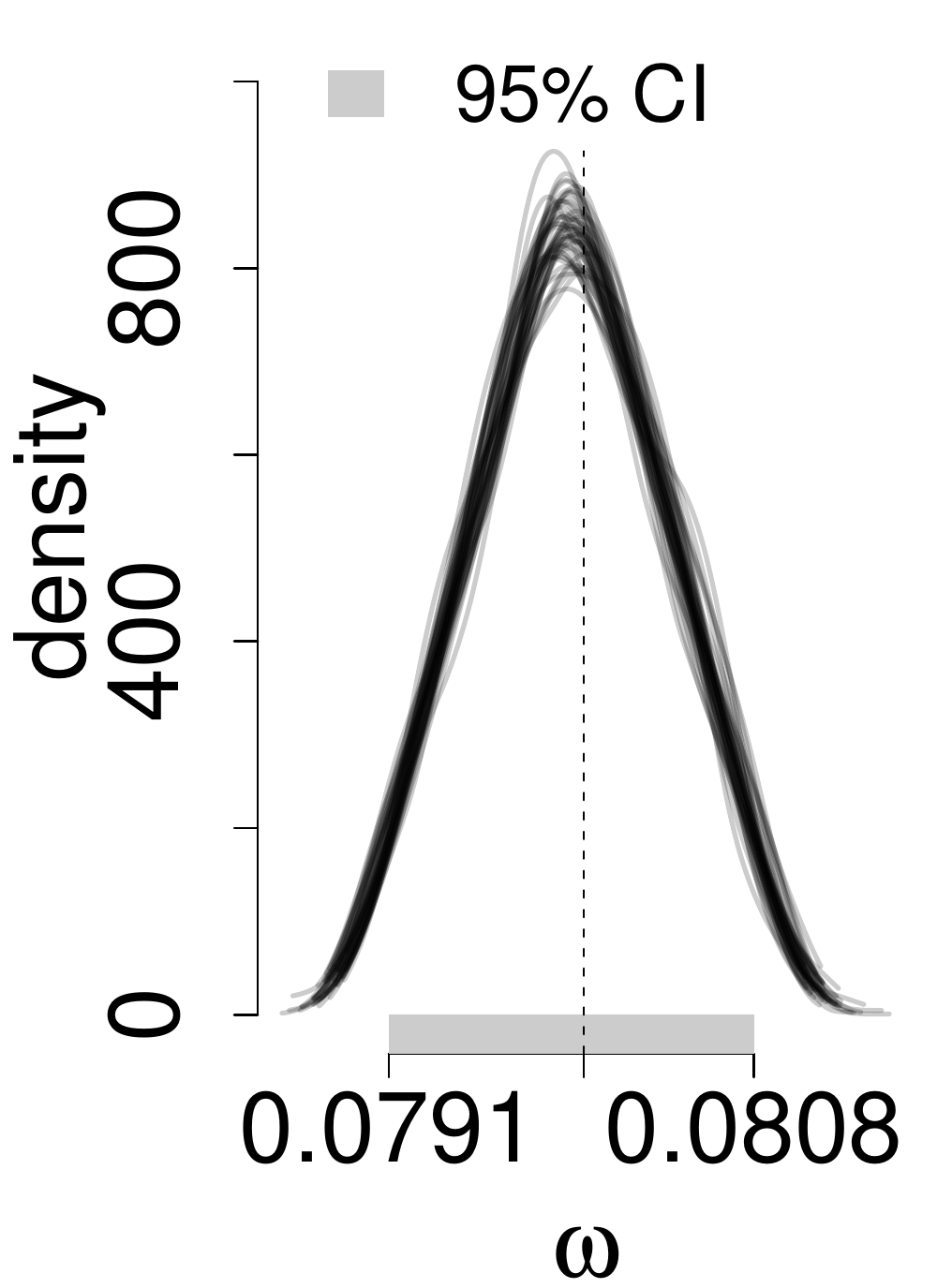}
				\end{minipage}								
		\end{minipage}		
	\end{minipage}			
	\end{minipage}					
	\caption{
{\bf Summary values for influenza A (H3N2) time series data and \nABC\ inference}
(A) Weekly ILI time series that is attributable to influenza A (H3N3) virus in the Netherlands under a regression model (gray), plus simulated incidence data under model \eqref{e:classofmodels} (black lines) with parameters $\Ro=3.5$, $1/\dincub=0.9$, $1/\di=1.8$, $1/\wane=10$, $\repR=0.08$, $N^\src=2\times 10^8$, $1/\birth^\src=50$, $\seas^\snk=0.4$, $\seas^\src=0.005$, $\trav^\snk=4\times 10^6$, $\trav^\src=0.01$. We considered three sets of summary values: (B) ILI annual attack rates (\sMATT, cumulated seasonal ILI data divided by population size) for the simulated and empirical ILI data, (C) the logarithm of the first order differences of \sMATT\ (\sVATT), and (not shown) population-level incidence attack rates (\sXATT). (D) QQ-plots for the odd values of \sMATT, \sVATT\ and \sXATT\ versus the normal distribution for a pseudo data set. (E) \nABC\ level sets were reconstructed as a by-product of the MCMC algorithm to check if the link function is bijective. (left) The reconstructed level set for the odd \sMATT\ (black) is cut orthogonally by the level set for the odd \sVATT\ (blue). (right) The resulting line (grey) is cut by the level set for the odd \sXATT. The remaining level sets did not reduce the resulting small ball close to $\theta_0$ any further. Thus, a very small subset of model parameters close to $\theta_0$ maps to $\rho^\star$, which suggests that $\Link$ is bijective around $\rho^\star$. The videos can be played if the pdf is opened in Acrobat Reader. (F) The marginal posterior densities of $\theta$,  estimated with standard ABC for $50$ pseudo data sets $x\sim\theta_0$. (G) The same using \nABC. The \nABC\ calibrated $\ABCthresh^-_k$, $\ABCthresh^+_k$ were substantially tighter than those used with standard ABC and we had no choice but to improve our MCMC sampler, leading to substantially more accurate parameter estimates.
}\label{f:SEIIRS_datasets}
\end{figure}

To illustrate that \nABC\ can be applied to complex inference problems, we consider weekly influenza-like-illness sentinel surveillance data that is attributable to the influenza A (H3N2) virus in the Netherlands (Figure~\ref{f:SEIIRS_datasets}). The magnitude, variation and correlation in annual attack rates (cumulative incidence per winter season divided by population size) are characteristic features of seasonal H3N2 dynamics \citep{Ratmann2012}. We considered annual attack rates of the reported ILI time series data (\sMATT), their first-order differences (\sVATT) and independent estimates of annual population-level attack rates in H3N2 seasons (\sXATT). The \sMATT, \sVATT\ and \sXATT\ of the simulated data are biennial, and those for the empirical data are weakly so (Figure~\ref{f:SEIIRS_datasets}). We formed two sets of summary values for each of the \sMATT, \sVATT\ and \sXATT\ by retaining odd and even values respectively (i.~e. $K=6$). Stationarity, independence and normality of these summary values could not be rejected (Figure~\ref{f:SEIIRS_datasets}). We therefore chose the TOSTs from \citep{Wellek2003} to test for location equivalence (A2 met).

We aim to fit an epidemiological compartment model which describes H3N2's disease dynamics in a population stratified into susceptible (S), infected but not yet infectious (E), infectious ($I_1$ and $I_2$) and immune (R) individuals across two large spatial areas. The transmission rate $\transtwo{t}{\snk}= \trans(1+\seas^\snk\sin(2\pi (t-t^\snk)))$ is seasonally forced in the sink population that represents the Netherlands  (indicated by $^\snk$), and only weakly so in the source population (indicated by $^\src$) where the virus persists and from where the winter epidemics in the sink population are seeded. $1/\wane$ is the average duration in years of host immunity to the virus. All parameters and accompanying prior distributions are described in Table~\ref{t:flu example full}, and the model is easily simulated from Markov transition probabilities derived from the deterministic ordinary differential equations
\begin{equation}\label{e:classofmodels}
\begin{split}
	&\frac{dS^\snk}{dt}=\:\birth^\snk(N^\snk-S^\snk) - \transtwo{t}{\snk}\frac{S^\snk}{N^\snk}(I_{1}^\snk+I_{2}^\snk+M^\snk) + \wane R^\snk\\
	&\frac{dE^\snk}{dt}=\:\transtwo{t}{\snk}\frac{S^\snk}{N^\snk}(I_{1}^\snk+I_{2}^\snk+M^\snk)-(\birth^\snk+\dincub)E^\snk\\
	&\frac{dI_1^\snk}{dt}=\: \dincub E^\snk-(\birth^\snk+2\di)I_1^\snk\\
	&\frac{dI_2^\snk}{dt}=\: 2\di I_1^\snk-(\birth^\snk+2\di)I_2^\snk\\
	&\frac{dR^\snk}{dt}=2\di I_2^\snk - (\birth^\snk + \wane) R^\snk	
\end{split}
\end{equation}
where $N^\snk=S^\snk+E^\snk+I_1^\snk+I_2^\snk+R^\snk$, and identically for the source population with $\seas^\src\ll\seas^\snk$ (equations not shown). Infected travellers from the source population are defined by $M^\snk= m^\snk ( I_{1}^\src+I_{2}^\src) / N^\src$ where the number of travelling individuals $m^\snk$ is a model parameter; see \cite{Ratmann2012} for details. To fit \eqref{e:classofmodels} to the ILI time series, we used a Poisson observation model of new weekly ILI cases with mean $\repR I_1^{\snk+}$, where $I_1^{\snk+}$ is the number of newly infected individuals in the sink population per week and $\repR$ is the reporting rate. 

The key parameters of interest to us are $\theta=(\Ro,1/\wane,\repR)$, where the baseline transmission rate $\trans$ is re-parameterised into the basic reproductive number $\Ro$ at disease equilibrium for ease of interpretation. We used the broad prior densities $\Ro\sim\mathcal{U}(1,8)$, $1/\wane\sim\mathcal{U}(2,30)$, $\repR\sim\mathcal{U}(0,1)$ throughout. The $\theta$ are unidentifiable under the exact likelihood of the ILI time series due to loss of host immunity to the virus after infection \citep{Truscott2011}. In particular, $\Ro$ and $1/\wane$ are strongly positively correlated over a large range of values.

We hypothesised that the $\theta$ are identifiable from  ILI time series data {\it and annual H3N2 incidence attack rates} \citep[between $10$ to $20\%$ for the Northern Hemisphere]{Cox2000} with ABC techniques, because the latter add absolute information on the burden of seasonal infection. Pseudo data sets $x$ (comprising both types of data) were generated for several realistic parameter combinations, e.~g. $\theta_0=(3.5,10,0.08)$ in Figure~\ref{f:SEIIRS_datasets}. The \sMATT\ and \sVATT\ were computed from the ILI cases simulated under the Poisson observation model, and \sXATT\ was directly computed from simulated incidence. Previously, we used a Markov Chain Monte Carlo (MCMC) standard ABC sampler based on \eqref{e:intersection}, and with the $d_k$'s set to the difference in the simulated and observed sample means of the even \sMATT, \sVATT\ and \sXATT. The $\ABCthresh^-_k$, $\ABCthresh^+_k$ were set such that our sampler produced a manageable acceptance probability of 5\%. We found that the $\theta$ are identifiable based on the ILI time series data and annual attack rate estimates. However, while the estimated 95\% credibility intervals contained $\theta_0$, the mean squared error of the posterior mean or $\map[\abc]$ to $\theta_0$ (averaged over $50$ pseudo data sets) was large and depended on the $\ABCthresh^-_k$, $\ABCthresh^+_k$ (top right in Table~\ref{t:uabc} and Figure~\ref{f:SEIIRS_datasets}). 

Now, we used an MCMC \nABC\ sampler, combining the six calibrated TOST statistics with the composite approach that is equivalent to \eqref{e:intersection}. The link function under the six test statistics for the odd and even \sMATT, \sVATT\ and \sXATT\  is a priori unknown. To verify whether $\Link$ is bijective around $\rho^\star$, we used the relation
\begin{equation}\label{e:refsettest}
{\Link}^{-1}(\rho^\star)= \bigcap_{k=1}^K\:{\Link}_k^{-1}(\rho_k^\star),\quad K=6,
\end{equation}
and estimated ${\Link}_k^{-1}$ numerically around $\rho_k^\star$ with local polynomial regression techniques as a by-product of \nABC\ output \citep{Loader1999}. The numerical estimates of the level sets ${\Link}_k^{-1}(\rho_k^\star)$ are highly complex and their intersection forms a very small ball, indicating that the link function is bijective around $\rho^\star$ (A3 met; Figures~\ref{f:SEIIRS_datasets}E-F). This justifies using the \sMATT, \sVATT\ and \sXATT\ for inference. The TOSTs depend on the sample standard deviation of the $\fsu[m](y)$. We therefore need to recalibrate $\ithresh^-_k$, $\ithresh^+_k$, $m_k$ and $\ABCthresh^-_k$, $\ABCthresh^+_k$ at every \nABC\ iteration (Figure~\ref{f:SEIIRS_simCali}). These $\ABCthresh^-_k$, $\ABCthresh^+_k$ were considerably tighter than those used in the previous standard ABC sampler for numerical convenience, which forced us to use a more sophisticated MCMC sampler (SOM~\ref{a:mcmc}). The \nABC\ posterior density was much tighter than the one obtained with standard ABC, and the mean square error of the posterior mean or $\map[\abc]$ to $\theta_0$ (averaged over $50$ pseudo data sets) was $\approx 20$-fold smaller than previously (bottom right in Table~\ref{t:uabc} and Figure~\ref{f:SEIIRS_datasets}). We could not verify A1 directly. However, adding further summary values such as the initial seasonal growth rates or peak timing did not change the $\pi_{\abc}(\theta|x)$ considerably, and removing any of the six statistics lead to less robust inference or larger mean square error. A4-A5 are not met, so that overall we expect that some approximation error remains even with the \nABC\ calibrations. A re-analysis of the empirical data will be reported elsewhere.

\section{Discussion}
Despite many recent methodological advances, the accuracy of the basic ABC accept/reject sampler with non-zero tolerances ($\ABCthresh>0$ or $\ABCthresh_k^-<\ABCthresh_k^+$) has not been understood even when sufficient statistics are available \citep{Grelaud2009, Dean2011, Drovandi2011, Barnes2012, Fearnhead2012, Filippi2013, Marin2013}. Here, we considered the accuracy of $\pi_{\abc}(\theta|x)$ in terms of its KL divergence to $\pi(\theta|x)$ and in terms of the closeness of the MAP estimate $\map[\abc]$ to the one of $\pi(\theta|x)$. We showed through three innovations how and when modelling distributions of summary values leads to accurate ABC parameter inference in the classical, computationally feasible setting that $\ABCthresh>0$ or $\ABCthresh^-_k<\ABCthresh^+_k$, without asymptotic arguments. 

First, we constructed a simple, yet appropriate auxiliary probability space through modelling the distribution of summary values, data points on a summary level. Our premise is that such stable and characteristic features can be extracted from high-dimensional data \citep{Wood2010}. Then, as in maximum-likelihood based indirect inference \citep{Gourieroux1993}, the rationale is to control the \nABC\ approximation on the auxiliary space and to provide regularity conditions under which $\pi_{\abc}(\theta|x)$ remains close to $\pi(\theta|x)$ on the original parameter space. In particular, the link function $\Link$ must be bijective and continuously differentiable, and the determinant of the Jacobian $\abs{\determinant\partial\!\Link}$ must be controlled by the amount of data available for accurate MAP estimates. Bijectivity determines the number of tests required ($K=\dim(\theta)$), and choosing different sets of $K$ summary values influences the smoothness of $\abs{\determinant\partial\!\Link}$. Monte Carlo output can be used to verify A3 based on \eqref{e:refsettest} as well as A4 and hence A5. We were less successful in verifying A6.

Second, the likelihood approximation under the binary \nABC\ accept/reject step can be interpreted as the power function of a suitable hypothesis test on the auxiliary space. Note that auxiliary data $y$ are repeatedly simulated  for given $\theta$ in ABC, so classical techniques can be naturally embedded within this Bayesian framework \citep{Rubin1984}. 
Test statistics satisfying A2.1-A2.4 can be constructed for a variety of simple probability models \citep{Wellek2003} and replace the arbitrary $d_k$ and $S_k$ in standard ABC.
It would be useful to extend A2 to more complex probability models. Such extensions could use the large sample theory for equivalence tests \citep{Lehmann2005}, or non-parametric tests \citep{Munk1998}, or once-and-for-all power simulations. 

Third, the free \nABC\ parameters of these tests are $\ithresh^-_k$, $\ithresh^+_k$, $m_k$, and they can be calibrated prior to the accept/reject step such that each power function is very close to the likelihood of each of the observed summary values on the auxiliary space. The critical regions $\ABCthresh^-_k$, $\ABCthresh^+_k$ of the tests are calculated from the $\ithresh^-_k$, $\ithresh^+_k$, $m_k$; plus further statistics of the simulated and observed summary values in case of a composite hypothesis test. 
They correspond to the standard ABC tolerances and \nABC\ specifies them exactly. 
In particular, the $\ABCthresh^-_k$, $\ABCthresh^+_k$ are not any longer \apostr{adjustable} to obtain a well-mixing algorithm at some unspecified loss in accuracy, as is common practice with standard ABC. It is harder to implement an efficient \nABC\ Monte Carlo sampler, but well worth the effort as we demonstrated on the flu example.
We also showed that combining the univariate, calibrated tests with the intersection approach leads to a composite likelihood approximation. It seems difficult to adjust for the correlations between summary values except in simple cases such as the Mahalanobis approach. This suggests to focus on identifying and modelling uncorrelated sets of summary values.

The set of assumptions A1-A6 set out when \nABC\ with non-zero tolerances ($\ABCthresh>0$ or $\ABCthresh_k^-<\ABCthresh_k^+$) can be as accurate as Bayesian computations for which the likelihood is known. Given the previous lack in understanding the basic ABC accept/reject sampler with non-zero tolerances, we feel these assumptions - together with the full specification of the test statistics, the tolerances and the number of simulated summary values - are overall a strength of \nABC. In our applications, we found that the accuracy of  $\pi_{\abc}(\theta|x)$ suffers most strongly when A3 is not met, strongly when A1 is not met, and considerably less when A2 or A4 are not met. Finally, while the arguments behind \nABC\ are at times subtle, our main observation is very simple and intuitive: unless we are comfortable with asymptotic arguments, the ABC accept/reject decision cannot be based on a single value. Understanding and modelling the data - now on a summary level with $n_k, m_k>1$ summary values - is key to ABC inference as for all Bayesian computations for which the likelihood is tractable.

\section*{Acknowledgments} We thank Ioanna Manolopoulou and Christian Robert for detailed and thoughtful comments. 
Epidemiological data was kindly provided by the Netherlands Institute for Health Services Research (NIVEL) and the National Institute for Public Health and the Environment (RIVM), The Netherlands. 
OR is supported by the Wellcome Trust (fellowship WR092311MF) and AC is supported by the Medical Research Council (fellowship MR/J01432X/1). 
Computations were performed at the Imperial College High Performance Computing Service, and we thank Simon Burbidge and Matt Harvey for their support.



\end{document}


		\renewcommand\thesection{S\arabic{section}}
		\renewcommand\theequation{S\arabic{equation}}
		\renewcommand\thefigure{S\arabic{figure}}
		\renewcommand\thetable{S\arabic{table}}
		\setcounter{equation}{0}
		\setcounter{figure}{0}
		\setcounter{table}{0}

		\section{Univariate equivalence statistics}\label{a:univariateequivalencestatistics}
		We abbreviate for convenience $s_y=\fsu[m](y)$ and $s_x=\fsu[n](x)$, their sample means $\bar{s}_y$, $\bar{s}_x$ and standard deviations $\hat{\sigma}_y$, $\hat{\sigma}_x$.

		\paragraph{Location equivalence, normal $s_x^i$, $s_y^j$.}
		Suppose the $s^i_x$ and $s^j_y$, $i=1,\dotsc,n$ and $j=1,\dotsc,m$, are iid normal with means $\mu_x$, $\mu(\theta)$ and common, unknown variance $\sigma^2$. Consider the maximum likelihood estimate $\hat{\mu}_x=\bar{s}_x$ of $\mu_x$. Following \cite{Schuirmann1981}, we reject the one-sample version of the two one-sided test statistics (TOST)
		\begin{equation*}
			T^-= \frac{\bar{s}_y-\hat{\mu}_x-\tau^-}{\hat{\sigma}_y/\sqrt{m}},\quad T^+= \frac{\bar{s}_y-\hat{\mu}_x-\tau^+}{\hat{\sigma}_y/\sqrt{m}}
		\end{equation*}
		when simultaneously $T^+<t_{\alpha,df}$ and $T^->t_{1-\alpha,df}$ in order to test
		\begin{equation*}
			H_{0}\colon\quad\mu(\theta)-\hat{\mu}_x\notin[\ithresh^-,\ithresh^+]\text{ versus }H_{1}\colon\quad\mu(\theta)-\hat{\mu}_x\notin[\ithresh^-,\ithresh^+].
		\end{equation*}
		Here, $t_{\alpha,df}$ is the lower $100\alpha$ percentile of a Student t-distribution with $df=m-1$ degrees of freedom. The test is size-$\alpha$ \citep{Berger1996} and centred at $\rho^\star=0$ when $\ithresh^-=-\ithresh^+$. In this case, the power of the TOST is
		\begin{equation*}
			P_x(R|\rho)= F_{t_{df,ncp}}\Big(\frac{\ithresh^+}{\hat{\sigma}_y/\sqrt{m}}+t_{\alpha,df}\Big)	-	F_{t_{df,ncp}}\Big(-\frac{\ithresh^+}{\hat{\sigma}_y/\sqrt{m}}-t_{\alpha,df}\Big)
		\end{equation*}
		with $ncp=\sqrt{m}\rho/\sigma$, and approximated by replacing $\sigma$ in $ncp$  by $\hat{\sigma}_y$ \citep{Owen1965}.

		\paragraph{Dispersion equivalence, normal $s_x^i$, $s_y^i$.} See main text.

		\paragraph{Equivalence in autocorrelations, normal $s_x^i$, $s_y^j$.}
		Suppose the pairs $(s_x^i,s_x^{i-1})$, $(s_y^j,s_y^{j-1})$, $i=2,\dotsc,n$ and $j=2,\dotsc,m$ are bivariate normal with correlations $\varrho_x$ and $\varrho(\theta)$ for fixed $i$ and $j$ respectively. Thin to $(\tilde{s}_x^i,\tilde{s}_x^{i-1})$, $(\tilde{s}_y^j,\tilde{s}_y^{j-1})$, $i=1,\dotsc,\tilde{n}$ and $j=1,\dotsc,\tilde{m}$, such that these pairs can be considered independent. Compute the sample Pearson correlation coefficients $r(\tilde{s}_x^i,\tilde{s}_x^{i-1})$, $r(\tilde{s}_y^j,\tilde{s}_y^{j-1})$ and their Z-transformations $z_x$, $z_y$, using $z(r)= \atanh(r)$, which are approximately normal with mean $\varrho_x$, $\varrho_y$ and variance $1/(\tilde{n}-3)$ \citep{Hotelling1953}. Let $\hat{\varrho}_x=z_x$, which is a slightly biased estimate of $\varrho_x$ \citep{Hotelling1953}. We reject the TOSZ
		\begin{equation*}
			T^-= \frac{z_y-\hat{\varrho}_x-\tau^-}{\sqrt{1/(\tilde{m}-3)}},\quad T^+= \frac{z_y-\hat{\varrho}_x-\tau^+}{\sqrt{1/(\tilde{m}-3)}}
		\end{equation*}
		for
		\begin{equation*}
			H_{0}\colon\quad\varrho(\theta)-\hat{\varrho}_x\notin [\ithresh^-,\ithresh^+]\text{ versus }H_{1}\colon\quad\varrho(\theta)-\hat{\varrho}_x\in [\ithresh^-,\ithresh^+]
		\end{equation*}
		when simultaneously $T^+<u_{\alpha}$ and $T^->u_{1-\alpha}$, where $u_{\alpha}$ is the lower $100\alpha$ percentile of a standard Normal. Since the one-sided tests are both approximately size-$\alpha$, the TOSZ is also approximately size-$\alpha$ \citep{Berger1996}. It is centred at $\rho^\star=0$ when $\ithresh^-=-\ithresh^+$. Under the normal approximation, the power of the TOSZ is in this case
		\begin{equation*}
			P_x(R|\rho)= F_{\mathcal{N}(0,1)}\Big(\frac{\ithresh^+-\rho}{\sqrt{1/(\tilde{m}-3)}}+u_\alpha\Big)	-	F_{\mathcal{N}(0,1)}\Big(-\frac{\ithresh^++\rho}{\sqrt{1/(\tilde{m}-3)}}-u_\alpha\Big).
		\end{equation*}

		\section{Calibration procedures}\label{a:calibration}
		For the simple auxiliary probability models considered here, test statistics can be found such that the power function is continuous in $\rho_k$, $\ithresh_k^-$, $\ithresh_k^+$ and enjoys monotonicity properties such that calibrations are particularly straightforward. First, for each $k$, it is possible to calibrate the tolerances so that the univariate power functions are maximised at the point of equality $\rho^\star_k$. 

		\begin{lemma}{{\bf (Univariate calibration of $\ithresh^-_k$)}}\label{th:tolerancecalibration_num}
			Suppose A2.1-A2.3 hold true. Consider critical regions $R_k(\ithresh^-)$ for tolerance regions $[\ithresh^-,\ithresh_k^+]$ with fixed $\ithresh_k^+$ and let $\rho_k^{\max}(\ithresh^-)=\argmax_{\rho_k} P_x(R_k(\ithresh^-)\:|\:\rho_k)$. Let $\ithresh^+_k>\rho^\star_k$ and suppose that $\varepsilon>0$ is small. There are $\ithresh^-_l$, $\ithresh^-_u$ such that $\rho_k^{\max}(\ithresh^-_l)<\rho^\star_k$ and $\rho_k^{\max}(\ithresh^-_u)\geq\rho^\star_k$, and $\ithresh^-_k$ such that  $\rho_k^{\max}(\ithresh^-_k)=\rho^\star_k$ can be found with the binary search procedure\\[-2mm]

			\begin{algorithmic}[1]
				\hspace{-1.32cm}{\bf Calibrate $\ithresh^-_k$}
				\Loop
				\State Set $\tilde{\ithresh}^-_k\leftarrow(\ithresh^-_l+\ithresh^-_u)/2$ and determine $\ABCthresh^-_k,\ABCthresh^+_k$ that satisfy \eqref{e:abcthresh}.
				\State Compute $\tilde{\rho}_k^{\max}=\rho_k^{\max}(\tilde{\ithresh}^-_k)$.
				\State If $\abs{\tilde{\rho}_k^{\max}-\rho^\star_k}<\varepsilon$, set $\ithresh^-_k\leftarrow\tilde{\ithresh}^-_k$ and stop.
				Else if $\tilde{\rho}_k^{\max}>\rho^\star_k$, set $\ithresh^-_u\leftarrow\tilde{\ithresh}^-_k$ and go to line 2.  
				\EndLoop
			\end{algorithmic}
		\end{lemma}

		\hspace{-0.53cm}Proof of Lemma~\ref{th:tolerancecalibration_num}: Let $\ithresh^-_u=\rho^\star_k$. By A2.1, we have $\rho^{\max}_k(\ithresh^-_u)\geq\rho^\star_k$. Using A2.3, $\rho^{\max}_k(\ithresh^-)$ decreases as $\ithresh^-$ decreases, so there is $\ithresh^-_l$ such that $\rho^{\max}_k(\ithresh^-_l)<\rho^\star_k$. Since $\ithresh^-\to \rho_k^{\max}(\ithresh^-)$ is continuous, there is exactly one solution $\ithresh^-_k$ such that $\rho^{\max}_k=\rho^\star_k$, and this solution can be found with a binary search algorithm.
		\vspace{0.2cm}\qed

		These calibrations determine $\ithresh^-_k$ as a function of $\ithresh^+_k$, $m_k$; and possibly further statistics $C$ of the simulated and observed summary values in case of a composite hypothesis test. Second, we calibrate $\ithresh^+_k$ for given $m_k$ (and $C$ if necessary) such that the univariate power functions are not flat around $\rho^\star_k$.
		\begin{lemma}{{\bf (Univariate calibration of $\ithresh^+_k$)}}\label{th:powercalibration_num}
			Suppose A2.1-A2.3 hold true. Consider rejection regions $R_k(\ithresh^+)$ for equivalence regions $[\ithresh^-,\ithresh^+]$ such that $\rho_k^{\max}=\rho_k^\star$ and denote the maximal power by $\gamma(\ithresh^+)= P_x(R_k(\ithresh^+)|\rho_k^\star)$. Suppose that $\varepsilon>0$ is small. Then, there are  $\ithresh^+_l, \ithresh^+_u$ such that $\gamma(\ithresh^+_l)<0.9$ and $\gamma(\ithresh^+_u)>0.9$ and $\ithresh^+_k$ can be found by the binary search procedure\\[-2mm]

			\begin{algorithmic}[1]
				\hspace{-1.32cm}{\bf Calibrate $\ithresh^+_k$}
				\Loop
				\State Set $\tilde{\ithresh}^+_k\leftarrow(\ithresh^+_l+\ithresh^+_u)/2$, calibrate $\tilde{\ithresh}^-_k$ as before and denote the corresponding rejection $\hphantom{xxi}$region by $R_k(\tilde{\ithresh}^+_k)$.
				\State Compute the maximal power, $\gamma(\tilde{\ithresh}^+_k)$.
				\State If $\abs{\gamma(\tilde{\ithresh}^+_k)-0.9}<\varepsilon$, set $\ithresh^+_k\leftarrow\tilde{\ithresh}^+_k$ and stop. 
				Else if $\gamma(\tilde{\ithresh}^+_k)<0.9$, set $\ithresh^+_l\leftarrow\tilde{\ithresh}^+_k$ and $\hphantom{xxi}$go to line 2. 
				Else if $\gamma(\tilde{\ithresh}^+_k)>0.9$, set $\ithresh^+_u\leftarrow\tilde{\ithresh}^+_k$ and go to line 2. 
				\EndLoop
			\end{algorithmic}
		\end{lemma}

		\hspace{-0.53cm}Proof of Lemma~\ref{th:powercalibration_num}: Let $\ithresh^+_l=\rho^\star_k$. By Lemma~\ref{th:tolerancecalibration_num}, the calibrated $\ithresh^-_l$ is also $\rho^\star_k$. Since $T_k$ is continuous and level-$\alpha$, we have $\gamma(\ithresh^+_l)\leq\alpha$ which is of course smaller than $0.9$. We next show that $\ithresh^+_k\to\gamma(\ithresh^+_k)$ is monotonically increasing with $\ithresh^+_k$. Consider $\ithresh^+_1<\ithresh^+_2$ along with two tests $\phi_1(y)=\Ind\{\ABCthresh^-_1\leq T(y) \leq\ABCthresh^+_1\}$, $\phi_2(y)=\Ind\{\ABCthresh^-_2\leq T(y) \leq\ABCthresh^+_2\}$ for equivalence regions $[\ithresh^-_1,\ithresh^+_1]$ and $[\ithresh^-_1,\ithresh^+_2]$. Let $\ithresh^-_1$ be calibrated for $\ithresh^+_1$. Let $\psi(y)=\phi_2(y)-\phi_1(y)$. By Lemmas 3.7.1 and 3.4.2(iv) in \citep{Lehmann2005}, $\psi\neq 0$ and $\mathbb{E}_\rho\psi(y)>0$ for all $\rho>\ithresh^-_1$. Consider now $\phi_3(y)=\Ind\{\ABCthresh^-_3\leq T(y) \leq\ABCthresh^+_3\}$ for $[\ithresh^-_2,\ithresh^+_2]$ such that $\ithresh^-_2$ is calibrated for $\ithresh^+_2$. We have $\ithresh^-_2<\ithresh^-_1$ by Lemma~\ref{th:tolerancecalibration_num}. Repeating the same argument as above, we obtain $\mathbb{E}_\rho\phi_3(y)>\mathbb{E}_\rho\phi_2(y)$ for all $\rho>\ithresh^-_2$. This implies in particular $\gamma(\ithresh^+_1)<\gamma(\ithresh^+_2)$. Thus, there is $\ithresh^+_u$ such that $\gamma(\ithresh^+_u)>0.9$. By A2.1, $\ithresh^+\to \gamma(\ithresh^+)$ is continuous. Hence, there is $\tau^+_k$ such that $\gamma(\ithresh^+_k)=0.9$ and this $\ithresh^+_k$ can be found with a binary search procedure.
		\vspace{0.2cm}\qed

		The ABC approximation is now correctly centred but $\pi_{\abc}(\theta|x)$ may still be broader than $\pi(\theta|x)$ due to the diluting effect of the tolerances $\ithresh^-_k<\ithresh^+_k$. In this case, we also calibrate the number $m_k$ of simulated data points used for each $T_k$ (given further statistics $C$ if a composite hypothesis test is used).
		\begin{lemma}{{\bf (Univariate calibration of $m_k$)}}\label{th:mcalibration_num}
			Suppose A2.1-A2.4 hold true. Consider rejection regions $R_k(m_k)$ for $m_k$ simulated and $n$ observed summary values such that  $\rho_k^{\max}=\rho_k^\star$ and $P_x(R_k(\ithresh^+_l)|\rho_k^\star)=0.9$. Denote the signed Kullback-Leibler divergence between the probability densities associated with the summary likelihood and the power function by
			\begin{equation*}
				\kappa(m_k)= \sign\big\{\KL(m_k+1)-\KL(m_k)\big\}\: \KL(m_k)
			\end{equation*}
			where 
			\begin{equation*}
				\KL(m_k)= \int \log\Big( \frac{\lkl(\fsu|\rho_k)/C_\lkl}{P_x(R_k(m_k)|\rho_k) / C_{\abc}}\Big) \: \lkl(\fsu|\rho_k)/C_\lkl\: d\rho_k
			\end{equation*}
			and $C_\lkl=\int \lkl(\fsu|\rho_k) d\rho_k$ and $C_{\abc}=\int P_x(R_k(m_k)|\rho_k) d\rho_k$. There is $m_u$ such that $\kappa(m_u)>0$ and $m_k$ can be found by the binary search procedure\\[-2mm]

			\begin{algorithmic}[1]
				\hspace{-1.3cm}{\bf Calibrate $m_k$}
				\State Set $m_l\leftarrow n$.
				\State If $\kappa(m_l)>0$, set $m_k\leftarrow m_l$ and stop.
				\For{$j = 1 \dotsc J$} 
				\State Set $\tilde{m}_k\leftarrow\integer((m_l+m_u)/2)$, calibrate $\tilde{\ithresh}^-_k$, $\tilde{\ithresh}^+_k$ as before and denote the corresponding $\hphantom{xxi}$rejection region by $R_k(\tilde{m}_k)$.
				\State Compute the signed Kullback-Leibler divergence, $\kappa(\tilde{m}_k)$.
				\State If $m_l=m_u$ or $j=J$, set $m_k\leftarrow\tilde{m}_k$ and stop. 
				Else if $\kappa(\tilde{m}_k)<0$, set $m_l\leftarrow\tilde{m}_k$ $\hphantom{xxi}$and go to line 4. 
				Else if $\kappa(\tilde{m}_k)>0$, set $m_u\leftarrow\tilde{m}_k$ and go to line 4. 
				\EndFor
			\end{algorithmic}
		\end{lemma}

		\hspace{-0.53cm}Proof of Lemma~\ref{th:mcalibration_num}: Consider the densities 
		\begin{equation*}
			\begin{split}
				&f(\rho_k)= \lkl(\fsu|\rho_k) / \int \lkl(\fsu|\rho_k) d\rho_k	\\
				&f_{\abc}(\rho_k;m)= P_x(R_k(m)|\rho_k)/\int P_x(R_k(m)|\rho_k) d\rho_k
			\end{split}
		\end{equation*}
		for calibrated $\ithresh^-_k$, $\ithresh^+_k$. If $\kappa(n_k)>0$, then $m_k\leftarrow n_k$ is found. We now suppose that $\kappa(n_k)<0$. We first show that $\kappa(m)$ is monotonically increasing with $m$. Since $T_k$ is consistent,  we have for fixed $\ithresh^-$, $\ithresh^+$ that $P_x(R_k(m+1)|\rho_k)$ is larger than $P_x(R_k(m)|\rho_k)$ for all $\rho_k$. This implies by Lemma~\ref{th:powercalibration_num} that the calibrated $[\ithresh_k^-(m+1),\ithresh_k^+(m+1)]$ is inside the calibrated $[\ithresh_k^-(m),\ithresh_k^+(m)]$. To compare the power of the calibrated tests $\phi_{m}(y)=\Ind\{\ABCthresh_k^-(m)\leq T_k(y;m)\leq \ABCthresh^+_k(m)\}$ and $\phi_{m+1}(y)$, note that $\psi(y)=\phi_{m+1}(y)-\phi_{m}(y)\leq 0$ is non-zero, and that $\mathbb{E}_{\rho_k^\star}\psi(y)=0$, $\mathbb{E}_{\ithresh_k^-(m+1)}\psi(y)<0$,  $\mathbb{E}_{\ithresh_k^+(m+1)}\psi(y)<0$. Along the lines of Lemma 3.4.2(iv) in \citep{Lehmann2005}, it follows that $\mathbb{E}_{\rho_k}\psi(y)<0$ for all $\rho_k\neq \rho^\star_k$. This implies $\kappa(m)<\kappa(m+1)$. In particular, there is $m_u$ such that $\kappa(m_u)>0$. Since $\kappa(n_k)<0$, there is $m_k$ that minimises $\abs{\kappa(m)}$ and this $m_k$ can be found with a binary search algorithm.
		\vspace{0.2cm}\qed

		\section{Proofs of the two Theorems}\label{a:proofs}
		
		Proof of Theorem~\ref{th:accurate}: Since the prior densities $\pi(\rho_k)$ are assumed flat, we have for all $k$ that $\KL( \pi(\rho_k|x) |\!| \pi_{\abc}(\rho_k|x) )=\varepsilon_k$. By A4-A5, we have $P_x(R|\rho)=\prod_{k=1}^K P_x(R_k|\rho_k)$ and similarly for $\lkl(x|\rho)$ so that $\KL( \pi(\rho|x) |\!| \pi_{\abc}(\rho|x) )=\sum_{k=1}^K \varepsilon_k$. Since the Kullback-Leibler divergence is invariant under parameter transformations, the claim follows with A1-A3.\qed

		\hspace{-0.53cm}Proof of Theorem~\ref{th:map}: Following the calibration of all $\ithresh^-_k$, the univariate power functions have a mode at $\rho^\star_k$. By A4, the mode of the multivariate $\rho\to P_x(R|\rho)$ is $\rho^\star=(\rho^\star_1,\dotsc,\rho^\star_K)$. By A5, the MLE of the multivariate $\rho\to\lkl(x|\rho)$ is also $\rho^\star$. Since $\Link$ is bijective, the \nABC\ MLE is also the same as the exact MLE. Next, we have that $\pi_\rho$ does not change the location of the modes of $P_x(R|\rho)$ and of $\lkl(x|\rho)$, so the modes of $\pi_{\abc}(\rho|x)\propto P_x(R|\rho)\pi_\rho(\rho)$ and $\pi(\rho|x)\propto \lkl(x|\rho)\pi_\rho(\rho)$ are again $\rho^\star$. Since $\abs{\partial\!\Link(\theta)}-\abs{\partial\!\Link(\theta^\star)}$ grows slower than $P_x(R|\rho)$ decays around $\rho^\star$, the mode of $\pi_{\abc}(\theta|x)$ is ${\Link}^{-1}(\rho^\star)$. We suppose that the power of the test is broader than $\lkl(s_k|\rho_k)$, so that $\lkl(s_k|\rho_k)$ also controls the change of variables. Therefore, the mode of $\pi(\theta|x)$ is also ${\Link}^{-1}(\rho^\star)$.
		\vspace{0.2cm}\qed

		\section{Further details on the moving average example}
		\label{a:ma}
		\subsection{Prior density $\pi(\theta)$}
		\label{a:ma_prior}
		We assume a uniform prior on $\rho=(\rho_{1},\rho_{2})\sim U([\rho_1^-,\rho_1^+]\times[\rho_2^-,\rho_2^+])$. To obtain the prior induced on $\theta=(a,\sigma^2)$ we decompose the joint $pdf$ as follows: $$f(a,\sigma^2)=f(\sigma^2|a)f(a)$$ where $f$ is a generic density specified by its arguments.

		\subsubsection{Calculation of $f(\sigma^2|a)$}

		We define the following link function, $\Link_{a,\mcnuo[1]}: \sigma^2 \rightarrow \rho_1= (1+a^2)\sigma^2/\mcnuo[1]$, which is monotonically increasing on $\mathbb{R}^+$ and whose inverse is $\Link^{-1}_{a,\mcnuo[1]}:\rho_1 \rightarrow \sigma^2= \mcnuo[1]\rho_1/(1+a^2)$. 
		Since $\Link^{-1}_{a,\mcnuo[1]}$ is linear on $\rho_1$ we have:
		\begin{equation}
			f(\sigma^2|a)=
			\begin{cases}
				\frac{1+a^2}{\mcnuo[1](\rho_1^+-\rho_1^-)}&\mbox{if }  \sigma^2\in [\Link^{-1}_{a,\mcnuo[1]}(\rho_1^-),\Link^{-1}_{a,\mcnuo[1]}(\rho_1^+)]\\
				0&\mbox{otherwise.}
			\end{cases}
			\label{eq:pdf_s2}
		\end{equation}

		\subsubsection{Calculation of $f(a)$}

		We first compute the $cdf$ $F(a)$ and then differentiate it to obtain the $pdf$ $f(a)$. We define the following link function, $\Link_{\mcnuo[2]}: a\rightarrow\rho_2=\atanh(a/(1+a^2))-\atanh(\mcnuo[2])$ which is monotonically increasing on $[-0.5,0.5]$ and whose inverse is 
		\begin{equation*}
			\link^{-1}_{\mcnuo[2]}:\rho_2\rightarrow a=
			\begin{cases}
				0 &\mbox{if }\rho_2=-\atanh(\mcnuo[2]),\\
				\frac{1-\sqrt{1-4\tanh(\rho_2+\atanh(\mcnuo[2]))}}{2\tanh(\rho_2+\atanh(\mcnuo[2]))} &\mbox{otherwise.}
			\end{cases}
		\end{equation*}

		We can now compute:

		\begin{align*}
			F(a)=
			\begin{cases}
				0&\mbox{if } a<\Link^{-1}_{\mcnuo[2]}(\rho_2^-),\\
				\frac{ \link_{\mcnuo[2]}(a)-\rho_2^-}{\rho_2^+-\rho_2^-}&\mbox{if } a\in [\Link^{-1}_{\mcnuo[2]}(\rho_2^-),\Link^{-1}_{\mcnuo[2]}(\rho_2^+)],\\
				1&\mbox{if } a>\Link^{-1}_{\mcnuo[2]}(\rho_2^+),
			\end{cases}
		\end{align*}

		and thus:
		\begin{align}
			f(a)=\frac{dF(x)}{dx}\bigg|_{a}=
			\begin{cases}
				\frac{1-a^2}{(1+a^2+a^4)(\rho_2^+-\rho_2^-)}&\mbox{if } a\in [\Link^{-1}_{\mcnuo[2]}(\rho_2^-),\Link^{-1}_{\mcnuo[2]}(\rho_2^+)],\\
				0&\mbox{otherwise}.
			\end{cases}
			\label{eq:pdf_fa}
		\end{align}

		\subsubsection{Calculation of $f(a,\sigma^2)$}

		Combining equations \eqref{eq:pdf_s2} and \eqref{eq:pdf_fa} we obtain:

		\begin{equation}
			f(a,\sigma^2)=
			\begin{cases}
				\frac{1-a^4}{(1+a^2+a^4)(\rho_2^+-\rho_2^-)\mcnuo[1](\rho_1^+-\rho_1^-)} &\mbox{if }  a\in [\Link^{-1}_{\mcnuo[2]}(\rho_2^-),\Link^{-1}_{\mcnuo[2]}(\rho_2^+)]\\
				& \mbox{and } \sigma^2\in [\Link^{-1}_{a,\mcnuo[1]}(\rho_1^-),\Link^{-1}_{a,\mcnuo[1]}(\rho_1^+)],\\
				0&\mbox{otherwise}.
			\end{cases}
			\label{fasig2}
		\end{equation}

		From a practical point of view it is more natural to parametrize the prior on $\theta$ by specifying boundaries for $a$ and $\sigma^2$ rather than for $\rho_1$ and $\rho_2$. For given boundaries $[a^-,a^+]\times[{\sigma^-}^2,{\sigma^+}^2]$ on $\theta$ we propose to choose the following boundaries for the uniform prior on $\rho$: 
		\begin{equation}
			\begin{cases}
				\rho_1^-=\argmin_{\theta}(\Link_{a,\mcnuo[1]}(\sigma^2))=\Link_{1,\underline{a},\mcnuo[1]}({\sigma^-}^2)\mbox{ with } \underline{a}=\argmin_{a\in[a^-,a^+]}(|a|),\\
				\rho_1^+=\argmax_{\theta}(\Link_{a,\mcnuo[1]}(\sigma^2))=\Link_{1,\overline{a},\mcnuo[1]}({\sigma^+}^2)\mbox{ with } \overline{a}=\argmax_{a\in[a^-,a^+]}(|a|),\\
				\rho_2^-=\argmin_{\theta}(\Link_{\mcnuo[2]}(a))=\Link_{\mcnuo[2]}(a^-),\\
				\rho_2^+=\argmax_{\theta}(\Link_{\mcnuo[2]}(a))=\Link_{\mcnuo[2]}(a^+),
			\end{cases}
			\label{eq:rho_bounds_natural}
		\end{equation}
		which ensures that the prior induced on $\theta$ contains the rectangle $[a^-,a^+]\times[{\sigma^-}^2,{\sigma^+}^2]$. With these bounds the expression of $f(a,\sigma^2)$ becomes:

		\begin{equation}
			f(a,\sigma^2)=
			\begin{cases}
				\frac{1-a^4}{(1+a^2+a^4)\atanh\big(\frac{a^+-a^-}{1-(a^+a^-)^2}\big) ((1+\overline{a}^2){\sigma^+}^2-(1+\underline{a}^2){\sigma^-}^2)}&\mbox{if }  a\in [a^-,a^+] \\
				& \mbox{and }\sigma^2\in \Big[\frac{1+\underline{a}^2}{1+a}{\sigma^-}^2,\frac{1+\overline{a}^2}{1+a}{\sigma^+}^2\Big],\\
				0,&\mbox{otherwise.}
			\end{cases}
			\label{eq:fasig2_natural}
		\end{equation}
		
\begin{figure}[!h]
		\begin{center}
			\includegraphics[width=0.9\linewidth]{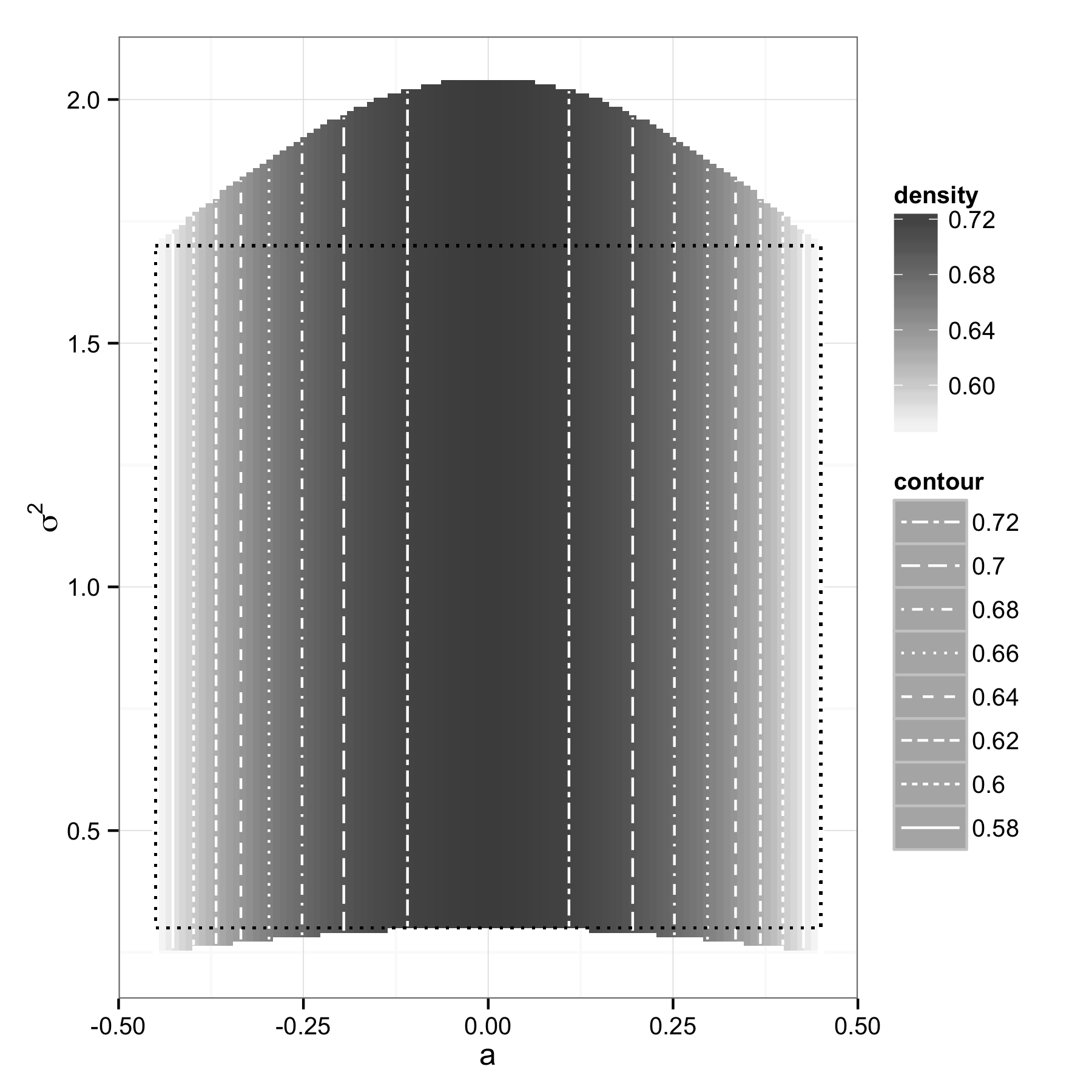}
			\caption{Prior on $\theta=(a,\sigma^2)$ induced by an uniform prior on $\rho=(\rho_1,\rho_2)$, as given by equation \eqref{eq:fasig2_natural}. The boundaries of the uniform prior on $\rho$ are computed using equations \eqref{eq:rho_bounds_natural} with $a^-=-0.43$, $a^+=0.43$, ${\sigma^-}^2=0.3$ and ${\sigma^+}^2=1.7$, which ensure that the prior induced on $\theta$ contains the rectangle $[a^-,a^+]\times[{\sigma^-}^2,{\sigma^+}^2]$ (black dotted line). Here, we have fixed the variance $\mcnuo[1]=(1+a^2)\sigma^2$ and autocorrelation $\mcnuo[2]=\atanh(a/(1+a^2))$ to their theoretical values and obtained the following uniform prior $(\rho_1,\rho_2)\sim U[(-0.493,0.294)\times(0.297,2.024)]$. Note that the prior induced on $\theta$ is rather uninformative.}
			\label{fig:induced_prior}
		\end{center}
	\end{figure}

		An example of prior induced on $\theta$ is shown in Figure~\ref{fig:induced_prior}.

		\subsection{Markov Chain Monte Carlo algorithm for estimating $\pi(\theta|x_{1:n})$}
		\label{a:ma_mcmc}

		When inferring the parameters $\theta=(a,\sigma^2)$ of $x_{1:n}\sim MA(1)$, the past white noise $u_0$ needs also to be inferred. However, since we used simulated data and were interested in the exact posterior of $\theta$ we simply fixed $u_0=0$ and used the likelihood of $x_{1:n}$ conditional on $u_0$ given by \cite{Marin:2007wg}:

		\begin{equation}
			l^c(a,\sigma^2|x_{1:n},u_0)\varpropto\sigma^{-n}\prod_{1}^{n}\exp(-\frac{\hat u_t^2}{2\sigma^2}),
		\end{equation}
		where the $\hat u_t (t>0)$ are given by the recursive formula: $$\hat u_t=x_t-a\hat u_{t-1}.$$

		We implemented a Metropolis-Hasting MCMC algorithm with a bivariate gaussian kernel proposal truncated to the natural support of $\theta$: $[-0.5,0.5]\times\mathbb{R}^+$ and with covariance matrix:
		\[
			\Sigma_\theta=\left(
			\begin{array}{ccc}
				5\times10^{-2} & 5\times10^{-4}    \\
				5\times10^{-4} &  5\times10^{-2}      
			\end{array}
			\right).
		\]
		leading to an acceptance rate of $\sim 20\%$. We ran and combined 6 chains for $2\times10^6$ iterations, starting near the true parameter values.

		\subsection{\nABC\ subsetting procedure}
		\label{a:ma_subset}

		We considered the following subsets for given time series data $x_{1:n}$. 
		First, autocorrelations in $x_{1:n}$ were ignored, leading to $s^i_1=x^i$, $i=1,\dotsc,n$ and $s_{2}=(x_1,x_2),(x_2,x_3),\dotsc$, $i=1,\dotsc,n$ for the variance and correlation test respectively. 
		Second, we thinned $x_{1:n}$ to $s^i_{1a}=x^{2i-1}$ and $s^i_{1b}=x^{2i}$, $i=1,\dotsc,n/2$ for two variance tests, and used $s_{2a}= (x_1,x_2),(x_4,x_5),(x_7,x_8),\dotsc$ and  $s_{2b}= (x_2,x_3),(x_5,x_6),(x_8,x_9),\dotsc$ and  $s_{2c}= (x_3,x_4),(x_6,x_7),(x_9,x_{10}),\dotsc$ for three correlation tests. 
		
		\subsection{Influence of the link function}		
\begin{figure}[!h]	
	\centering			
	\begin{minipage}[c]{0.32\textwidth}		
	\begin{minipage}[b]{0.001\textwidth}
		{\bf A}\newline\vspace{-0.5cm}

	\end{minipage}	
	\begin{minipage}[b]{0.99\textwidth}
		\includegraphics[type=pdf,ext=.pdf,read=.pdf,width=\textwidth]{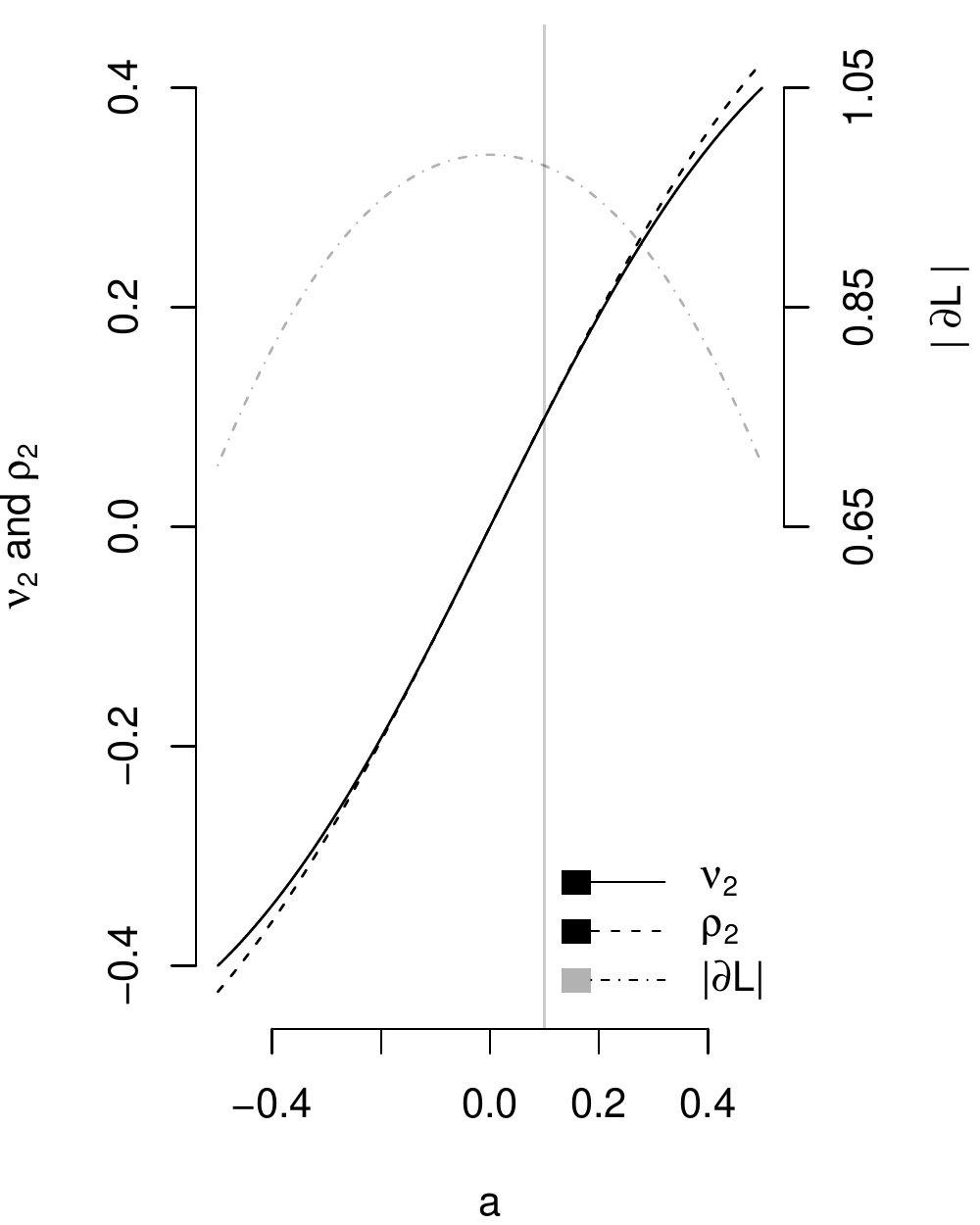}
	\end{minipage}	
\end{minipage}	
\begin{minipage}[c]{0.32\textwidth}		
\begin{minipage}[b]{0.001\textwidth}
	{\bf B}\newline\vspace{-0.9cm}

\end{minipage}	
\begin{minipage}[b]{0.99\textwidth}
	\includegraphics[type=pdf,ext=.pdf,read=.pdf,width=\textwidth]{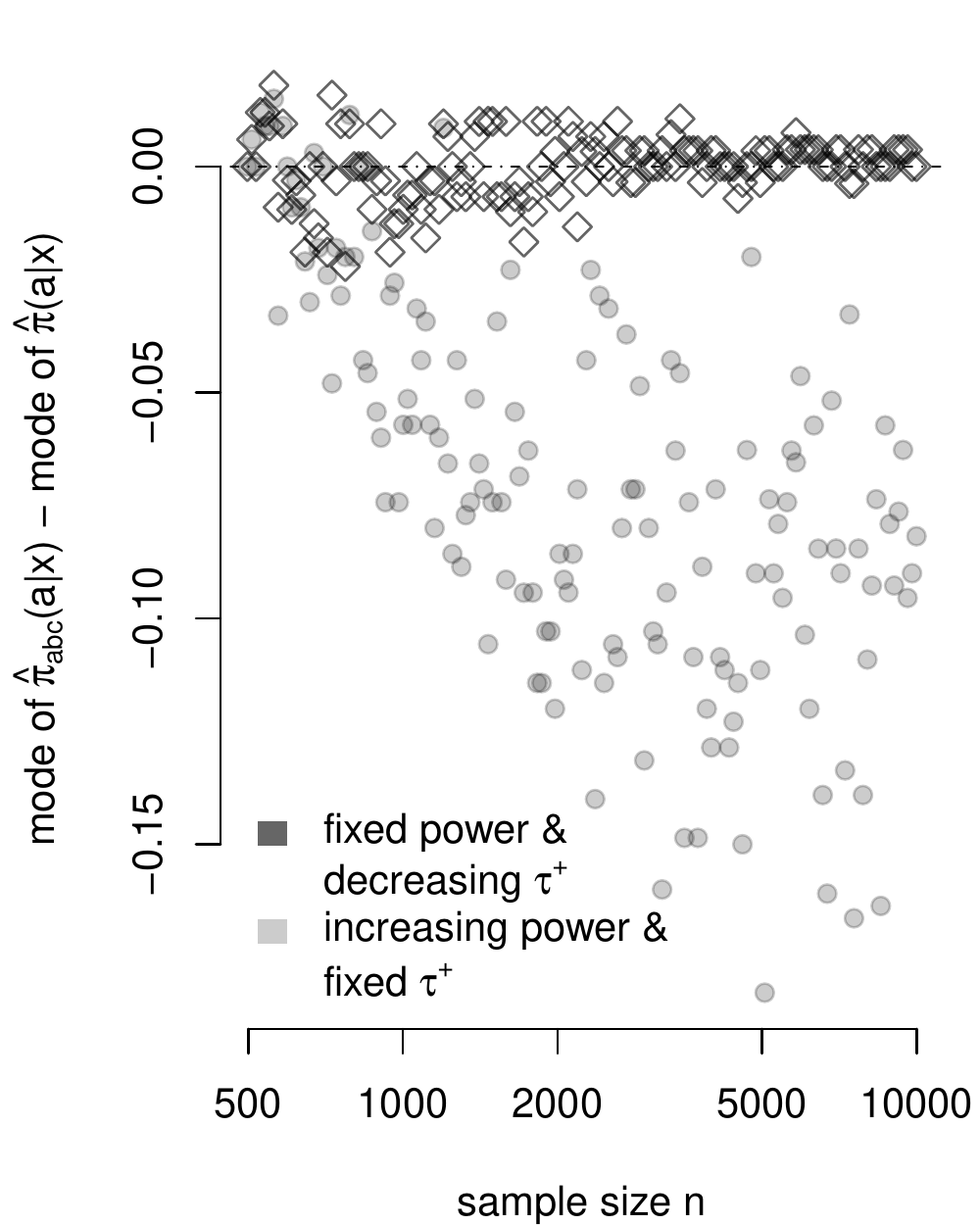}
\end{minipage}	
\end{minipage}	
\begin{minipage}[c]{0.32\textwidth}		
\begin{minipage}[b]{0.001\textwidth}
	{\bf C}\newline\vspace{-0.5cm}

\end{minipage}	
\begin{minipage}[b]{0.99\textwidth}
	\includegraphics[type=pdf,ext=.pdf,read=.pdf,width=\textwidth]{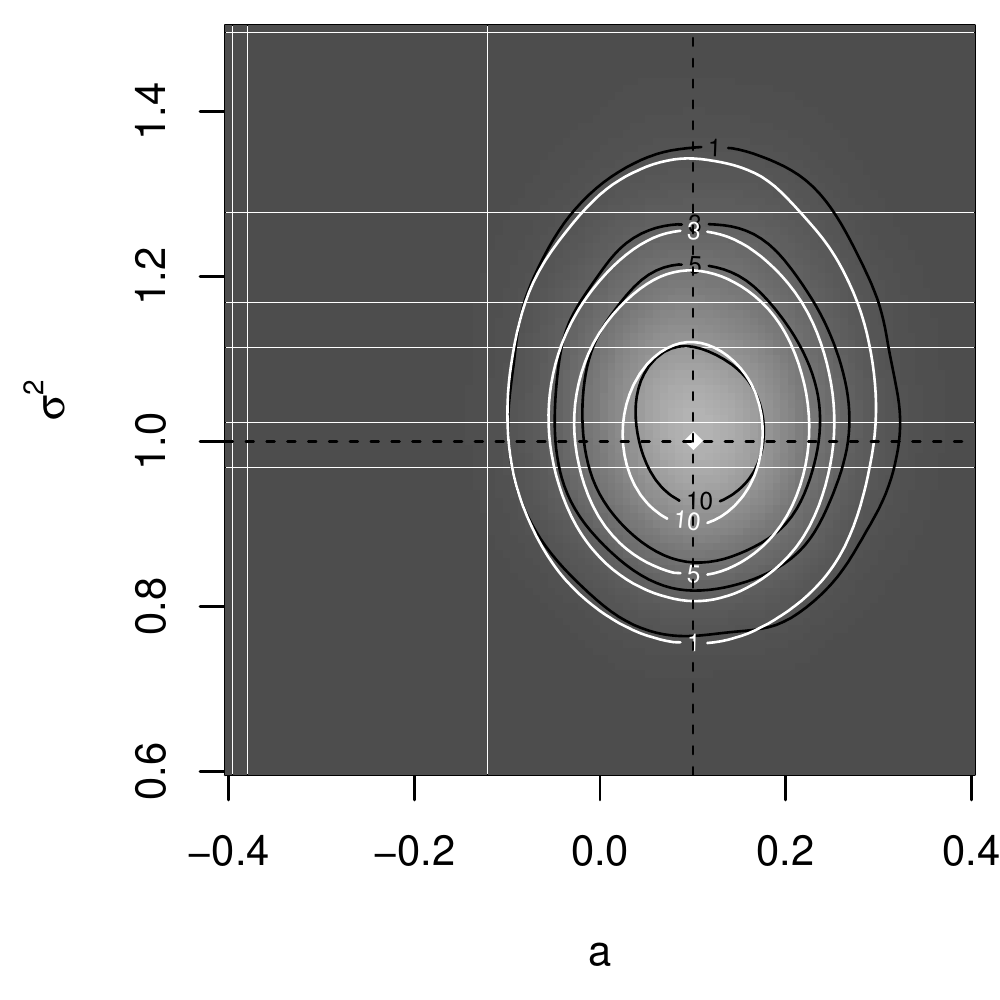}
\end{minipage}			
\end{minipage}			
\caption{{\bf Link function and \nABC\ inference for the MA(1) model.} (A) The link function (black) and $\abs{\determinant\partial\!\Link}$ (gray) as a function of $a$. $\abs{\determinant\partial\!\Link}$ increases to the left of $a_0=0.1$ (vertical grey line). (B) We increased the sample size $n$ but did not re-calibrate $\ithresh^+_k$, expecting that $\map[\abc]$ is increasingly inaccurate as the power function plateaus at one. The $\map[\abc]$ was indeed increasingly inaccurate when the $\ithresh^+_k$ are not re-calibrated for each $m$ (light gray dots). We repeated inference, now with the $\ithresh^+_k$ re-calibrated so that power peaks at $0.9$. There was no systematic difference between $\map[\abc]$ and the exact MAP estimate (A6 met, dark gray dots). (C) Standard ABC posterior density (black contour) as described in the main text, with tolerances set so that \rejABC\ acceptance probability was 0.5\%. Monte Carlo error becomes noticeable and otherwise ABC approximation is comparable to that of \nABC.}\label{f:MA1_infs}	
\end{figure}

		The rate of change $\abs{\determinant\partial\!\Link}$ is non-linear and may compromise the accuracy of point estimates of $\pi_{\abc}(\theta|x_{1:n})$. This is particularly so when $P_x(R|\rho)$ is broad so that $\abs{\determinant\partial\!\Link}$ has considerable support to act on. To illustrate, we increased the sample size $n$ but did not re-calibrate $\ithresh^+_k$, expecting that $\map[\abc]$ is increasingly inaccurate as the power function plateaus at one. We ran \nABC\ for different pseudo data sets that increase from $n=m=500$ to $n=m=10^5$. The $\map[\abc]$ was indeed increasingly inaccurate when the $\ithresh^+_k$ are not re-calibrated for each $m$ (light gray dots in Figure~\ref{f:MA1_infs}D). We repeated inference, now with the $\ithresh^+_k$ re-calibrated so that power peaks at $0.9$. There was no systematic difference between $\map[\abc]$ and the exact MAP estimate (A6 met, dark gray dots in Figure~\ref{f:MA1_infs}D). The amount of data available controlled $\abs{\determinant\partial\!\Link}$ and the calibrations ensured that the \nABC\ MAP estimate was very close to $\map$ (A6 met; Figure~\ref{f:MA1_infs}D-E). 

		\section{Advanced \nABC\ algorithms}\label{a:mcmc}
		\subsection{Markov Chain Monte Carlo algorithm (MCMC)}
		The MCMC algorithm follows from \citep{Marjoram2003}. Set initial values $\theta^0$, compute $y^0\sim\lkl(\,\cdot\,|\theta^0)$ and $z^0_k=T_k\big(\fsu[m](y^0),\fsu(x)\big)$ for all $k$. We suppose that $\ABCthresh^-_k\leq z^0_k\leq\ABCthresh^+_k$ for all $k$.\\[-5mm]
		\begin{singlespaceddescription}
			\item[\mnABC 1] If now at $\theta$, propose $\theta^\prime$ according to a proposal density $q(\theta\rightarrow\theta^\prime)$.
			\item[\mnABC 2] Simulate $y^\prime\sim\lkl(\,\cdot\,|\theta^\prime)$, extract $\fsu[m](y^\prime)$ for all $k=1,\dotsc,K$.
			\item[\mnABC 3] Compute $z^\prime_k=T_k\big(\fsu[m](y^\prime),\fsu(x)\big)$ for all $k$.
			\item[\mnABC 4] Accept $(\theta^\prime,z^\prime)$ with probability
			\begin{equation*}
				\begin{split}
					&mh(\theta,z;\theta^\prime,z^\prime)=\\
					&\quad\min\Bigg\{1\:,\frac{q(\theta^\prime\rightarrow\theta)}{q(\theta\rightarrow\theta^\prime)}\times\frac{\pi(\theta^\prime)}{\pi(\theta)}\times\prod_{k=1}^{K}\Ind\{\ABCthresh^-_k\leq z_k^\prime\leq\ABCthresh^+_k\} \Bigg\},
				\end{split}
			\end{equation*}
			and otherwise stay at $(\theta,x)$. Return to \mnABC 1.
		\end{singlespaceddescription}

		Throughout, we used a Gaussian proposal kernel. Annealing procedures were added to the covariance matrix of the Gaussian proposal kernel and the tolerances $\ithresh^-_k$, $\ithresh^+_k$ during burn-in.

		For the influenza time series example, we previously used a standard ABC MCMC algorithm with annealing schemes on the covariance matrix of a Gaussian proposal matrix and the tolerances. The covariance matrix was diagonal. For \nABC, the calibrated $\ABCthresh^-_k$, $\ABCthresh^+_k$ were considerably smaller than those used previously in the standard ABC routine, and we were forced to improve the MCMC sampler. We estimated a more suitable covariance matrix for the proposal density from a sequence of pilot runs, and also employed an annealing scheme on the covariance matrix as well as the tolerances $\ithresh^-_k$, $\ithresh^+_k$.

		\subsection{Sequential Importance sampling algorithm}
		This algorithm follows in analogy to the above from \citep{Toni2008}.

		\section{Supplementary Figures}\label{s:extrafigs}

		\begin{figure}[!h]
			\centering
			\includegraphics[type=pdf,ext=.pdf,read=.pdf,width=0.5\textwidth]{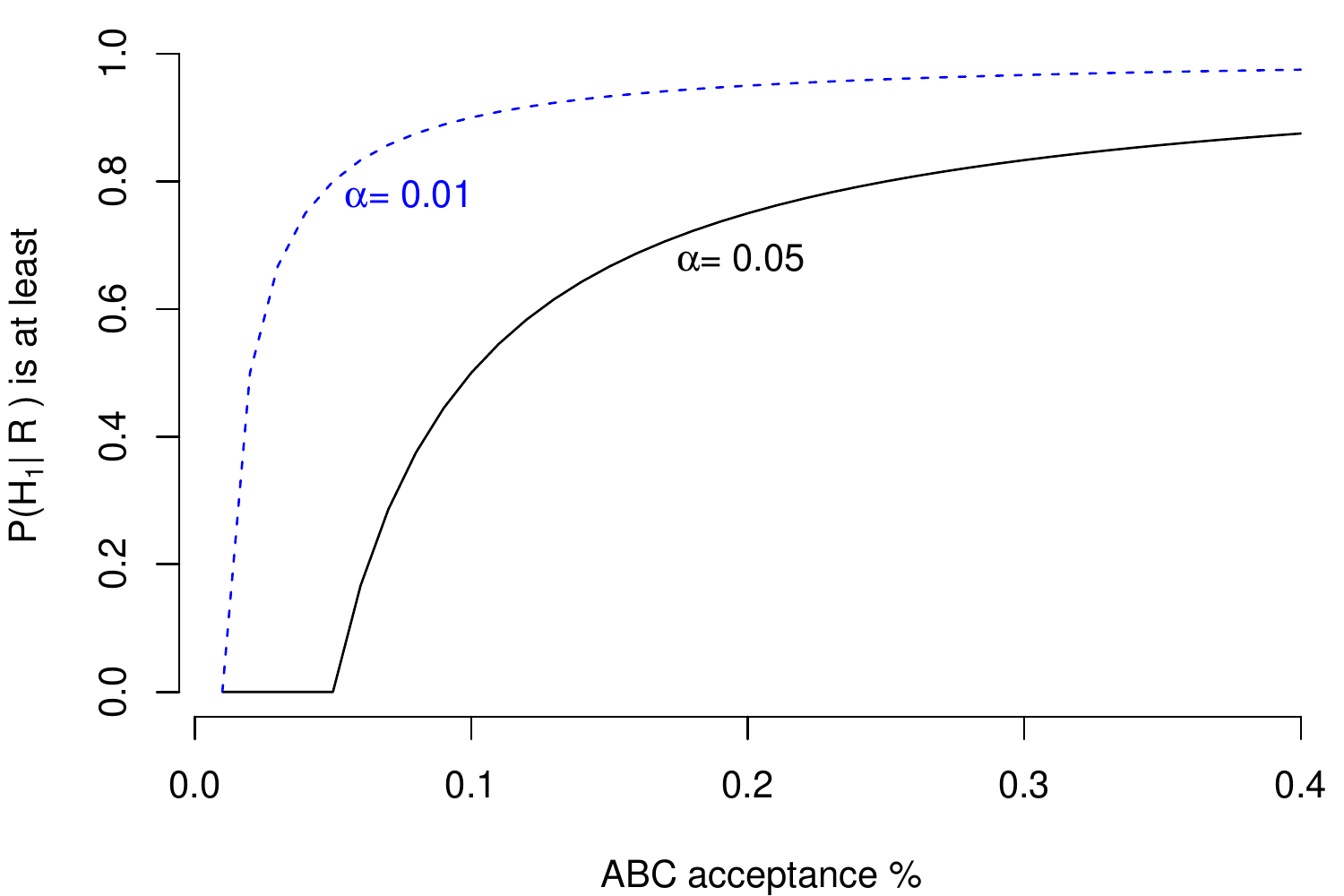}
			\caption{{\bf Lower bound of ABC true positives.}
			The true positive \nABC\ samples among all accepted \nABC\ samples after completion of the algorithm are those $\theta$ for which $\rho=\Link(\theta)$ is in the equivalence region $H_1$. We obtain from Bayes Theorem that the percentage of true positive \nABC\ samples is bounded below by $P_x(H_{1}|R)=1-P_x(H_0|R)\geq 1 - \alpha / P_x(R)$. The lower bound on the probability of correctly rejecting a level-$\alpha$ equivalence test statistic (i.~e. $1-P_x(H_0|R)$) is plotted as a function of the \nABC\ acceptance probability $P_x(R)$ for $\alpha=0.01$ (blue) and $\alpha=0.05$ (black). Since the \nABC\ acceptance probability rarely exceeds $20\%$, we fix $\alpha=0.01$. In this case, if the \nABC\ acceptance probability is smaller than 2\%, then the percentage of true positive \nABC\ samples could be as low as 50\%}\label{f:tpr}
		\end{figure}

		\begin{figure}[!h]
			\centering		
			\begin{minipage}[c]{0.35\textwidth}	
			\includegraphics[type=pdf,ext=.pdf,read=.pdf,width=\textwidth]{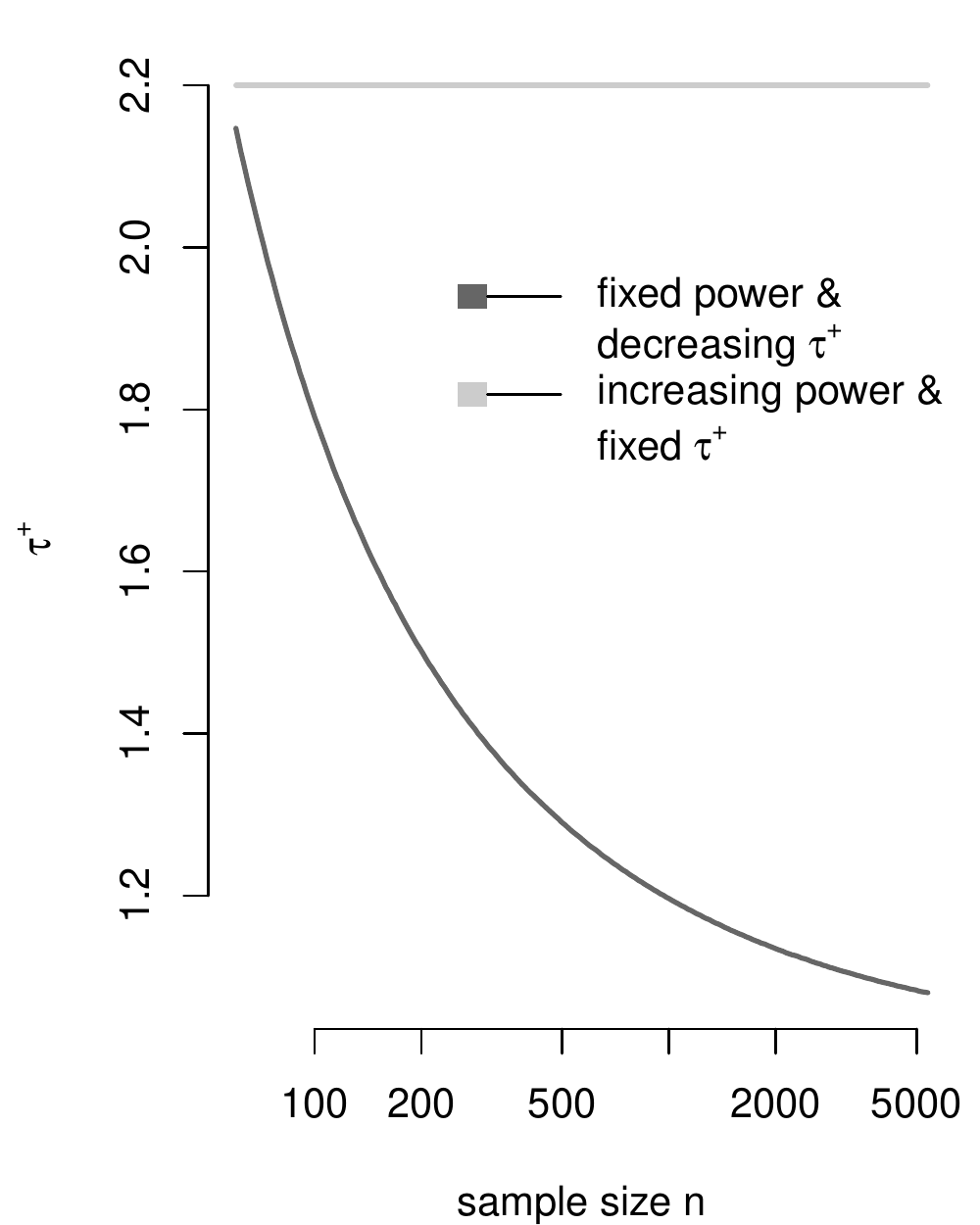}			
		\end{minipage}	
		\caption{
			{\bf \nABC\ inference of $\sigma^2$ under calibrated tolerances $\ithresh^-<\ithresh^+$.}
			Calibrated upper tolerance $\ithresh^+$ as a function of $m$ (dark gray). The dark gray $\map[\abc]$ estimates in Figure~\ref{f:scaledFpower}F were obtained from an \nABC\ run with these calibrated upper tolerances. The light gray $\map[\abc]$ estimates in Figure~\ref{f:scaledFpower}F were obtained from an ABC run with fixed upper tolerance (shown here also in light gray).
		}\label{f:nABCscaledFb}
	\end{figure}

\begin{figure}[!h]
	\centering		
	\begin{minipage}[c]{0.7\textwidth}	
	\begin{minipage}[t]{0.48\textwidth}		
	\begin{minipage}[b]{0.001\textwidth}
		{\bf A}\newline\vspace{4.5cm}

	\end{minipage}	
	\begin{minipage}[b]{0.99\textwidth}
		\includegraphics[type=pdf,ext=.pdf,read=.pdf,width=\textwidth]{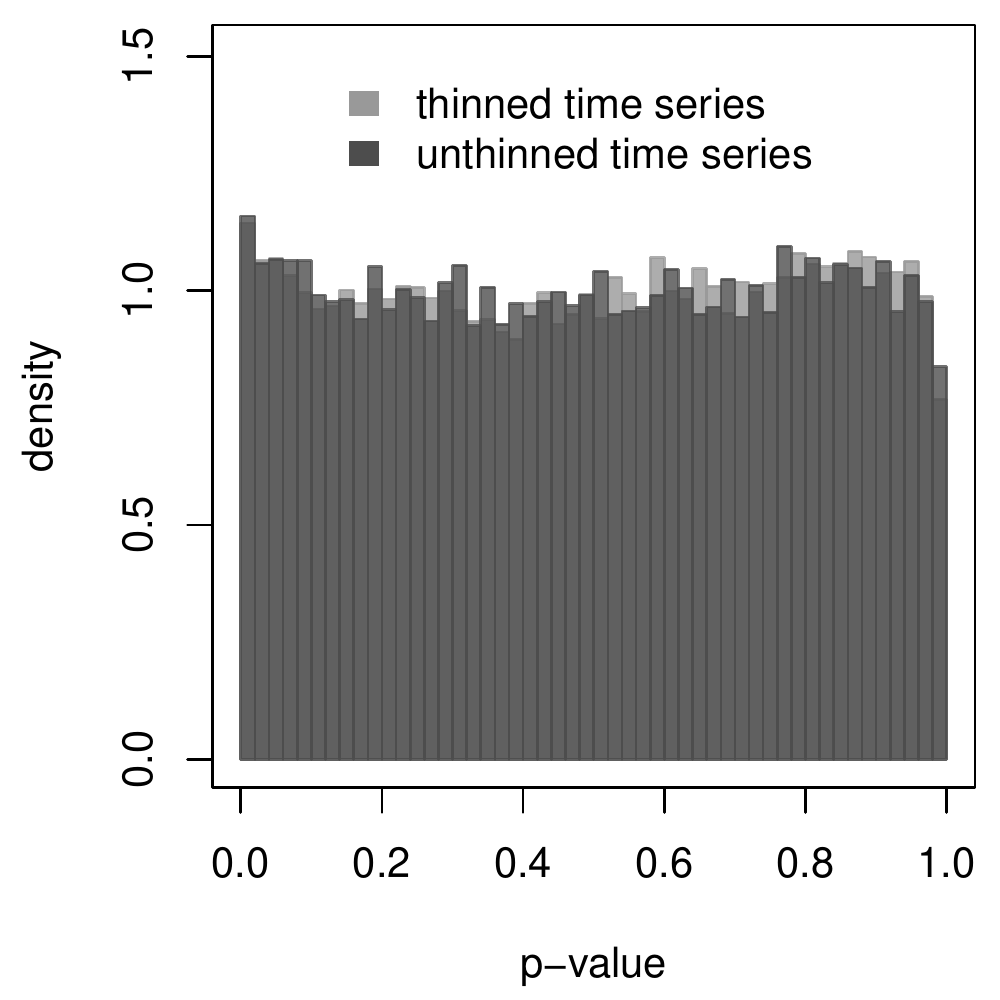}				
	\end{minipage}	
\end{minipage}	
\begin{minipage}[t]{0.48\textwidth}		
\begin{minipage}[b]{0.001\textwidth}
	{\bf B}\newline\vspace{4.5cm}

\end{minipage}	
\begin{minipage}[b]{0.99\textwidth}
	\includegraphics[type=pdf,ext=.pdf,read=.pdf,width=\textwidth]{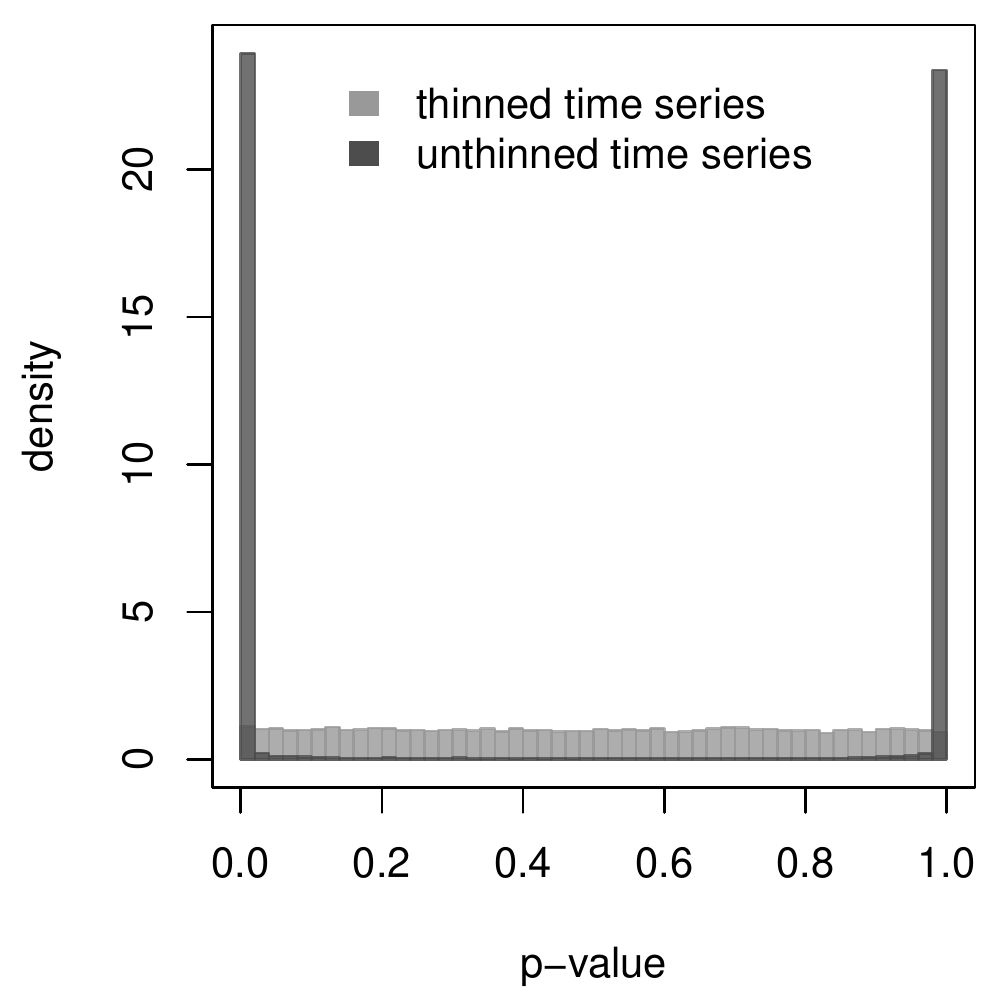}
\end{minipage}	
\end{minipage}	
\begin{minipage}[t]{0.48\textwidth}		
\begin{minipage}[b]{0.001\textwidth}
	{\bf C}\newline\vspace{4cm}

\end{minipage}	
\begin{minipage}[b]{0.99\textwidth}
	\includegraphics[type=pdf,ext=.pdf,read=.pdf,width=\textwidth]{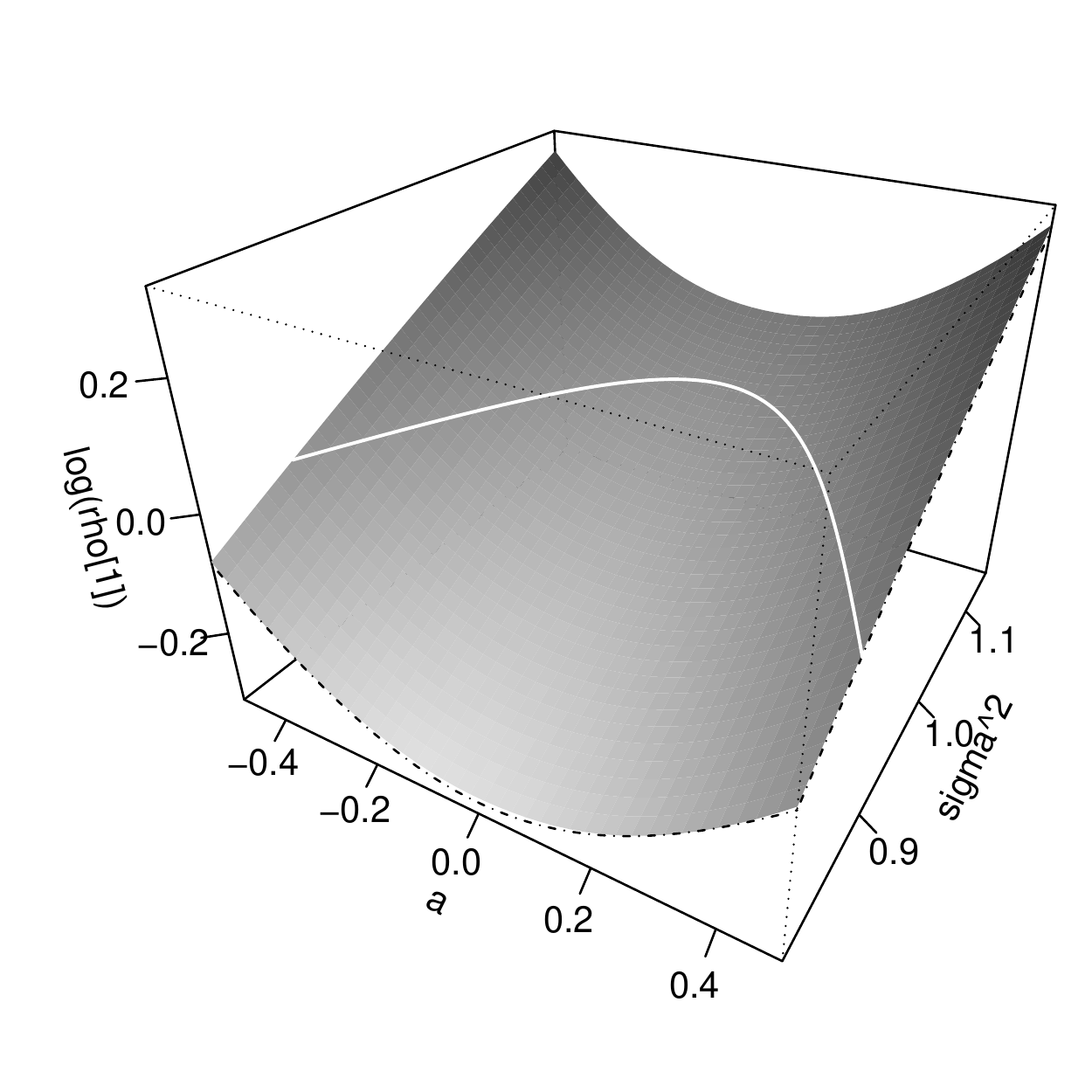}
\end{minipage}	
\end{minipage}								
\begin{minipage}[t]{0.48\textwidth}		
\begin{minipage}[b]{0.001\textwidth}
	{\bf D}\newline\vspace{4cm}

\end{minipage}	
\begin{minipage}[b]{0.99\textwidth}
	\includegraphics[type=pdf,ext=.pdf,read=.pdf,width=\textwidth]{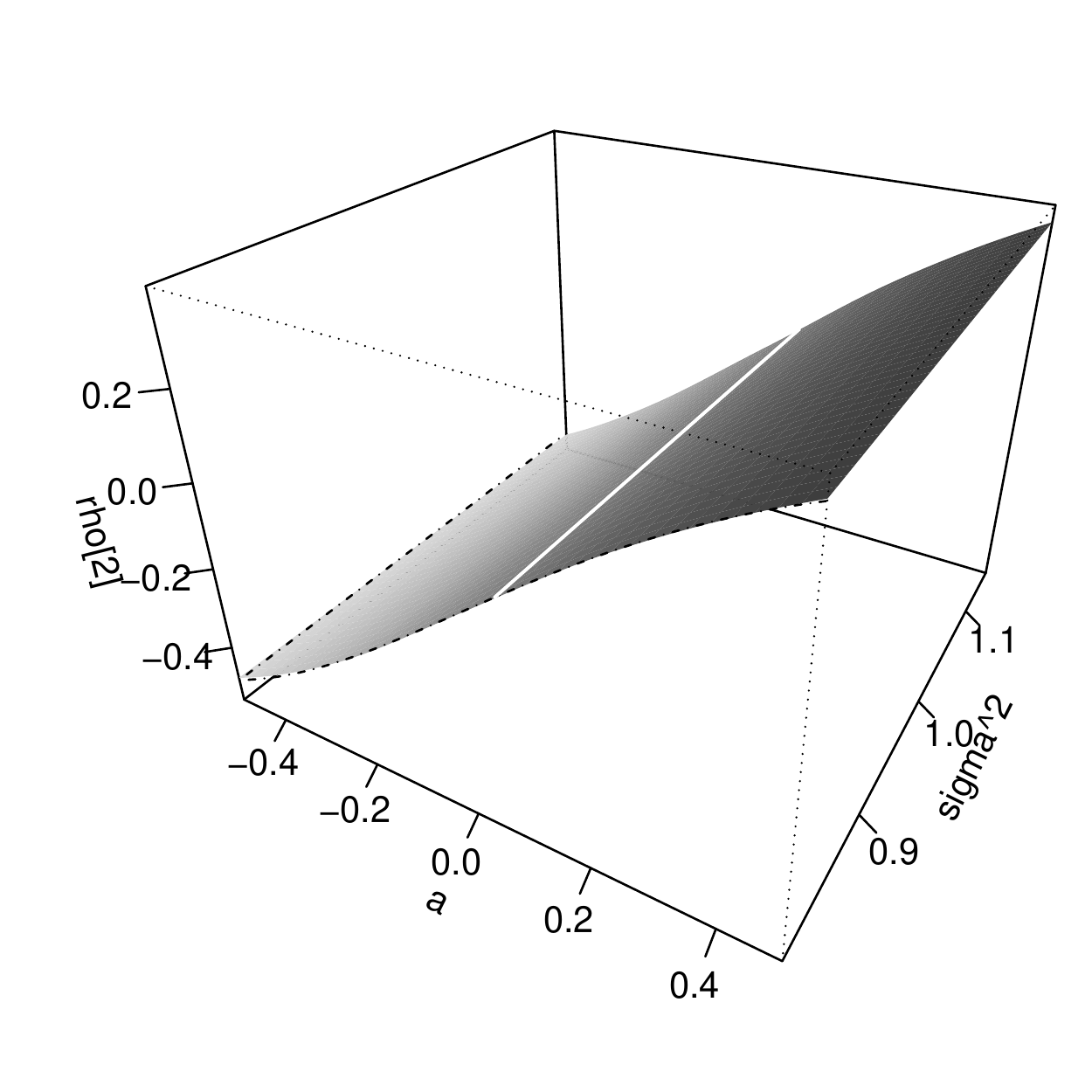}
\end{minipage}	
\end{minipage}	
\end{minipage}	
\caption{
	{\bf ABC self-assessment for ABC inference on the MA(1) model.} 
	We verified during ABC runtime that the test assumptions on the summary values are met. (A) To assess normality, we used the Shapiro-Francia test \citep{Royston1993} to compute a p-value at every ABC iteration for each set of summary values. After runtime, the distribution of p-values was tested for departures from $\mathcal{U}(0,1)$. Histograms of p-values are shown for two cases: summary values set to the thinned time series (light gray) and to the time series without thinning (dark gray). (B) To assess if summary values are uncorrelated, we tested for non-zero partial autocorrelations \citep{Box2011}. Histograms are shown for the same two cases. Summary values can be considered uncorrelated only after thinning. Further, we evaluated if the link function can be numerically reconstructed from ABC output with local polynomial regressions. (C) The first dimension of the reconstructed link function $\hat{\Link}_1$ (gray) and true link function $\Link_1$ (black, dashed-dotted, on the sides), with added level set $\Link_1^{-1}(\rho_1^\star)$ (white). (D) The second dimension of the reconstructed link function $\hat{\Link}_2$ (gray) and true link function $\Link_2$ (black, dashed-dotted, on the sides), with added level set $\Link_2^{-1}(\rho_1^\star)$ (white).
}\label{f:MA1a}
\end{figure}

\begin{table}
	\caption{\label{t:flu example full} SEIIRS model parameters, priors, and standard ABC and \nABC\ posterior densities}	
	{\footnotesize
	\begin{tabular}{lllll}
		\toprule
		&{\bf description}					&{\bf prior density}						&\multicolumn{2}{c}{{\bf mean$\pm$std. dev., 95\% conf. interval of}}\\[2mm]	
		&								&									&{\bf standard ABC}				&{\bf \nABC}\\
		&								&									&{\bf posterior density}			&{\bf posterior density}\\[1mm]
		&								&									&\multicolumn{2}{c}{$\theta_0=(\Ro, 1/\wane, \repR)=(3.5, 10, 0.08)$}\\
		\noalign{\smallskip}\cmidrule(r){4-5}\noalign{\smallskip}			
		$\Ro$		&Basic reproductive					&$\mathcal{U}(1,8)$						&3.74$\pm$0.39, [3.01, 4.37]			&3.51$\pm$0.02, [3.47, 3.54]\\ 
		&number							&									&								&\\[2mm]	
		$1/\dincub$	&Avg incubation					&0.9	$^\S$							&								&\\
		&period [day]						&									&								&\\[2mm]							
		$1/\di$		&Avg infectiousness					&1.8	$^\S$							&								&\\
		&period [day]						&									&								&\\[2mm]	
		$1/\wane$	&Avg duration						&$\mathcal{U}(2,30)$					&11.1$\pm$1.58, [8.15, 13.76]			&9.93$\pm$0.07, [9.79, 10.07]\\ 
		&of immunity [year]					&									&								&\\[2mm]				
		$\repR$ 		&Reporting rate					&$\mathcal{U}(0,1)$						&0.092$\pm$0.02,					&0.0799$\pm 0.0004$,\\
		&								&									&[0.058, 0.133]					&[0.0791, 0.0808]\\[2mm]
		$N^\snk$		&Size of sink 						&$\approx15\times$10$^6$				&								&\\
		&population						&varies in time$^\dagger$				&								&\\[2mm]
		$N^\src$		&Size of source 					&$\mathcal{U}(1,10)$					&$^\ast$							&$^\ast$\\
		&population						&$\times$10$^8$						&								&\\[2mm]	
		$\birth^\snk$	&Birth/death rate of					&fixed$^\dagger$						&								&\\ 
		&sink population					&									&								&\\[2mm]	
		$\birth^\src$	&Birth/death rate of					&1/50 $^\P$							&								&\\ 
		&source population,					&									&								&\\
		&[1/year]							&									&								&\\[2mm]
		$\seas^\snk$	&Seasonal forcing of				&$\mathcal{U}(0.07,0.6)$					&$^\ast$							&$^\ast$\\
		&$\transtwo{t}{\snk}$					&									&								&\\[2mm]		
		$\seas^\src$	&Seasonal forcing of				&$\mathcal{U}(0,0.03)$					&$^\ast$							&$^\ast$\\
		&$\transtwo{t}{\src}$					&									&								&\\[2mm]	
		$\trav^\snk$	&Number of travelers				&$\mathcal{U}(3,15)$					&$^\ast$							&$^\ast$\\
		&visiting the sink					&$\times 10^6$	$^\dagger$				&								&\\
		&population$^\dagger$				&									&								&\\[2mm]
		$\trav^\src$	&Fraction of $\hat{I}^\src$ re-			&$\mathcal{U}(0,0.1)$					&$^\ast$							&$^\ast$\\
		&seeding the source 				&									&								&\\	
		&population						&									&								&\\
		\midrule	
		\multicolumn{5}{l}{$M^\src$ in the analogue of \eqref{e:classofmodels} for the source population is defined by $M^\src=\trav^\src\hat{I}^\src$ where $\hat{I}^\src$ is the}\\
		\multicolumn{5}{l}{number of infected individuals at disease equilibrium in the source. $^\dagger$Fixed to demographic data}\\
		\multicolumn{5}{l}{http://statline.cbs.nl. Number of travellers encompass annual records. $^\S$Fixed to match influenza}\\
		\multicolumn{5}{l}{A (H3N2)'s estimated generation time $^\P$Assuming an average lifespan of 60 years, adjusted by}\\
		\multicolumn{5}{l}{net fertility rate in SE Asia $^\ast$For the simulated data set, these parameters were fixed to the}\\
		\multicolumn{5}{l}{values reported in Figure~\ref{f:SEIIRS_datasets}.}\\
		\bottomrule
	\end{tabular}
}
\end{table}

\begin{figure}[ht]
	\centering		
	\begin{minipage}[c]{0.8\textwidth}	
	\begin{minipage}[c]{\textwidth}	
	\begin{minipage}[b]{0.04\textwidth}
		{\bf A}\newline\vspace{4.7cm}

	\end{minipage}	
	\begin{minipage}[b]{0.95\textwidth}
		\includegraphics[type=pdf,ext=.pdf,read=.pdf,width=\textwidth]{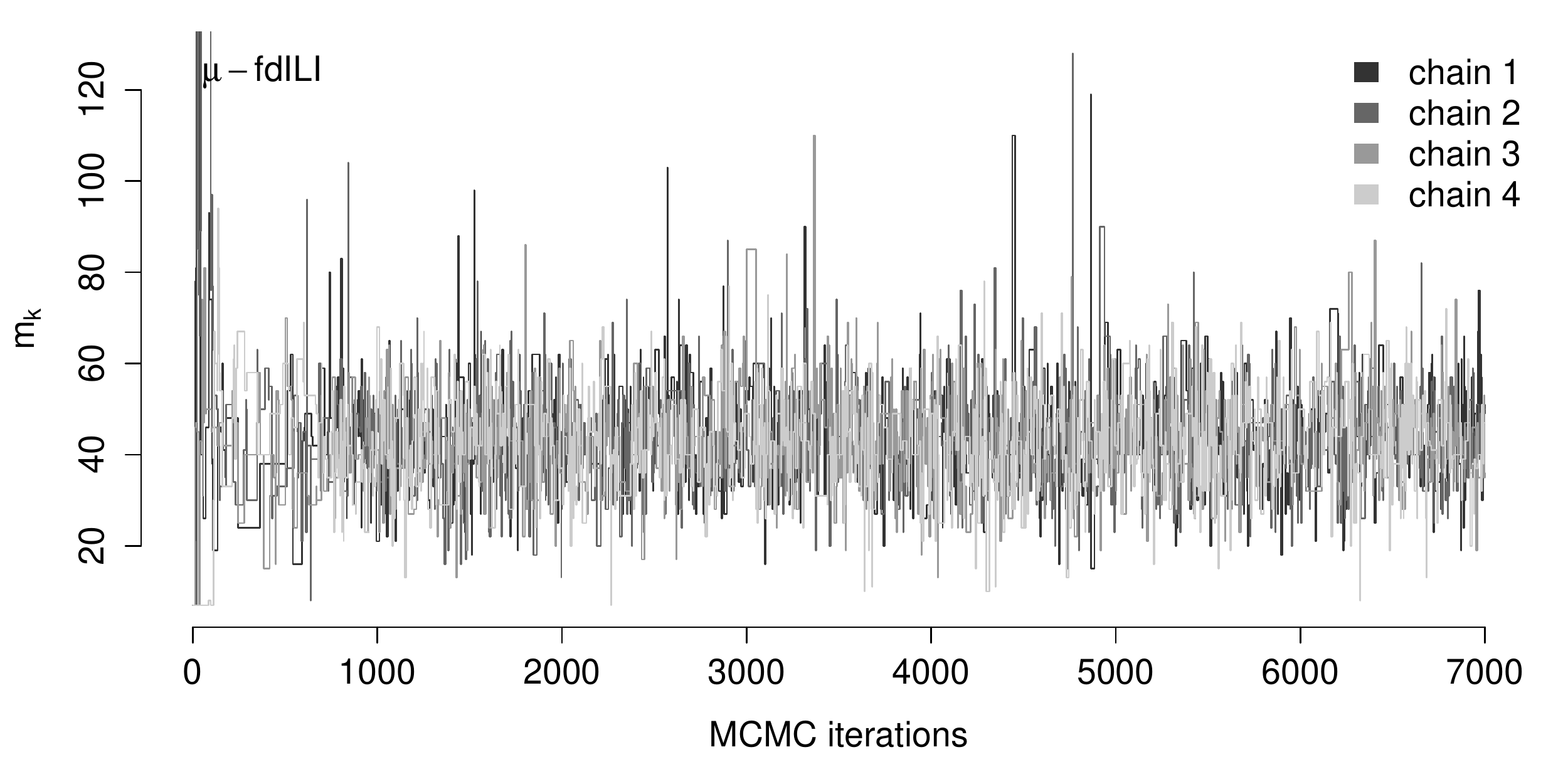}
	\end{minipage}			
\end{minipage}		
\begin{minipage}[c]{\textwidth}	
\begin{minipage}[b]{0.04\textwidth}
	{\bf B}\newline\vspace{4.7cm}

\end{minipage}	
\begin{minipage}[b]{0.95\textwidth}
	\includegraphics[type=pdf,ext=.pdf,read=.pdf,width=\textwidth]{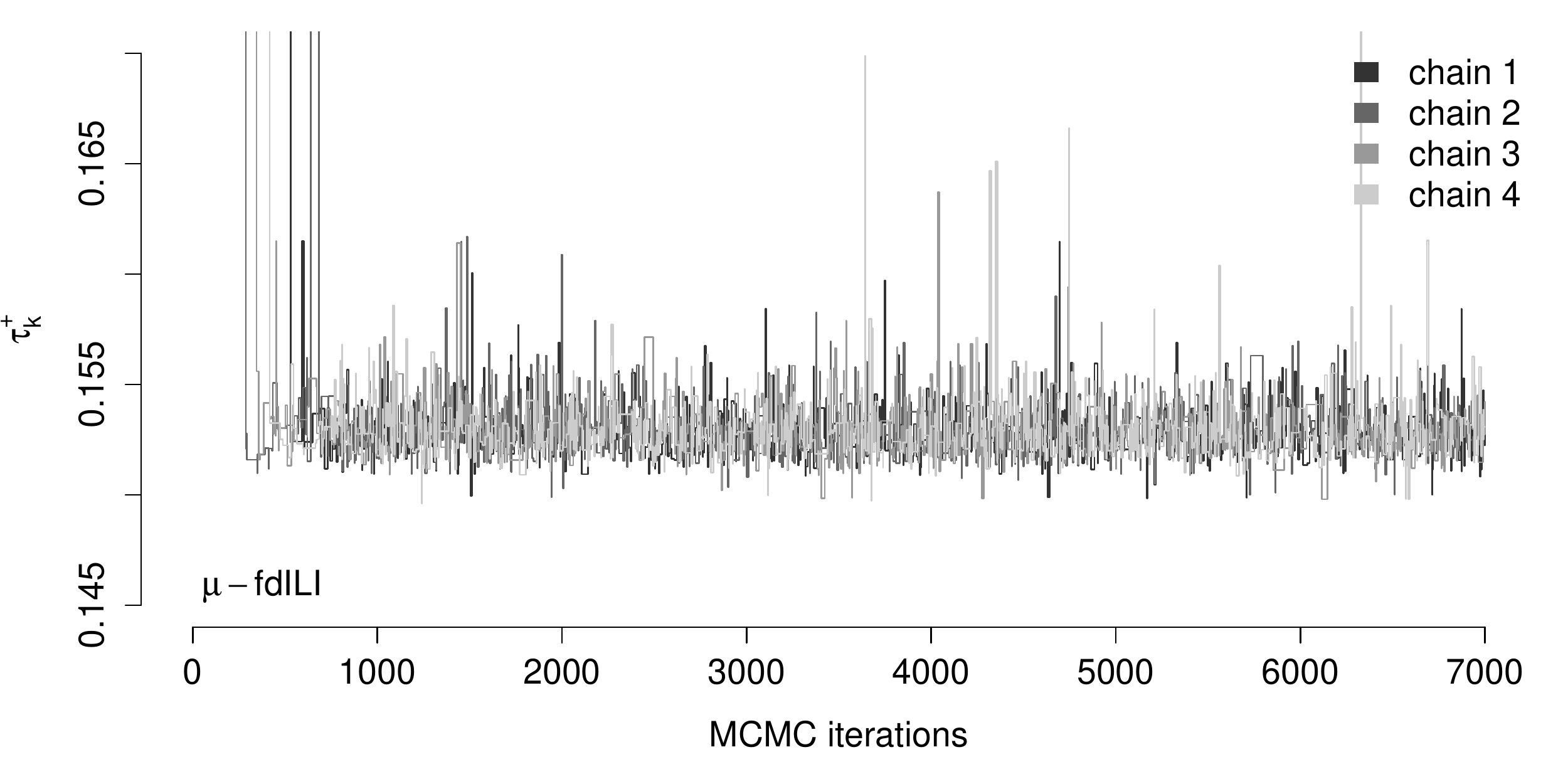}
\end{minipage}			
\end{minipage}	
\end{minipage}	
\caption{
	{\bf ABC inference on the influenza A (H3N2) data on simulated data - calibrations.} \nABC\ inference based on the summary values in Figure~\ref{f:SEIIRS_datasets}. TOST test statistics were used based on the distribution of the summary vales, and MCMC chains were run for 10.000 iterations. The free ABC parameters were re-calibrated at every iteration because the TOST also depends on the standard deviation of the summary values. (A) Calibrated $m_k$ and (B) calibrated $\ithresh^+_k$ for four MCMC chains that were run in parallel from overdispersed starting values. }\label{f:SEIIRS_simCali}
\end{figure}